\documentclass[11pt]{article}
\usepackage{soul}
\usepackage{epsfig,subfigure}
\usepackage{amssymb, amsmath}
\usepackage{color} 
\usepackage{hyperref}  
\newcommand{\define}{\stackrel{\triangle}{=}}
\def\QED{\mbox{\rule[0pt]{1.5ex}{1.5ex}}}
\def\proof{\noindent\hspace{2em}{\it Proof: }}

\usepackage{xcolor,cancel}
\usepackage{graphicx}
\usepackage{color}

\usepackage{amssymb}
\usepackage{amsmath,amsfonts,amssymb}

\usepackage{verbatim}
\usepackage{stfloats}
\usepackage[bookmarks=false]{}

\usepackage{tikz}
\usetikzlibrary{patterns}
\usepackage[normalem]{ulem}
\usepackage{subfigure}

\usepackage{hyperref}
\usepackage{epsfig,subfigure}
\usepackage{color}
\usetikzlibrary{positioning}
\usetikzlibrary{decorations.markings}

\usepackage[font=small,skip=0pt]{caption}

\usepackage{enumerate}

\usepackage{xcolor,cancel}

\newtheorem{theorem}{Theorem}

\newtheorem{definition}{Definition}
\newtheorem{lemma}{Lemma}

\begin{document}
\tikzset{->-/.style={decoration={
  markings,
  mark=at position #1 with {\arrow{>}}},postaction={decorate}}}

\date{}
\title{Optimality of Simple Layered Superposition Coding in the $3$ User MISO BC with Finite Precision CSIT}
\author{ \normalsize Arash Gholami Davoodi and Syed A. Jafar \\
{\small Center for Pervasive Communications and Computing (CPCC)}\\
{\small University of California Irvine, Irvine, CA 92697}\\
{\small \it Email: \{gholamid, syed\}@uci.edu}
}
\maketitle

\begin{abstract}
We study the $K=3$ user multiple input single output (MISO) broadcast channel (BC) with $M=3$ antennas at the transmitter and $1$ antenna at each receiver, from the generalized degrees of freedom (GDoF) perspective, under the assumption that the channel state information at the transmitter (CSIT) is limited to finite precision. In particular, our goal is to identify a parameter regime where a simple layered superposition (SLS) coding scheme achieves the entire GDoF region.  With $\alpha_{ij}$ representing the channel strength parameter for the link from the $j^{th}$ antenna of the transmitter to the $i^{th}$ receiver, we prove that SLS is GDoF optimal without the need for time-sharing if  $\max(\alpha_{ki},\alpha_{im})\leq\alpha_{ii}$ and $\alpha_{ki}+\alpha_{im}\le\alpha_{ii}+\alpha_{km}$ for all $i,k\in[3],m\in[M]$. The GDoF region under this condition is a convex polyhedron. The result generalizes to arbitrary $M\geq 3$.
\end{abstract}

\section{Introduction}
Capacity characterizations of broadcast and interference channels are among the most important open problems in network information theory. Especially significant for wireless networks are the $K$ user Gaussian interference channel (IC) and the corresponding $K$ user MISO BC that is obtained by allowing full cooperation between all the transmitters of the $K$ user interference channel. Macroscopic insights into the performance limits of wireless networks can be obtained through generalized degrees of freedom (GDoF) studies. These studies often lead to sophisticated but fragile schemes such as dirty paper coding, zero forcing, and interference alignment which have limited practical relevance. Arguably what matters most in practice is \emph{robustness} and \emph{simplicity}. 

For \emph{robust} insights it is desirable to restrict the channel state information at the transmitter(s) (CSIT) to finite precision. However, in spite of the tremendous practical significance of the finite precision CSIT assumption, finding tight information theoretic bounds under this model has been surprisingly challenging even in the DoF sense. The difficulty is exemplified by the conjecture of Lapidoth, Shamai, and Wigger \cite{Lapidoth_Shamai_Wigger_BC} made at Allerton 2005 (also a featured open problem at the inaugural ITA workshop in 2006) that the DoF of a MISO BC should collapse to unity under finite precision CSIT. The conjecture remained open for nearly a decade in spite of a variety of efforts that include --- employing the Csiszar sum lemma in the original work by Lapidoth, Shamai and Wigger \cite{Lapidoth_Shamai_Wigger_BC} which produced a loose outer bound; harnessing extremal inequalities in \cite{Clerckx_hybrid} by Rassouli and Clerckx  which could not effectively accommodate channel uncertainty; extension to a stronger conjecture in \cite{Weingarten_Shamai_Kramer} by Weingarten, Shamai and Kramer under a compound setting, where the channel states are drawn from a set of large but finite cardinality (the conjecture under the compound setting was shown to be false by Gou, Jafar and Wang in \cite{Gou_Jafar_Wang} and by Maddah-Ali in \cite{Maddah_Compound}); extension to a stronger conjecture under the PN setting in \cite{Tandon_Jafar_Shamai_Poor} by Tandon, Jafar, Shamai and Poor, where perfect CSIT is available for one user and no CSIT for another (this conjecture also remained open); and blind interference alignment schemes by Jafar \cite{Jafar_corr} that achieve more than $1$ DoF under finite precision CSIT but only if different users experience different channel coherence patterns.  The conjectures were settled in the affirmative in 2016 in \cite{Arash_Jafar_PN} based on an aligned image sets (AIS) argument. The approach taken by the AIS argument is essentially a combinatorial accounting of the number of codewords that can align at one receiver while remaining resolvable at another receiver, under finite precision CSIT. Over $n$ channel uses, this number is bounded by $O(\log(\mbox{SNR})^n)$ so that its contribution to rate is bounded by $O(\log\log(\mbox{SNR}))$ which is negligible in the DoF sense, thus proving that the DoF do collapse as conjectured. Since its introduction,  generalizations of the AIS argument have produced GDoF characterizations for various canonical settings that include --- $2$ user fully asymmetric MISO BC parameterized by arbitrary channel strength levels and arbitrary channel uncertainty levels in \cite{Arash_Bofeng_Jafar_BC}; $K$ user symmetric MISO BC under arbitrary cross channel strength and channel uncertainty levels also in \cite{Arash_Bofeng_Jafar_BC}; $K$ user MIMO interference channel under finite precision CSIT and symmetric channel strengths in \cite{Arash_Jafar_KMIMOIC}; and the $2$ user symmetric MIMO interference channel under arbitrary cross channel strength and channel uncertainty levels \cite{Arash_Jafar_MIMOICGDoF}. AIS has also been employed recently in the context of topological interference management \cite{Jafar_TIM} to settle open problems highlighted by Naderializadeh and Avestimehr in \cite{Naderi_Avestimehr} and a conjecture by Gou et al. in \cite{Gou_TIM}. In order to facilitate direct applications of  AIS arguments in the future, a collection of basic sumset inequalities based on AIS is presented in \cite{Arash_Jafar_sumset} as essential instruments for robust GDoF bounds. To illustrate their utility, in this work will directly utilize these  sumset inequalities to prove our outer bounds.

In addition to robustness, the second issue that motivates this work is the need for \emph{simple} schemes. In particular, the need for simplicity motivates the search for broad regimes where simple coding schemes are provably optimal.  As a case in point, for the $K$ user interference channel, this approach is exemplified by recent studies that have found broad regimes where simple schemes such as orthogonal access \cite{Maleki_Jafar_Convex, Yi_Sun_Jafar_Gesbert} or treating interference as noise (TIN) \cite{Geng_TIN} are optimal in a GDoF sense. Reference \cite{Yi_Sun_Jafar_Gesbert} shows that in a partially connected $K$ user interference network, orthogonal access (such as TDMA) is DoF optimal for all unicast message sets if and only if the network topology is chordal bipartite. Remarkably this also solves the corresponding class of index coding problems due to an equivalence between index coding and topological interference management identified in \cite{Jafar_TIM}. Reference \cite{Geng_TIN} shows that joint power control and treating interference as noise is GDoF optimal in an interference network where the strength of each desired link is stronger than the sum of the strengths of the strongest interference that can be caused by the corresponding transmitter and the strongest interference that can be heard by the corresponding receiver. Notably, these insights have  found use in information-theoretically inspired scheduling algorithms \cite{Naderi_Avestimehr_ITLinQ, Yi_Caire_TIN}. In contrast, for the corresponding $K$ user MISO BC, much less is known about the optimality of simple schemes under finite precision CSIT. This is the motivation for our work. 

Our goal  is to identify broad regimes where simple\footnote{There is no non-trivial regime where TIN is GDoF optimal in the $K$ user MISO BC under finite precision CSIT \cite{Arash_Jafar_TC}. SLS is therefore the natural choice for the simplest scheme of interest.} layered superposition (SLS) coding schemes are GDoF-optimal for the $K$ user MISO BC under finite precision CSIT. By simple layered superposition coding schemes we mean the following. In the $K$ user MISO BC there are $K$ independent messages, one for each receiver. Let us partition each message  into several independent sub-messages,  intended to be decoded by  various subsets of users that must always include the desired user of the original message (cf. Han-Kobayashi scheme for the interference channel \cite{Han_Kobayashi}).  These sub-messages are independently coded. Each transmit antenna sends a weighted sum (\emph{superposition}) of these independent codewords. The weights assigned to the codewords are primarily  for power control. In the GDoF sense, the codewords transmitted from an antenna are mapped to various partitions (\emph{layers}) of the signal dimension according to power levels (cf. ADT deterministic models \cite{Avestimehr_Diggavi_Tse}).   Furthermore, we restrict the codebook design to single-letter\footnote{This rules out multi-letter schemes such as space-time rate-splitting schemes of \cite{Tandon_Jafar_Shamai_Poor, Clerckx_hybrid, Hao_Rassouli_Clerckx} that can potentially outperform single-letter coding schemes.} Gaussian (\emph{simple}) codebooks, over the input random variables corresponding to one channel use. This is the class of coding schemes that we call simple layered superposition, or SLS in short, in this work. 

The possibility that SLS could be GDoF-optimal in the $K$ user MISO BC over a potentially large regime under finite precision CSIT is intriguing. For example, consider the $K=2$ user case. Reference \cite{Arash_Jafar_TC} has shown that SLS achieves the entire GDoF region of the $2$ user MISO BC under finite precision CSIT for \emph{all} choices of channel strength parameters. The optimality of SLS remains unexplored for $K\geq 3$. As the next step forward, in this work we  focus primarily on the $K=3$ user MISO BC setting with finite precision CSIT. The main technical challenge is two-fold. First, we apply recent generalizations of the aligned image sets  \cite{Arash_Jafar_PN,Arash_Bofeng_Jafar_BC,Arash_Jafar_MIMOICGDoF,Arash_Jafar_sumset,Arash_Jafar_Coherence} argument to generate an outer bound. Then, we  prove that in the appropriate parameter regime, the bound is achievable by SLS.\footnote{We also show through an example that strictly tighter GDoF outer bounds may be found outside this regime.} Our main result, reported in Theorem \ref{theorem:GDoF3}, identifies a broad parameter regime where SLS achieves the entire GDoF region. This parameter regime is  significantly larger than the parameter regime where the GDoF optimality of TIN was established for the corresponding $K$ user IC in \cite{Geng_TIN}. A direct representation of the GDoF region in this regime is also presented, which eliminates all power control and rate partitioning variables, automatically optimizing over all such choices within the scope of SLS. In this parameter regime, the GDoF region shows a surprising duality property, i.e., it remains unchanged if the roles of all transmit and receive antennas are switched. Finally, a natural extension of the GDoF outer bounds from the $K=3$ user MISO BC to the $K>3$ user MISO BC is presented in Theorem \ref{theorem:GDoF2}. These bounds may be useful to find a a corresponding parameter regime where SLS is GDoF-optimal in the $K$ user setting.

{\it Notation:} For $n\in\mathbb{N}$, we use the notation $[n]=\{1,2,\cdots,n\}$ and $X^{[n]}=\{X(1), X(2), \cdots, X(n)\}$. The cardinality of a set $A$ is denoted as $|A|$. If $A$ is a set of random variables, then $H(A)$ refers to the joint entropy of the random variables in $A$. Conditional entropies, mutual information and joint and conditional probability densities of sets of random variables are similarly interpreted.  The notation $f(x)=o(g(x))$ denotes that $\limsup_{x\rightarrow\infty}\frac{|f(x)|}{|g(x)|}=0$.  We define $(x)^+=\max(x,0)$. The transpose of a matrix $M$ is represented by $M^{\dagger}$.


\section{Definitions}
The following definitions are needed for aligned image sets arguments.
\begin{definition}\label{bd}{\normalfont (Bounded Density Channel Coefficients \cite{Arash_Jafar_PN})} Define a set of real valued random variables, $\mathcal{G}$ such that the magnitude of each random variable $g\in\mathcal{G}$ is bounded away from  infinity, $ |g|\leq\Delta<\infty$, for some positive constant $\Delta$, and there exists a finite positive constant $f_{\max}$, such that for all finite cardinality disjoint subsets $\mathcal{G}_1, \mathcal{G}_2$ of $\mathcal{G}$, the joint probability density function of all random variables in $\mathcal{G}_1$, conditioned on all random variables in $\mathcal{G}_2$, exists and is bounded above by $f_{\max}^{|\mathcal{G}_1|}$. Without loss of generality we will assume that $f_{\max}\geq 1, \Delta\geq 1$.
\end{definition}

\begin{definition}  Define a set of real valued random variables, $\mathcal{H}$ where each random variable $h\in\mathcal{H}$ is bounded away from  infinity, $ |h|\leq\Delta<\infty$. 
 \end{definition}

\begin{definition}[Power Levels] An integer valued random variable $X$ has power level not more  than $\lambda$ if it takes values over alphabet $\mathcal{X}_{\lambda}$,
\begin{eqnarray}
\mathcal{X}_{\lambda}&\triangleq&\{0,1,2,\cdots,\bar{P}^{\lambda}-1\}
\end{eqnarray}
where $\bar{P}^{\lambda}$ is a compact notation for $\left\lfloor\sqrt{P^{\lambda}}\right\rfloor$.
\end{definition}
Note that if $X\in \mathcal{X}_\lambda$, then it is also true that $X\in\mathcal{X}_{\lambda+\epsilon}$ for all $\epsilon >0$.

\begin {definition}\label{powerlevel} For any nonnegative real numbers $X$,  $\lambda_1$ and $\lambda_2$, define  $(X)_{\lambda_1}$ and $(X)^{\lambda_2}_{\lambda_1}$  as,
 \begin{eqnarray}
(X)_{\lambda_1}&\triangleq& X-\bar{P}^{\lambda_1} \left \lfloor \frac{X}{\bar{P}^{\lambda_1}} \right \rfloor\\
(X)^{\lambda_2}_{\lambda_1}&\triangleq&\left \lfloor \frac{X-\bar{P}^{\lambda_2}\left \lfloor\frac{X}{ {\bar{P}}^{\lambda_2}}\right \rfloor }{{\bar{P}}^{\lambda_1}} \right \rfloor\label{mid}
\end{eqnarray}
\end {definition}
In words, for any $X\in\mathcal{X}_{\lambda_1+\lambda_2}$, $(X)^{\lambda_1+\lambda_2}_{\lambda_1}$ retrieves the top $\lambda_2$ power levels of $X$, while $(X)_{\lambda_1}$ retrieves the bottom $\lambda_1$ levels of $X$.  $(X)^{\lambda_3}_{\lambda_1}$ retrieves only the part of $X$ that lies between power levels $\lambda_1$ and $\lambda_3$.  Note that $X\in \mathcal{X}_\lambda$ can be expressed as $X={\bar{P}^{\lambda_1}}{(X)}_{\lambda_1}^{\lambda}+{(X)}_{\lambda_1}$ for $0\leq\lambda_1<\lambda$.  Equivalently, suppose $X_1\in\mathcal{X}_{\lambda_1}$, $X_2\in\mathcal{X}_{\lambda_2}$, $0<\lambda_2$ and $X=X_1+X_2\bar{P}^{\lambda_1}$. Then $X_1={(X)}_{\lambda_1}$, $X_2={(X)}^{\lambda_1+\lambda_2}_{\lambda_1}$. Also note that if $X\in\mathcal{X}_\lambda$ then $(X)_{\lambda_1}^\lambda=(X)_{\lambda_1}^{\lambda+\epsilon}$ for all $\epsilon>0$.
\begin {definition}\label{deflc}
 For  $x_1,x_2,\cdots,x_k\in\mathcal{X}_{\lambda}$, define  the notations $L_j^b(x_i,1\le i\le k)$ and $L_j(x_i,1\le i\le k)$ as,
\begin {eqnarray}
L^b_j(x_1,x_2,\cdots,x_k )&=&\sum_{1\le i\le k} \lfloor g_{j_i}x_i\rfloor\\
L_j(x_1,x_2,\cdots,x_k)&=&\sum_{1\le i\le k} \lfloor h_{j_i}x_i\rfloor
\end{eqnarray}
for  distinct random variables $g_{j_i}\in\mathcal{G}$,  and $h_{j_i}\in\mathcal{H}$. The subscript $j$ is used to distinguish among various linear combinations. We refer to the {{}$L$ and} $L^b$ functions as the {{}arbitrary linear combinations and}  bounded\footnote{{}Note that throughout this paper, the superscript $(\cdot)^b$ is used to signify the {\bf b}ounded density assumption, which is the most critical assumption about the channel model. Thus, wherever the superscript $(\cdot)^b$ is present, the channel coefficients involved in those expressions are drawn from $\mathcal{G}$ and only their probability density functions are known to the transmitters.} density linear combinations, respectively. 
\end {definition}

\section{System Model} 
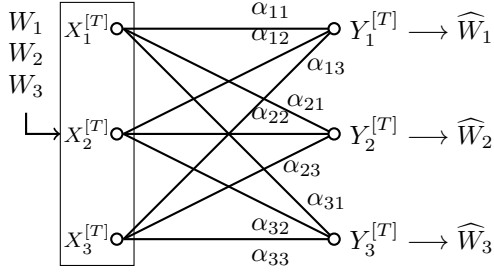
\begin{figure}[h] 
\centering
\begin{tikzpicture}[scale=0.7]
\foreach \m in {1,2,3}
{
\coordinate (M\m) at (1,1.5-2*\m);
\coordinate (N\m) at (5,1.5-2*\m){};
};

\draw [thick] (M1)--(N1) node[above, pos=0.7]{\small $\alpha_{11}$};
\draw[thick] (M2)--(N1) node[above=2pt, pos=0.7]{\small $\alpha_{12}$};
\draw[thick] (M3)--(N1) node[right, pos=0.82]{\small $\alpha_{13}$};

\draw[thick] (M1)--(N2) node[right=2pt, pos=0.7]{\small $\alpha_{21}$};
\draw[thick] (M2)--(N2) node[above, pos=0.7]{\small $\alpha_{22}$};
\draw[thick] (M3)--(N2) node[below, pos=0.85]{\small $\alpha_{23}$};

\draw[thick] (M1)--(N3) node[right, pos=0.82]{\small $\alpha_{31}$};
\draw[thick] (M2)--(N3) node[below, pos=0.7]{\small $\alpha_{32}$};
\draw[thick] (M3)--(N3) node[below, pos=0.7]{\small $\alpha_{33}$};

\draw [thin] (-0.2,0) rectangle (1.2,-5);

\foreach \m in {1,2,3}
{
\node[thick, circle, draw=black, fill=white, inner sep = 1.5, left] at (M\m){};
\node[thick, circle, draw=black, fill=white, inner sep = 1.5] at (N\m){};
}
\node [left=1pt of M1] (X1) {\scriptsize $X_1^{[T]}$};
\node [left=1pt of M2] (X2) {\scriptsize $X_2^{[T]}$};
\node [left=1pt of M3] (X3) {\scriptsize $X_3^{[T]}$};
\node [right=2pt of N1] (Y1) {\small $Y_1^{[T]}\longrightarrow \widehat{W}_1$};
\node [right=2pt of N2] (Y2) {\small $Y_2^{[T]}\longrightarrow \widehat{W}_2$};
\node [right=2pt of N3] (Y3) {\small $Y_3^{[T]}\longrightarrow \widehat{W}_3$};
\node at (-0.85,-1) (W) {\small $\begin{array}{l}W_1\\ W_2\\ W_3\end{array}$};
\draw [thick, ->] (W)|-(-0.2,-2.5);

\end{tikzpicture}
\caption[]{$K=3$ user MISO BC with $M=3$ antennas at the transmitter.}
\label{pco+}
\end{figure}

{\label{sec-sys}}
While in this section we define the system model for arbitrary $K,M$, note that our focus is primarily on the $K=3$ user MISO BC with $M=3$ antennas at the transmitter as shown in Fig. \ref{pco+}.
The channel is defined by the following input-output equation over $T$ channel uses, $t\in[T]$.
\begin{eqnarray}
{Y}_{k}(t)&=&\sum_{m=1}^M\sqrt{P^{\alpha_{km}}}{ G}_{km}(t)X_m(t)+{Z}_{k}(t).
\end{eqnarray}
Over the $t^{th}$ channel use,  $Y_k(t)$ is the signal observed by the $k^{th}$ receiver (user), $k\in[K]$, $Z_k(t)$ is the zero mean unit variance additive white Gaussian noise (AWGN),  $X_m(t)$ is the symbol sent from the $m^{th}$ transmit antenna, $m\in[M]$, $G_{km}(t)$ are  random variables representing the fading coefficient for the channel between the $m^{th}$ antenna of the transmitter and the $k^{th}$ receiver,  $\alpha_{km}$ is the channel strength parameter for the same channel, and $P$ is the nominal power parameter that is allowed to approach infinity in the GDoF limit while the $\alpha_{km}$ parameters are held fixed. The transmitted signals $X_m(t)$ are each subject to unit power constraint. All symbols take real values.

%
%
%

\subsection{CSIT and CSIR}
The channel coefficients are distinct random variables from the bounded density set, i.e., $G_{km}(t)\in\mathcal{G}$, $\forall k\in[K],m\in[M], t\in[T]$. Channel state information at the receivers (CSIR) is assumed to be perfect, while the CSIT is limited to finite precision. Thus, the transmitter is only aware of the joint probability density functions of the channel fading coefficients $G_{km}(t)$ and not the actual realizations of the channel coefficients. The receivers know all channel realizations. 
\subsection{GDoF}
Achievable rates $R_i(P)$ and capacity region $\mathcal{C}(P)$ are defined in the standard Shannon-theoretic sense. The GDoF region is defined as
\begin{align}
\mathcal{D}=&\{(d_1,d_2,\cdots, d_K): \exists (R_1(P),R_2(P), \cdots, R_K(P)) \in\mathcal{C}(P),\nonumber\\
&\mbox{ s.t. } d_k=\lim_{P\rightarrow\infty}\frac{R_k(P)}{\frac{1}{2}\log(P)}, \forall k\in[K]\}.
\end{align}

\subsection{Simple Layered Superposition (SLS) Coding}\label{layer}
Let us partition User $k$'s message as $W_k=(\bar{W}_{S}^k:S\subset[K],k\in S)$. Here, $\bar{W}_{\{k\}}^k$ acts as a  private sub-message to be decoded only
by user $k$ while $\bar{W}_{S}^k$ for $|S|>1$ acts a common sub-message to be decoded by  each User $j$, such that $j\in S$. Further, define $\bar{W}_S=(\bar{W}_{S}^k:k\in S)$. The message $\bar{W}_{S}$ carries $d_{S}$ DoF which may be arbitrarily divided among the users in $S$, so that a fraction $\mu_S^kd_S$ is assigned to user $k$, for each $k\in S$.
 \begin{align}
d_k&=\sum_{S:k\in S}\mu_{S}^kd_{S}\\
\sum_{k:k\in S}\mu_{S}^k&=1\\
\mu_S^k&\geq 0, &&\forall k\in[K], \forall S\subset[K].
\end{align}
 For example, when $K=3$, we have 
 $$W_1=(\bar{W}_{\{1\}}^1,\bar{W}_{\{1,2\}}^1,\bar{W}_{\{1,3\}}^1,\bar{W}_{\{1,2,3\}}^1)$$ 
 $$\bar{W}_{\{1,2,3\}}=(\bar{W}_{\{1,2,3\}}^1,\bar{W}_{\{1,2,3\}}^2,\bar{W}_{\{1,2,3\}}^3)$$
 \begin{align}
d_1&=d_{\{1\}}+\mu_{\{1,2\}}^1d_{\{1,2\}}+\mu_{\{1,3\}}^1d_{\{1,3\}}+\mu_{\{1,2,3\}}^1d_{\{1,2,3\}}\\
d_2&=d_{\{2\}}+\mu_{\{1,2\}}^2d_{\{1,2\}}+\mu_{\{2,3\}}^2d_{\{2,3\}}+\mu_{\{1,2,3\}}^2d_{\{1,2,3\}}\\
d_3&=d_{\{3\}}+\mu_{\{1,3\}}^3d_{\{1,3\}}+\mu_{\{2,3\}}^3d_{\{2,3\}}+\mu_{\{1,2,3\}}^3d_{\{1,2,3\}}\\
1&=\mu_{\{1,2\}}^1+\mu_{\{1,2\}}^2\\
1&=\mu_{\{1,3\}}^1+\mu_{\{1,2\}}^3\\
1&=\mu_{\{2,3\}}^2+\mu_{\{2,3\}}^3\\
1&=\mu_{\{1,2,3\}}^1+\mu_{\{1,2,3\}}^2+\mu_{\{1,2,3\}}^3\\
0&\leq \mu_{\{1,2\}}^1, \mu_{\{1,2\}}^2, \mu_{\{1,3\}}^1, \mu_{\{1,3\}}^3, \mu_{\{2,3\}}^2, \mu_{\{2,3\}}^3, \mu_{\{1,2,3\}}^1, \mu_{\{1,2,3\}}^2, \mu_{\{1,2,3\}}^3
\end{align}


 Messages $\bar{W}_{\{1\}}$, $\bar{W}_{\{2\}}$, $\bar{W}_{\{1,2\}}, \cdots,$ $\bar{W}_{[K]}$ are encoded according to independent Gaussian codebooks into ${X}_{\{1\}},{X}_{\{2\}},{X}_{\{1,2\}},\cdots,{X}_{[K]}$ with powers ${P}^{-\lambda_{\{1\}}},{P}^{-\lambda_{\{2\}}},{P}^{-\lambda_{\{1,2\}}},$ $\cdots,{P}^{-\lambda_{[K]}}$, respectively, such that,
 \begin{align}
\sum_{S\subset [K],k\in S}{P}^{-\lambda_S}&\le 1,~\forall k\in[K].
 \end{align}
The transmitted and received signals are,
 \begin{eqnarray}
X_m&=&\sum_{S\subset[K]}\sqrt{P^{-\gamma_{m,S}}}{X}_S, \forall m\in[M],\\
Y_k&=&\sum_{m\in[M]}\sqrt{P^{\alpha_{km}}}G_{km}X_m+Z_k, \forall k\in[K],
\end{eqnarray}
where $\lambda_S$, $\gamma_{k,S}$ and $d_S$  are some arbitrary non-negative numbers depending on $S$ which should be optimized for each point in the GDoF region separately.  Note that power control is integral to SLS.

\section{Main Result}
\begin{definition}\label{defdelta_}Define the parameters $\delta_{i}$, $\delta_{i,j}$ and $\delta$ as follows.
\begin{eqnarray}
\delta_{i}&=&\max_{m\in[M]}\alpha_{im}, \forall i\in[K],\\
\delta_{i,j}&=&\max_{m\in[M]}{(\alpha_{im}-\alpha_{jm})}^+, \forall i,j\in[K], i\neq j,\label{deltaij0}\\
\delta&=&\min_{\{i,j,k\}=[3]}\min\left(\delta_{i}+\delta_{j,i}+\delta_{k,j},\frac{\delta_{i}+\delta_{k}+\delta_{i,j}+\delta_{j,i}+\delta_{j,k}+\delta_{k,i}}{2}\right).\label{def_delta}
\end{eqnarray}
\end{definition}
\subsection{Three User MISO BC}
\begin{theorem}\label{theorem:GDoF3} In the $K=3$ user MISO BC with $M=3$ transmit antennas defined in Section \ref{sec-sys}, if the following conditions are satisfied for all $i,k\in[3],m\in[M]$,
\begin{eqnarray}
\max(\alpha_{im},\alpha_{ki})\le\alpha_{ii}, \label{con1}\\
\alpha_{ki}+\alpha_{im}\le\alpha_{ii}+\alpha_{km}, \label{con}
\end{eqnarray}
then simple layered superposition (SLS) coding achieves the whole GDoF region, which is described as follows.
\begin{eqnarray}
\mathcal{D}=\bigg\{(d_1,d_2,d_3)\in\mathbb{R}^3_+, \mbox{ such that }~~ \forall  \mbox{ distinct } i,k\in[3],\nonumber\\
\begin{array}{lll}
d_i&\leq&\delta_{i},\\
d_i+d_k&\leq&\min\left(\delta_{i}+\delta_{k,i}, \delta_{k}+\delta_{i,k}\right),~~~~\\
d_1+d_2+d_3&\leq&\delta\hfill\bigg\},
\end{array}\label{BB}
\end{eqnarray}

\end{theorem}

The following remarks are in order.
\begin{enumerate}
\item The result of Theorem \ref{theorem:GDoF3} generalizes to $M>3$ transmit antennas. The converse proof of Theorem \ref{theorem:GDoF3}, provided in Section \ref{proof3}, allows  $M\geq 3$, and since the achievability proof, presented in Section \ref{achiev}, utilizes only the first three transmit antennas, it applies to $M\geq 3$ as well, simply by switching off the remaining antennas. Note that if \eqref{con1}, \eqref{con} are satisfied, then the GDoF region in \eqref{BB} does not depend on $\alpha_{km}$ for $m>3$.


\item The converse proof of Theorem \ref{theorem:GDoF3} shows  that the region described by \eqref{BB} is a valid outer bound on the GDoF region for all $\alpha_{ij}$ values. The parameter regime identified by \eqref{con1} and \eqref{con} is the regime where the outer bound is  tight, and is achieved by SLS. In this parameter regime,  $\delta_{i}=\alpha_{ii}$ and $\delta_{i,j}=\alpha_{ii}-\alpha_{ji}$. Condition \eqref{con} is illustrated in Fig. \ref{figcon}.
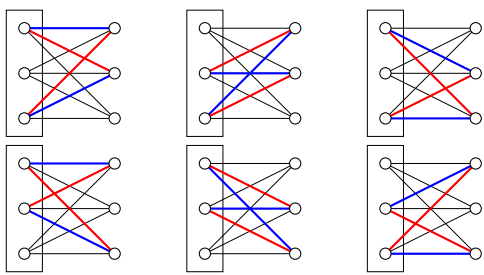
\begin{figure}[h] 
\centering
\begin{tikzpicture}[scale=1.2]
\begin{scope}[shift={(0,0)}]
\foreach \m in {1,2,3}
{
\coordinate (M\m) at (1,-0.5*\m);
\coordinate (N\m) at (2,-0.5*\m){};
};

\foreach \m in {1,2,3}
{ 
	\foreach \n in {1,2,3}
	{
	\draw[ultra thin](M\m)--(N\n);
	}
}

\draw [ultra thin] (0.8,-0.3) rectangle (1.2,-1.7);

\draw[thick, color=blue](M1)--(N1);
\draw[thick, color=blue](M3)--(N2);
\draw[thick, color=red](M3)--(N1);
\draw[thick, color=red](M1)--(N2);

\foreach \m in {1,2,3}
{
\node[circle, draw=black, fill=white, inner sep = 1.5] at (M\m){};
\node[circle, draw=black, fill=white, inner sep = 1.5] at (N\m){};
}
\end{scope}

\begin{scope}[shift={(0,-1.5)}]
\foreach \m in {1,2,3}
{
\coordinate (M\m) at (1,-0.5*\m);
\coordinate (N\m) at (2,-0.5*\m){};
};

\draw [ultra thin] (0.8,-0.3) rectangle (1.2,-1.7);
\foreach \m in {1,2,3}
{ 
	\foreach \n in {1,2,3}
	{
	\draw[ultra thin](M\m)--(N\n);
	}
}
\draw[thick, color=blue](M1)--(N1);
\draw[thick, color=blue](M2)--(N3);
\draw[thick, color=red](M2)--(N1);
\draw[thick, color=red](M1)--(N3);

\foreach \m in {1,2,3}
{
\node[circle, draw=black, fill=white, inner sep = 1.5] at (M\m){};
\node[circle, draw=black, fill=white, inner sep = 1.5] at (N\m){};
}
\end{scope}

\begin{scope}[shift={(2,0)}]
\draw [ultra thin] (0.8,-0.3) rectangle (1.2,-1.7);
\foreach \m in {1,2,3}
{
\coordinate (M\m) at (1,-0.5*\m);
\coordinate (N\m) at (2,-0.5*\m){};
};

\foreach \m in {1,2,3}
{ 
	\foreach \n in {1,2,3}
	{
	\draw[ultra thin](M\m)--(N\n);
	}
}
\draw[thick, color=blue](M2)--(N2);
\draw[thick, color=blue](M3)--(N1);
\draw[thick, color=red](M2)--(N1);
\draw[thick, color=red](M3)--(N2);

\foreach \m in {1,2,3}
{
\node[circle, draw=black, fill=white, inner sep = 1.5] at (M\m){};
\node[circle, draw=black, fill=white, inner sep = 1.5] at (N\m){};
}
\end{scope}

\begin{scope}[shift={(2,-1.5)}]
\draw [ultra thin] (0.8,-0.3) rectangle (1.2,-1.7);
\foreach \m in {1,2,3}
{
\coordinate (M\m) at (1,-0.5*\m);
\coordinate (N\m) at (2,-0.5*\m){};
};

\foreach \m in {1,2,3}
{ 
	\foreach \n in {1,2,3}
	{
	\draw[ultra thin](M\m)--(N\n);
	}
}
\draw[thick, color=blue](M2)--(N2);
\draw[thick, color=blue](M1)--(N3);
\draw[thick, color=red](M1)--(N2);
\draw[thick, color=red](M2)--(N3);

\foreach \m in {1,2,3}
{
\node[circle, draw=black, fill=white, inner sep = 1.5] at (M\m){};
\node[circle, draw=black, fill=white, inner sep = 1.5] at (N\m){};
}
\end{scope}

\begin{scope}[shift={(4,0)}]
\draw [ultra thin] (0.8,-0.3) rectangle (1.2,-1.7);
\foreach \m in {1,2,3}
{
\coordinate (M\m) at (1,-0.5*\m);
\coordinate (N\m) at (2,-0.5*\m){};
};

\foreach \m in {1,2,3}
{ 
	\foreach \n in {1,2,3}
	{
	\draw[ultra thin](M\m)--(N\n);
	}
}
\draw[thick, color=blue](M3)--(N3);
\draw[thick, color=blue](M1)--(N2);
\draw[thick, color=red](M1)--(N3);
\draw[thick, color=red](M3)--(N2);

\foreach \m in {1,2,3}
{
\node[circle, draw=black, fill=white, inner sep = 1.5] at (M\m){};
\node[circle, draw=black, fill=white, inner sep = 1.5] at (N\m){};
}
\end{scope}

\begin{scope}[shift={(4,-1.5)}]
\draw [ultra thin] (0.8,-0.3) rectangle (1.2,-1.7);
\foreach \m in {1,2,3}
{
\coordinate (M\m) at (1,-0.5*\m);
\coordinate (N\m) at (2,-0.5*\m){};
};

\foreach \m in {1,2,3}
{ 
	\foreach \n in {1,2,3}
	{
	\draw[ultra thin](M\m)--(N\n);
	}
}
\draw[thick, color=blue](M3)--(N3);
\draw[thick, color=blue](M2)--(N1);
\draw[thick, color=red](M3)--(N1);
\draw[thick, color=red](M2)--(N3);

\foreach \m in {1,2,3}
{
\node[circle, draw=black, fill=white, inner sep = 1.5] at (M\m){};
\node[circle, draw=black, fill=white, inner sep = 1.5] at (N\m){};
}
\end{scope}

\end{tikzpicture}
\caption[]{The six conditions implied by \eqref{con} are illustrated. The sum of blue channel strengths ($\alpha_{ij}$) must be greater than or equal to the sum of the red channel strengths in each case.}
\label{figcon}
\end{figure}
\item Subject to  conditions \eqref{con1} and \eqref{con}, the GDoF region shows a surprising duality property. Specifically, the GDoF region remains unchanged if the roles of transmitters and receivers are switched, i.e., if $\alpha_{ij}$ and $\alpha_{ji}$ values are switched. The top of Fig. \ref{fig_region} shows an example of a $3$ user MISO BC and its dual. It is easy to verify that conditions \eqref{con1} and \eqref{con} are satisfied and the GDoF region (sketched at the bottom of Fig. \ref{fig_region}) is the following.
\begin{align}
&\mathcal{D}=\{(d_1,d_2,d_3):\nonumber\\
& 0\le d_1\le 1.2,~~0\le d_2\le 1.3,~~0\le d_3\le 1,\nonumber\\
&d_1+d_2\le1.4,~~d_1+d_3\le 1.3,~~d_2+d_3\le 1.4,\nonumber\\
&d_1+d_2+d_3\le 1.6\}\nonumber
\end{align}
It is also easy to construct examples where such a duality does not hold and \eqref{con1} and \eqref{con} are not satisfied. As a simple example, consider the case where all channel strength parameters $\alpha_{ij}=0$ except, $\alpha_{12}=1, \alpha_{21}=2$. Note that \eqref{con1} is not satisfied because $\max(\alpha_{12}, \alpha_{21})=2> 0=\alpha_{11}$. For this example, the GDoF region is $\{(d_1, d_2, d_3): d_1\leq 1, d_2\leq 2, d_3=0\}$, but in its dual (reciprocal) setting the GDoF region is $\{(d_1, d_2,d_3): d_1\leq 2, d_2\leq 1, d_3=0\}$. Thus, the dual setting does not have the same GDoF region.
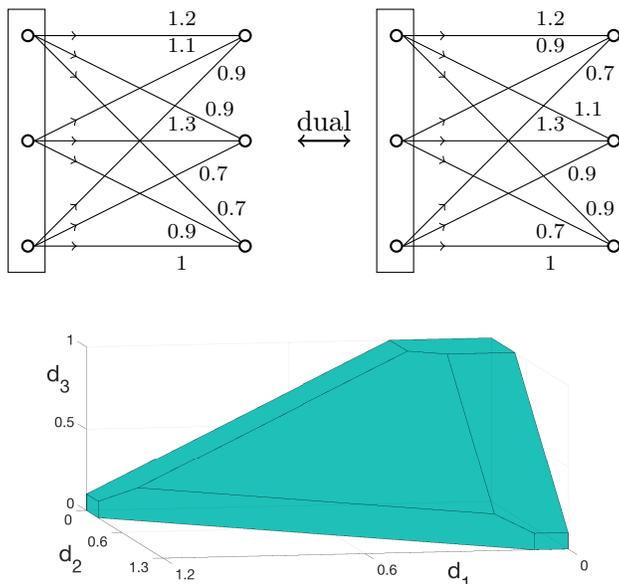
\begin{figure}[h] 
\centering
\begin{tikzpicture}[scale=0.7]

\begin{scope}[shift={(0,0)}]
\foreach \m in {1,2,3}
{
\coordinate (M\m) at (1,1.5-2*\m);
\coordinate (N\m) at (5,1.5-2*\m){};
};

\draw [thin, ->-=0.2] (M1)--(N1) node[above, pos=0.7]{\scriptsize $1.2$};
\draw[thin, ->-=0.2] (M2)--(N1) node[above=2pt, pos=0.7]{\scriptsize $1.1$};
\draw[thin, ->-=0.2] (M3)--(N1) node[right, pos=0.82]{\scriptsize $0.9$};

\draw[thin, ->-=0.2] (M1)--(N2) node[right=5pt, pos=0.7]{\scriptsize $0.9$};
\draw[thin, ->-=0.2] (M2)--(N2) node[above, pos=0.7]{\scriptsize $1.3$};
\draw[thin, ->-=0.2] (M3)--(N2) node[below, pos=0.85]{\scriptsize $0.7$};

\draw[thin, ->-=0.2] (M1)--(N3) node[right, pos=0.82]{\scriptsize $0.7$};
\draw[thin, ->-=0.2] (M2)--(N3) node[below, pos=0.7]{\scriptsize $0.9$};
\draw[thin, ->-=0.2] (M3)--(N3) node[below, pos=0.7]{\scriptsize $1$};

\draw [thin] (0.5,0) rectangle (1.2,-5);

\foreach \m in {1,2,3}
{
\node[thick, circle, draw=black, fill=white, inner sep = 1.5, left] at (M\m){};
\node[thick, circle, draw=black, fill=white, inner sep = 1.5] at (N\m){};
}
\end{scope}

\begin{scope}[shift={(7,0)}]

\foreach \m in {1,2,3}
{
\coordinate (M\m) at (1,1.5-2*\m);
\coordinate (N\m) at (5,1.5-2*\m){};
};

\draw [thin, ->-=0.2] (M1)--(N1) node[above, pos=0.7]{\scriptsize $1.2$};
\draw[thin, ->-=0.2] (M2)--(N1) node[above=2pt, pos=0.7]{\scriptsize $0.9$};
\draw[thin, ->-=0.2] (M3)--(N1) node[right, pos=0.82]{\scriptsize $0.7$};

\draw[thin, ->-=0.2] (M1)--(N2) node[right=5pt, pos=0.7]{\scriptsize $1.1$};
\draw[thin, ->-=0.2] (M2)--(N2) node[above, pos=0.7]{\scriptsize $1.3$};
\draw[thin, ->-=0.2] (M3)--(N2) node[below, pos=0.85]{\scriptsize $0.9$};

\draw[thin, ->-=0.2] (M1)--(N3) node[right, pos=0.82]{\scriptsize $0.9$};
\draw[thin, ->-=0.2] (M2)--(N3) node[below, pos=0.7]{\scriptsize $0.7$};
\draw[thin, ->-=0.2] (M3)--(N3) node[below, pos=0.7]{\scriptsize $1$};

\draw [thin] (0.5,0) rectangle (1.2,-5);

\foreach \m in {1,2,3}
{
\node[thick, circle, draw=black, fill=white, inner sep = 1.5, left] at (M\m){};
\node[thick, circle, draw=black, fill=white, inner sep = 1.5] at (N\m){};
}
\end{scope}

\draw[thick, <->] (6,-2.5)--(7,-2.5) node [midway, above] {dual};

\end{tikzpicture}
\includegraphics[width=0.5\textwidth]{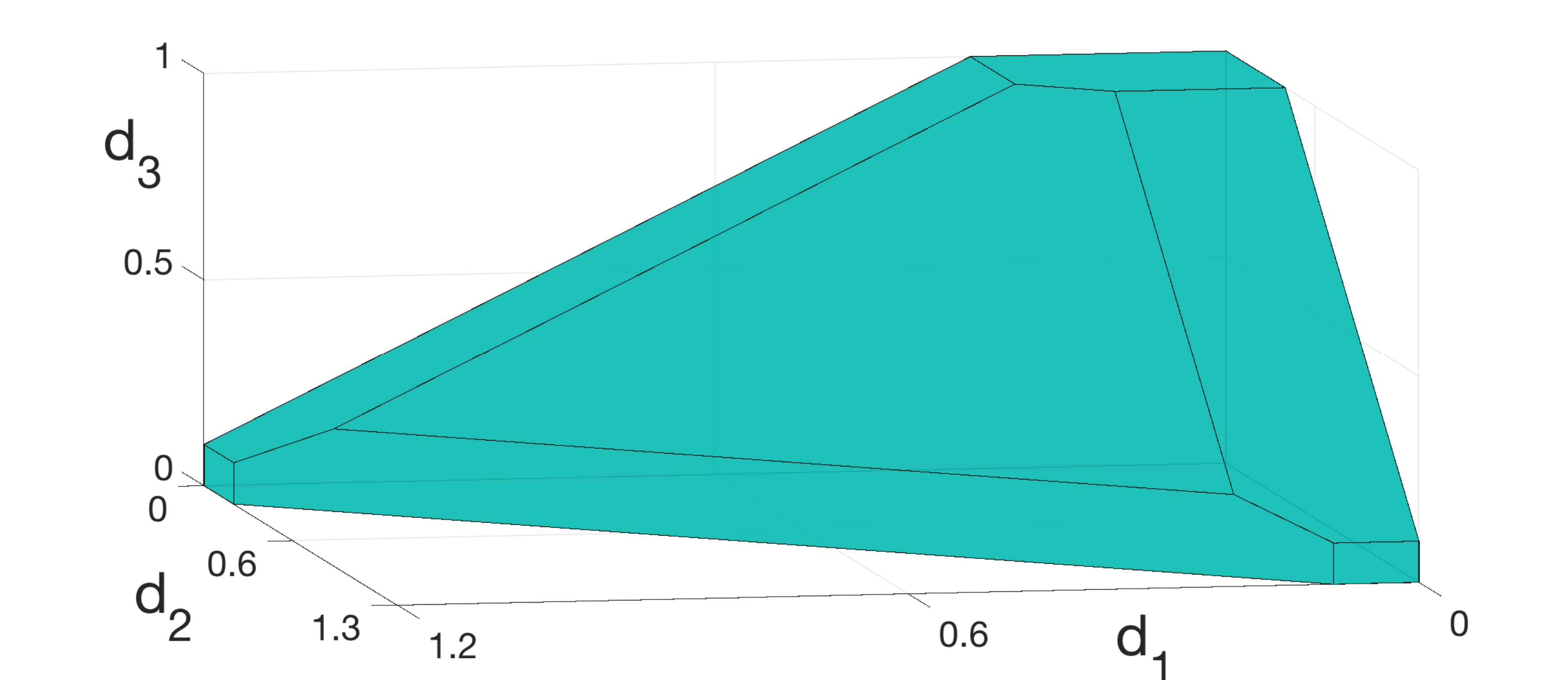}
\caption[]{(Top): A 3-user MISO BC that satisfies the SLS-optimality conditions \eqref{con1} and \eqref{con}, and its dual channel where values of $\alpha_{ij}$ and $\alpha_{ji}$ are switched. The duality property implies that both have the same GDoF region. (Bottom): The GDoF region.}
\label{fig_region}
\end{figure}

\item The GDoF region of a MISO BC does not depend on the labeling of transmit antennas, i.e., it remains the same  if we substitute each $\alpha_{km}$ with $\alpha_{k\pi(m)}$ where $\pi$ is any permutation on $[M]$. However, note that the conditions \eqref{con1} and \eqref{con} do depend on the labeling of transmit antenna indices. Therefore, in order to determine if a given MISO BC setting satisfies the SLS optimality conditions \eqref{con1} and \eqref{con}, it is necessary to check these conditions for all permutations of transmit antenna indices. Furthermore, if the conditions are satisfied for one of these permutations, say $\pi_1$, and not for another permutation, say $\pi_2$, then the duality property described above is claimed for the labeling of transmit antenna according to $\pi_1$, but not for $\pi_2$.

\item It would be useful to consider as a special case of Theorem \ref{theorem:GDoF3}, a $3$ user cyclic $(1,a,b)$ MISO BC  sketched in the left half of Fig. \ref{figcyclic}. In the parameter regime $0\le a\le b\le1$ and $b-a\le 1-b$, the GDoF region for this channel is achieved by SLS and is represented as follows.
\begin{align}
&\mathcal{D}=\{(d_1,d_2,d_3): 0\le d_i\le 1,d_i+d_j\le 2-b,\nonumber\\
&d_1+d_2+d_3\le 3-2b,\forall i,j\in[3],i\neq j\}
\end{align}
\begin{figure}[h] 
\centering
\begin{tikzpicture}
\begin{scope}[shift={(0,3.6)}, yscale=0.8]
\foreach \m in {1,2,3}
{
\coordinate (M\m) at (1,1.5-2*\m);
\coordinate (N\m) at (3,1.5-2*\m){};
};

\draw [thick] (M1)--(N1) node[above, pos=0.7]{\scriptsize $1$};
\draw[thick] (M2)--(N1) node[above, pos=0.7]{\scriptsize $a$};
\draw[thick] (M3)--(N1) node[right, pos=0.87]{\scriptsize $b$};

\draw[thick] (M1)--(N2) node[above, pos=0.8]{\scriptsize $b$};
\draw[thick] (M2)--(N2) node[above, pos=0.7]{\scriptsize $1$};
\draw[thick] (M3)--(N2) node[above, pos=0.75]{\scriptsize $a$};

\draw[thick] (M1)--(N3) node[right, pos=0.85]{\scriptsize $a$};
\draw[thick] (M2)--(N3) node[below, pos=0.7]{\scriptsize $b$};
\draw[thick] (M3)--(N3) node[below, pos=0.7]{\scriptsize $1$};

\draw [thin] (0.8,-0.3) rectangle (1.2,-4.7);

\foreach \m in {1,2,3}
{
\node[circle, draw=black, fill=white, inner sep = 1.5] at (M\m){};
\node[circle, draw=black, fill=white, inner sep = 1.5] at (N\m){};
}
\end{scope}

\begin{scope}[scale=2.5, shift={(1.75,0)}]
\draw[draw=black!70, fill=gray!30!white]
plot[smooth,samples=100,domain=0:1](\x, 0.5+\x/2)--
plot[smooth,samples=100,domain=1:0.5](\x, 2*\x-1)--(0,0);
\draw[draw=gray!50, fill, pattern= north west lines]
	plot[smooth,samples=100,domain=0:0.5](\x,{0})--
	plot[smooth,samples=100,domain=0.5:0](\x,{0.5});
\foreach \x in {0, 0.5, 1}
    \draw (\x,1pt)--(\x,-1pt) node[below] {\scriptsize $\x$};
\foreach \y in {0.5, 1}
    \draw (1pt,\y)--(-1pt,\y) node[left] {\scriptsize $\y$};
\draw[thick,->] (0,0)--(1.2,0) node[below]{ $b$};
\draw[thick,->] (0,0)--(0,1.2) node[left]{ $a$};
\draw[help lines] (0,1)--(1,1)--(1,0);
\draw[help lines](0,0)--(1,1);

\end{scope}
\end{tikzpicture}
\caption[]{(Left): $3$ user cyclic $(1,a,b)$ MISO BC with channel strength levels ($\alpha_{ij}$) shown for each link. (Right): Gray shaded region shows the regime where SLS is optimal in the $3$ user cyclic $(1,a,b)$ MISO BC. The slanted line pattern is the regime where TIN is optimal for the corresponding $3$ user IC.}
\label{figcyclic}
\end{figure}
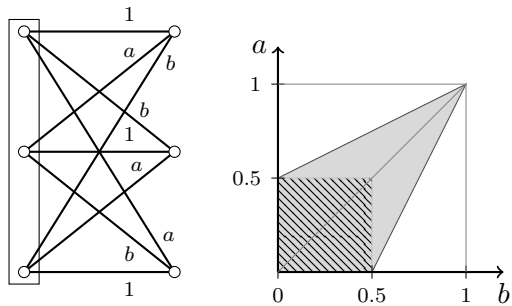
\item  From \cite{Geng_TIN}, TIN  is optimal in the  $3$ user IC if $\max_{j\in[3], j\neq i}\alpha_{ij}+\max_{k\in[3],k\neq i}\alpha_{ki}\le \alpha_{ii}, \forall i\in[3]$.
It is easily observed that the region of $\alpha_{ij}$ where SLS is optimal in the $3$ user MISO BC is larger than the one where TIN is optimal in the corresponding $3$ user IC.  For instance, as shown on the right half of Fig. \ref{figcyclic}, in the $3$ user cyclic $(1,a,b)$ MISO BC, the region in the ($a$,$b$) plane where TIN is optimal has an area of $1/4$ while the region where SLS is optimal has an area of $1/2$. 
\item For a challenging example outside the parameter regime identified by \eqref{con1} and \eqref{con}, consider  the three user cyclic $(1,2,2)$ MISO BC where the condition \eqref{con} is not satisfied. From Theorem \ref{theorem:GDoF3}, the best sum GDoF bound for this channel is equal to $4$. However, this bound is not tight because we are able to establish a tighter bound of $\frac{15}{4}$, see Appendix \ref{154}. Thus, $\eqref{BB}$ does not describe the GDoF region when conditions \eqref{con1}, \eqref{con} are not satisfied.
\end{enumerate}

\subsection{Proof of Theorem  \ref{theorem:GDoF3}: Converse}\label{proof3}
The bounds $d_i\le\delta_{i}$  follow  from the single user bounds. For the remaining bounds, the first step in the converse proof is the transformation into a deterministic setting such that a GDoF outer bound
on the deterministic setting is also a GDoF outer bound on the original setting. This step is identical to \cite{Arash_Jafar_IC}. To avoid repetition, let us start our proof after this step. 

\subsubsection{Deterministic Model}\label{DM_1}
The following input-output relationship holds in the deterministic model,
\begin{align}
\bar{Y}_{k}(t)=&\sum_{m=1}^M\left\lfloor G_{km}(t)\lfloor\bar{P}^{\alpha_{km}-\lambda_m} \bar{X}_m(t)\rfloor\right\rfloor\label{dm1}
\end{align}
for all $k\in[3],t\in[T]$,   where  $\forall m\in[M],t\in[T]$, we have 
\begin{eqnarray}
\bar{X}_{m}(t)&\in&\mathcal{X}_{\lambda_m},\\
\lambda_m&\triangleq&{\max_{k\in[3]}\alpha_{km}},\\
\lambda&\triangleq&{\max_{m\in[M]}\lambda_{m}}.
\end{eqnarray}
Thus, the signal from the $m^{th}$ transmit antenna, $\bar{X}_m$, has power level no more than $\lambda_m$, which is the highest power level with which $\bar{X}_m$ can be heard by any receiver $k$, $k\in[3]$. Furthermore, $\lambda$ is the maximum of all $\lambda_m$, so that for all $m\in[M]$, we can also write that $\bar{X}_{m}(t)\in\mathcal{X}_{\lambda}$.  Note that  \eqref{dm1} can be equivalently\footnote{From  \eqref{mid} and \eqref{dm1}, we have
$\bar{Y}_{k}(t)=\sum_{m=1}^M\left\lfloor G_{km}(t) \left \lfloor\frac{\bar{X}_m(t)}{\bar{P}^{\lambda_m-\alpha_{km}}}\right \rfloor\right\rfloor=\sum_{m=1}^M\left\lfloor G_{km}(t) (\bar{X}_m(t))_{\lambda_m-\alpha_{km}}^{\lambda_m} \right\rfloor=\sum_{m=1}^M\left\lfloor G_{km}(t) (\bar{X}_m(t))_{\lambda_m-\alpha_{km}}^{\lambda} \right\rfloor$.}
 expressed as follows.
\begin{eqnarray}
\bar{Y}_{k}(t)=&\sum_{m=1}^M\left\lfloor G_{km}(t)(\bar{X}_m(t))_{\lambda_m-\alpha_{km}}^\lambda\right\rfloor.
\end{eqnarray}

\subsubsection{A Key Lemma and an Observation} \label{secdelta+}
To invoke the aligned image sets argument, we need the following lemma from \cite{Arash_Jafar_KMIMOIC}.
\begin{lemma}\label{lemmaxc}{\normalfont (\cite{Arash_Jafar_KMIMOIC}, Lemma $1$)} Define the two random variables $\bar{\bf U}_1$ and  $\bar{\bf U}_2$ as,
\begin{eqnarray}
\bar{\bf U}_1&=&\left({U}_{11}^{[T]},{U}_{12}^{[T]},\cdots,{U}_{1N}^{[T]}\right)\label{lemmamimox1}\\
\bar{\bf U}_2&=&\left({U}_{21}^{[T]},{U}_{22}^{[T]},\cdots,{U}_{2N}^{[T]}\right)\label{lemmamimox2}
\end{eqnarray}
where for all $n\in[N]$, $t\in[T]$, $U_{1n}(t)$ and $U_{2n}(t)$ are defined as,
\begin{eqnarray}
U_{1n}(t)&=&L_{1n}^b(t)\left((\bar{{V}}_1(t))^{\eta}_{\eta-\lambda_{11}},(\bar{{V}}_2(t))^{\eta}_{\eta-\lambda_{12}},\cdots,(\bar{{V}}_M(t))^{\eta}_{\eta-\lambda_{1M}}\right),\label{lemmamimox3}\\
U_{2n}(t)&=&L_{2n}^b(t)\left((\bar{{V}}_1(t))^{\eta}_{\eta-\lambda_{21}},(\bar{{V}}_2(t))^{\eta}_{\eta-\lambda_{22}},\cdots,(\bar{{V}}_M(t))^{\eta}_{\eta-\lambda_{2M}}\right). \label{lemmamimox4}
\end{eqnarray}
The $\bar{V}_{m}(t)\in\mathcal{X}_{\eta}$, $m\in[M]$,  are all independent of $\mathcal{G}$, and $\forall m\in[M]$, $0\le\lambda_{1m},\lambda_{2m}\le\eta$.    Without loss of generality,  $(\lambda_{1m}-\lambda_{2m})^+$ are sorted in descending order, i.e., $(\lambda_{1m}-\lambda_{2m})^+\ge(\lambda_{1m'}-\lambda_{2m'})^+$ if $1\le m< m'\le M$.  For any acceptable\footnote{Let $\mathcal{G}(Z)\subset\mathcal{G}$ denote the set of all bounded density channel coefficients that appear in $\bar{\bf U}_1,\bar{\bf U}_2$.  $W$ is acceptable if  conditioned on any $\mathcal{G}_o\subset (\mathcal{G}/\mathcal{G}(Z))\cup \{W\}$, the channel coefficients $\mathcal{G}(Z)$ satisfy the bounded density assumption. For instance, any random variable $W$ independent of $\mathcal{G}$ can be utilized in  Lemma \ref{lemmaxc}.} random variable ${W}$, if $N\le M$, then we have,
\begin{eqnarray}
H({\bar{\bf U}}_1\mid {W},\mathcal{G})-H({\bar{\bf U}}_2\mid {W},\mathcal{G})&\le&T\left(\sum_{m=1}^N(\lambda_{1m}-\lambda_{2m})^+\right)\log{\bar{P}}+T~o~(\log{\bar{P}}).\label{lemmamimox5}
\end{eqnarray}  
\end{lemma} 
{{}Lemma \ref{lemmaxc} is a simple generalization from $M=2$ to $M>2$ of the bound in \cite{Arash_Jafar_TC}. For proof of Lemma \ref{lemmaxc}. see \cite{Arash_Jafar_KMIMOIC}. The proof is presented in  Appendix \ref{lemmaxcp} for the sake of completeness}. Lemma \ref{lemmaxc} may be intuitively understood as follows. Consider a transmitter with $M$ antennas, with transmit symbol $\bar{V}_m$ originating at the $m^{th}$ antenna, $m\in[M]$. The transmitted signals have power levels no more than $\eta$. Consider $2$ receivers, $\bar{\bf U}_1, \bar{\bf U}_2$, each equipped with $N$ receive antennas, that see different bounded density linear combinations of the $M$ transmitted symbols, scaled by channels of different strengths, so that the $n^{th}$ receive antenna of the $k^{th}$ receiver, $k\in[2]$, sees only the power levels above $\eta-\lambda_{km}$ of the transmitted signal $\bar{V}_m$. If the CSIT is limited to finite precision, CSIR is perfect, and $N\leq M$, then the greatest difference in entropies that can be created between the two receivers in the GDoF sense is no more than the sum of the $N$ largest terms of the pairwise differences between strengths of signals seen at the two receivers from the same transmit antenna. The random variable $W$ generalizes this statement to conditional entropies provided that the bounded density character of the linear combinations is maintained even after conditioning on $W$.

Now consider the specialization of Lemma \ref{lemmaxc} to the system model in this paper. Our transmitter has $M$ antennas, each receiver has $N=1$ antenna, all transmitted signals $\bar{V}_m=\bar{X}_m, m\in[M]$, have power levels no more than $\eta=\lambda$, and the $k^{th}$ receiver sees only the power levels above $\lambda_m-\alpha_{km}$ from $\bar{X}_m$, so that 
\begin{eqnarray}
\eta-\lambda_{km}&=&\lambda_m-\alpha_{km}\\
\Rightarrow \lambda_{km}&=&\eta-\lambda_m+\alpha_{km}\\
&=&\lambda-\lambda_m+\alpha_{km}.\label{eq:lkm}
\end{eqnarray} 
Furthermore, $M> N$, the CSIT is limited to finite precision, and the CSIR is perfect.
Therefore, for any $k_1, k_2\in[3]$, and for any acceptable $W$, from Lemma \ref{lemmaxc}, we conclude,
\begin{eqnarray}
H(\bar{Y}_{k_1}^{[T]}\mid W,\mathcal{G})-H(\bar{Y}_{k_2}^{[T]}\mid W,\mathcal{G})
&\le&T\max_{m\in[M]}(\alpha_{k_1m}-\alpha_{k_2m})^+\log{\bar{P}}+T~o~(\log{\bar{P}}).\label{deltaij}
\end{eqnarray}
where we used the fact that based on \eqref{eq:lkm}, we have $(\lambda_{k_1m}-\lambda_{k_2m})^+=(\alpha_{k_1m}-\alpha_{k_2m})^+$.

\subsubsection{Proof of bound: $d_1+d_2+d_3\le \delta_{1}+\delta_{2,1}+\delta_{3,2}$}
Suppressing $o(T)$and $o(\log(P))$ terms that are inconsequential for GDoF,
\begin{eqnarray}
TR_1&\le& I(\bar{Y}_{1}^{[T]};W_1\mid \mathcal{G})\label{fano1}\\
TR_2&\le& I(\bar{Y}_{2}^{[T]};W_2\mid W_1,\mathcal{G})\label{fano2}\\
TR_3&\le& I(\bar{Y}_{3}^{[T]};W_3\mid W_1,W_2,\mathcal{G})\label{fano3}
\end{eqnarray}
Summing over \eqref{fano1}, \eqref{fano2} and \eqref{fano3}, we have,
\begin{eqnarray}
\sum_{k=1}^3TR_k
&\le& H(\bar{Y}_{1}^{[T]}\mid \mathcal{G})+H(\bar{Y}_{2}^{[T]}\mid W_1,\mathcal{G})-H(\bar{Y}_{1}^{[T]}\mid W_1,\mathcal{G})\nonumber\\
&&+H(\bar{Y}_{3}^{[T]}\mid W_1,W_2,\mathcal{G})-H(\bar{Y}_{2}^{[T]}\mid W_1,W_2,\mathcal{G})\\
&\le& H(\bar{Y}_{1}^{[T]}\mid \mathcal{G})+(\delta_{2,1}+\delta_{3,2})T\log{\bar{P}}\label{fn4}\\
&\le& (\delta_{1}+\delta_{2,1}+\delta_{3,2})T\log{\bar{P}}\label{fn5}
\end{eqnarray}
\eqref{fn4} follows from \eqref{deltaij}, and for \eqref{fn5} we use the fact that $|\bar{Y}_1(t)|\le M\Delta\bar{P}^{\delta_{1}}$ and that the entropy of any discrete random variable is bounded by the logarithm of the  cardinality of its support. From \eqref{fn5}  we obtain the GDoF bound $d_1+d_2+d_3\le \delta_{1}+\delta_{2,1}+\delta_{3,2}$. Similarly, the bound $d_1+d_2\le \delta_{1}+\delta_{2,1}$ follows by summing \eqref{fano1} and \eqref{fano2}, 
\begin{eqnarray}
TR_1+TR_2&\le& H(\bar{Y}_{1}^{[T]}\mid \mathcal{G})+H(\bar{Y}_{2}^{[T]}\mid W_1,\mathcal{G})-H(\bar{Y}_{1}^{[T]}\mid W_1,\mathcal{G})\\
&\le& (\delta_{1}+\delta_{2,1})T\log{\bar{P}}.
\end{eqnarray}

\subsubsection{Proof of bound: $d_1+d_2+d_3\le(\delta_{1}+\delta_{3}+\delta_{1,2}+\delta_{2,1}+\delta_{3,1}+\delta_{2,3})/{2}$} \label{bound/2}
\begin{eqnarray}
TR_1&\le& I(\bar{Y}_{1}^{[T]};W_1\mid \mathcal{G})\label{fano11}\\
TR_1&\le& I(\bar{Y}_{1}^{[T]};W_1\mid W_2,\mathcal{G})\label{fano12}\\
TR_2&\le& I(\bar{Y}_{2}^{[T]};W_2\mid \mathcal{G})\label{fano21}\\
TR_2&\le& I(\bar{Y}_{2}^{[T]};W_2\mid W_3,\mathcal{G})\label{fano22}\\
TR_3&\le& I(\bar{Y}_{3}^{[T]};W_3\mid \mathcal{G})\label{fano31}\\
TR_3&\le& I(\bar{Y}_{3}^{[T]};W_3\mid W_1,W_2,\mathcal{G})\label{fano32}
\end{eqnarray}
Using the fact that $I(A;B)\le I(A;B\mid C)$ if $B$ and $C$ are independent of each other, we have
\begin{align}
H(\bar{Y}_{2}^{[T]}\mid \mathcal{G})-H(\bar{Y}_{2}^{[T]}\mid W_2,W_3,\mathcal{G})\le H(\bar{Y}_{2}^{[T]}\mid W_1, \mathcal{G})\label{fnn}
\end{align}
Moreover, applying \eqref{deltaij}   we have,
\begin{eqnarray}
H(\bar{Y}_{1}^{[T]}\mid \mathcal{G})&\le&T\delta_{1}\log{\bar{P}}\\
H(\bar{Y}_{3}^{[T]}\mid \mathcal{G})&\le&T\delta_{3}\log{\bar{P}}\\
H(\bar{Y}_{2}^{[T]}\mid W_1, \mathcal{G})-H(\bar{Y}_{1}^{[T]}\mid W_1, \mathcal{G})&\le&T\delta_{2,1}\log{\bar{P}}\\
H(\bar{Y}_{1}^{[T]}\mid W_2, \mathcal{G})-H(\bar{Y}_{2}^{[T]}\mid W_2, \mathcal{G})&\le&T\delta_{1,2}\log{\bar{P}}\\
H(\bar{Y}_{3}^{[T]}\mid W_1,W_2, \mathcal{G})-H(\bar{Y}_{1}^{[T]}\mid W_1,W_2, \mathcal{G})&\le&T\delta_{3,1}\log{\bar{P}}\\
H(\bar{Y}_{2}^{[T]}\mid W_3, \mathcal{G})-H(\bar{Y}_{3}^{[T]}\mid W_3, \mathcal{G})&\le&T\delta_{2,3}\log{\bar{P}}\label{fm,}
\end{eqnarray}
Summing over (\eqref{fano11}-\eqref{fm,}), the bound $d_1+d_2+d_3\leq(\delta_{1}+\delta_{3}+\delta_{1,2}+\delta_{2,1}+\delta_{3,1}+\delta_{2,3})/{2}$ is obtained.

\subsection{$K>3$ User MISO BC}
In this section we generalize the outer bounds of Theorem \ref{theorem:GDoF3} to the MISO BC with $K>3$ users. For ease of exposition, we will introduce the main elements one by one before combining them into a general theorem. Let us start with some definitions.
\begin{definition}\label{defpp} For any $S\subset [K]$, let ${\bf p}=(k_1, k_2, \cdots, k_m)$ denote an ordered arrangement of the elements of $S=\{k_1, k_2, \cdots, k_m\}$. Then ${\bf p}$ is called a permutation of the set $S$. The $i^{th}$ element of ${\bf p}$ is denoted by ${p}(i)$, and the number of terms in ${\bf p}$ is denoted equivalently as $|{\bf p}|=|S|=m$. Since the set $S$ is a function of ${\bf p}$ we may denote it as $S({\bf p})$. Furthermore, the set  $[K]/S$ is denoted equivalently as $[K]/S = S^c = {\bf p}^c$. 
\end{definition}

For example,  suppose $K=6$ and $S=\{2, 4, 5\}\subset [6]$, then ${\bf p}_1=(2,4, 5)$, ${\bf p}_2=(4,5,2)$ are two of the six possible permutations of  $S$,  $|{\bf p}_1|=|{\bf p}_2|=3$,  ${p}_1(1)=2, {p}_2(2)=5$, and $S^c={\bf p}_1^c={\bf p}_2^c=\{1,3,6\}$.

\begin{definition} For any permutation ${\bf p}=(k_1, k_2, \cdots, k_m)$ of $S\subset[K]$, define the function
\begin{eqnarray}
f({\bf p})&=&\left\{
\begin{array}{ll}
\delta_{k_2, k_1}+\delta_{k_3,k_2}+\cdots+\delta_{k_m,k_{m-1}}, & \mbox{ if } |{\bf p}|\geq 2\\
0, &\mbox{ if } |{\bf p}|=1.
\end{array}
\right.
\end{eqnarray}
\end{definition}
For example, if ${\bf p}=(2,4,5, 6)$, then $f({\bf p})=\delta_{4,2}+\delta_{5,4}+\delta_{6,5}$.
\begin{definition} Define the notation
\begin{eqnarray}
\bar{H}\left(Y_k\mid  W_{\{k_1, k_2, \cdots, k_m\}}\right)&=&\lim_{\bar{P}\rightarrow\infty}\lim_{T\rightarrow\infty}\frac{H(\bar{Y}^{[T]}_k\mid W_{k_1}, W_{k_2}, \cdots, W_{k_m}, \mathcal{G})}{T\log(\bar{P})}
\end{eqnarray}
\end{definition}

\begin{lemma} \label{lemma1} For any  $k\in[K]$,
\begin{eqnarray}
d_k&\leq& \delta_k - \bar{H}(Y_k\mid  W_k) .\label{eq:lemma1}
\end{eqnarray}
\end{lemma}
\noindent\proof Lemma \ref{lemma1} is trivially obtained from Fano's inequality, $TR_k\leq I(W_k; \bar{Y}_k^{[T]}\mid \mathcal{G})+To(T) $ and bounding $\bar{H}(Y_k)$ by $\delta_k$.
\hfill\QED\\
\bigskip

Lemma \ref{lemma1} can be used for the immediate bound $d_k\leq \delta_k$, by simply dropping the negative entropy term in \eqref{eq:lemma1}. However, it can also be combined with other bounds that produce corresponding positive entropy terms that can be cancelled by the negative terms from \eqref{eq:lemma1}. This is facilitated by the next lemma.

\begin{lemma}\label{lemma2} If ${\bf p}=(k_1, k_2, \cdots, k_m)$ is a permutation of $S\subset[K]$, such that $|{\bf p}|>1$, then
\begin{eqnarray}
d_{k_2}+d_{k_3}+\cdots+d_{k_m}&\leq& \bar{H}(Y_{k_1}\mid W_{k_1}, W_{S'})+f({\bf p}) - \bar{H}(Y_{k_m}\mid W_S, W_{S'}).
\end{eqnarray}
for any $S' \subset S^c$.
\end{lemma}
\noindent \proof Adding the chain of Fano's inequalities: $TR_{k_j}\leq I(W_{k_j}; \bar{Y}_{k_j}^{[T]}\mid W_{\{ k_{l},1\le l\le j-1\}}, W_{S'},\mathcal{G})+To(T)$ for $j\in [2:m]$, and applying GDoF limits, we have,
\begin{eqnarray}
\sum_{j=2}^m d_{k_j}&\leq&\sum_{j=2}^m\left(\bar{H}(Y_{k_j}\mid W_{\{k_{l},1\le l\le j-1\}}, W_{S'})-\bar{H}(Y_{k_j}\mid W_{\{k_{l},1\le l\le j\}}, W_{S'})\right)\\
&=&\bar{H}(Y_{k_2}\mid  W_{k_1},W_{S'})+\sum_{j=3}^m\left(\underbrace{\bar{H}(Y_{k_j}\mid W_{k_1, k_2, \cdots, k_{j-1}}, W_{S'})-\bar{H}(Y_{k_{j-1}}\mid W_{k_1, k_2, \cdots, k_{j-1}}, W_{S'})}_{\leq \delta_{j, j-1}}\right)\nonumber\\
&&-\bar{H}(Y_{k_m}\mid W_S, W_{S'})\label{eq:sumdif}\\
&\leq&\bar{H}(Y_{k_2}\mid W_{k_1},W_{S'}) +f({\bf p})-\delta_{k_2,k_1} -\bar{H}(Y_{k_m}\mid W_S, W_{S'})\label{eq:delta21s}\\
&\leq&\bar{H}(Y_{k_1}\mid W_{k_1},W_{S'}) +f({\bf p}) -\bar{H}(Y_{k_m}\mid W_S, W_{S'})\label{eq:delta21s}
\end{eqnarray}
where each of the difference of entropy terms inside the summation in \eqref{eq:sumdif} is bounded by $\delta_{j, j-1}$ by applying the result of Lemma 1 in  \cite{Arash_Jafar_KMIMOIC}, (reproduced in this work as Lemma \ref{lemmaxc} for convenience).\hfill \QED \\

Note that for any permutation ${\bf p}=(k_1, k_2, \cdots, k_m)$ such that $|{\bf p}|>1$, the two lemmas can be combined to cancel the negative entropy term of Lemma \ref{lemma1} with the positive entropy term of Lemma \ref{lemma2}, and dropping the negative entropy term of Lemma \ref{lemma2}, to produce the bound
\begin{eqnarray}
d_{k_1}+d_{k_2}+\cdots+d_{k_m}&\leq& \delta_{k_1}+f({\bf p})
\end{eqnarray}
However, instead of dropping the remaining negative entropy terms, it is possible to combine them with other bounds that produce corresponding positive entropy terms. These new bounds utilize the notion of \emph{merging} two permutations, defined as follows.

\begin{definition} \label{defpp} {\normalfont {\bf (Merge)}} Consider two permutations ${\bf p}=({ p}(1),\cdots,{ p}(k))$ and  ${\bf q}=({ q}(1),\cdots,{ q}(l))$, such that $k>1, l>1$, and $p(k')=q(l')$ for some $k'\in[k], l'\in[l]$. A merge of {\bf p} and {\bf q} at $p(k')$ produces   four permutations ${\bf u}_1, {\bf u}_2, {\bf u}_3,{\bf u}_4$ such that
\begin {eqnarray}
{\bf u}_1&=&({ p}(1),\cdots,{ p}(k'))\\
{\bf u}_2&=&({ q}(1),\cdots,{ q}(l'))\\
{\bf u}_3&=&({ p}(k'),i_1,i_2,\cdots,i_{|{\bf p}_{+}\cap {\bf q}_{+}|}),\\
{\bf u}_4&=&({ q}(l'),j_1,j_2,\cdots,j_{|{\bf p}_{+}\cup {\bf q}_{+}|}),
\end{eqnarray}
where
\begin{eqnarray}
{\bf p}_{+}&=&\{{ p}(k'+1),\cdots,{ p}(k)\},\\
{\bf q}_{+}&=&\{{ q}(l'+1),\cdots,{ q}(l)\},\\
{\bf p}_{+}\cap {\bf q}_{+}&=&\{i_1,\cdots,i_{|{\bf p}_{+}\cap {\bf q}_{+}|}\},\\
{\bf p}_{+}\cup {\bf q}_{+}&=&\{j_1,\cdots,j_{|{\bf p}_{+}\cup {\bf q}_{+}|}\}.
\end{eqnarray}
\end{definition}
There may be more than one possible {\it merge} for the same {\bf p} and {\bf q} even with the same choice of $p(k'), q(l')$. For instance, suppose we merge the two permutations $(1,2,3,4)$ and $(4,3,2,1)$, at ${p}(k')={ q}(l')=2$. Then one possible merge is 
${\bf u_1}, {\bf u}_2, {\bf u_3}, {\bf u}_4 = (1,2),(4,3,2),(2),(2,3,4,1)$
while another possible merge is ${\bf u_1}, {\bf u}_2, {\bf u_3}, {\bf u}_4 = (1,2),(4,3,2),(2),(2,1,4,3)$. In fact in this case there are $6$ possible merges corresponding to $6$ different choices for ${\bf u}_4=(2,a,b,c)$, where $\{a,b,c\}=\{1,3,4\}$.

Applying Lemma \ref{lemma2} to the merge of two permutations produces the next set of bounds, represented in Lemma \ref{lemma3}.

\begin{lemma}\label{lemma3} If ${\bf p}, {\bf q}$ are permutations whose merge produces ${\bf u}_1, {\bf u}_2, {\bf u}_3, {\bf u}_4$ as stated in Definition \ref{defpp}, then
\begin{eqnarray}
\sum_{n=2}^{k}d_{p(n)}+\sum_{n=2}^{l}d_{q(n)}&\leq&
\bar{H}(Y_{p(1)}\mid W_{p(1)}, W_{S'})
+\bar{H}(Y_{q(1)}\mid W_{q(1)}, W_{S'})+\sum_{n=1}^4f({\bf u}_n)  \nonumber\\
&&-\bar{H}(Y_{i_{|{\bf p}_+\cap {\bf q}_+|}}\mid W_{S({\bf u}_3)}, W_{S({\bf u}_1)\cup S({\bf u}_2)}, W_{S'})\nonumber\\
&&-\bar{H}(Y_{i_{|{\bf p}_+\cup {\bf q}_+|}}\mid W_{S({\bf u}_4)}, W_{S({\bf u}_1)\cap S({\bf u}_2)}, W_{S'})
\end{eqnarray}
for any $S'\subset (S({\bf p})\cup S({\bf q}))^c$.
\end{lemma}
\noindent\proof
Applying Lemma \ref{lemma2} to each of the permutations ${\bf u}_1, {\bf u}_2, {\bf u}_3, {\bf u}_4$, we obtain,
\begin{align}
{\bf u}_1: &&\sum_{n=2}^{k'}d_{p(n)}&\leq\bar{H}(Y_{p(1)}\mid W_{p(1)}, W_{S'})+f({\bf u}_1)-\bar{H}(Y_{p(k')}\mid W_{S({\bf u}_1)}, W_{S'})\label{eq:neg1}\\
{\bf u}_2: &&\sum_{n=2}^{l'}d_{q(n)}&\leq\bar{H}(Y_{q(1)}\mid W_{q(1)}, W_{S'})+f({\bf u}_2)-\bar{H}(Y_{q(l')}\mid W_{S({\bf u}_2)}, W_{S'})\label{eq:neg2}\\
{\bf u}_3: &&\sum_{i=1}^{|{\bf p}_+\cap {\bf q}_+|}d_{i}&\leq\bar{H}(Y_{p(k')}\mid W_{S({\bf u}_1)\cup S({\bf u}_2)}, W_{S'})+f({\bf u}_3)-\bar{H}(Y_{i_{|{\bf p}_+\cap {\bf q}_+|}}\mid W_{S({\bf u}_3)}, W_{S({\bf u}_1)\cup S({\bf u}_2)}, W_{S'})\label{eq:neg3}\\
{\bf u}_4: &&\sum_{j=1}^{|{\bf p}_+\cup {\bf q}_+|}d_{i}&\leq\bar{H}(Y_{q(l')}\mid  W_{S({\bf u}_1)\cap S({\bf u}_2)}, W_{S'})+f({\bf u}_4)-\bar{H}(Y_{i_{|{\bf p}_+\cup {\bf q}_+|}}\mid W_{S({\bf u}_4)}, W_{S({\bf u}_1)\cap S({\bf u}_2)}, W_{S'})\label{eq:neg4}
\end{align}
Note that $p(k')=q(l')$. From Definition \ref{defpp} it is easily verified that $S({\bf u}_1)\cup S({\bf u}_2)$ has no elements in common with ${\bf p}_+\cap {\bf q}_+$, and that $S({\bf u}_1)\cap S({\bf u}_2)$ has no elements in common with ${\bf p}_+\cup {\bf q}_+$, facilitating the application of Lemma \ref{lemma2}. Adding all four inequalities,  and using the submodularity property of entropy, $H(X|A)+H(X|B)\geq H(X|A\cup B)+H(X|A\cap B)$, to cancel the positive entropy terms of \eqref{eq:neg3} and \eqref{eq:neg4} with the negative entropy terms of \eqref{eq:neg1} and \eqref{eq:neg2}, we obtain the result of Lemma \ref{lemma3}.\hfill\QED\\

By dropping the negative entropy terms in Lemma \ref{lemma3} and canceling the positive entropy terms in Lemma \ref{lemma3} with the corresponding negative entropy terms from Lemma \ref{lemma1}, we obtain the bound
\begin{eqnarray}
\sum_{n=1}^{k}d_{p(n)}+\sum_{n=1}^{l}d_{q(n)}&\leq&
\delta_{p(1)}+\delta_{q(1)}+\sum_{n=1}^4f({\bf u}_n)
\end{eqnarray}
For example, consider a $K=7$ user setting, and let us merge the permutations $(1,2,3,4,5,6,7)$ and $(1,2,5,4,3,6,7)$ at $4$ to obtain ${\bf u}_1=(1,2,3,4)$, ${\bf u}_2=(1,2,5,4)$, ${\bf u}_3=(4,6,7)$ and ${\bf u}_4=(4,5,3,6,7)$. According to Lemma \ref{lemma3}, corresponding to this merge we obtain the following bounds.
\begin{align}
{\bf u}_1=(1,2,3,4):&\sum_{i\in\{2,3,4\}}d_i\leq \bar{H}(Y_1\mid W_1)+f((1,2,3,4))-\bar{H}(Y_4\mid W_{\{1,2,3,4\}})\label{eq:ex1}\\
{\bf u}_2=(1,2,5,4):&\sum_{i\in\{2,5,4\}}d_i\leq \bar{H}(Y_1\mid W_1)+f((1,2,5,4))-\bar{H}(Y_4\mid W_{\{1,2,5,4\}})\label{eq:ex2}\\
{\bf u}_3=(4,6,7):&\sum_{i\in\{6,7\}}d_i\leq \bar{H}(Y_4\mid W_{\{1,2,3,4,5\}})+f((4,6,7))-\bar{H}(Y_7\mid W_{\{1,2,3,4,5,6,7\}})\label{eq:ex3}\\
{\bf u}_4=(4,5,3,6,7):&\sum_{i\in\{5,3,6,7\}}d_i\leq \bar{H}(Y_4\mid W_{\{1,2,4\}})+f((4,5,3,6,7))-\bar{H}(Y_7\mid W_{\{1,2,3,4,5,6,7\}})\label{eq:ex4}
\end{align}
By dropping the negative entropy terms in \eqref{eq:ex3}, \eqref{eq:ex4} and canceling the positive entropy terms in \eqref{eq:ex1}, \eqref{eq:ex2} with the corresponding negative entropy terms from Lemma \ref{lemma1}, i.e., $d_1\leq \delta_1-\bar{H}(Y_1|W_1)$, we obtain the bound
\begin{eqnarray}
2\sum_{n=1}^7d_n&\leq& 2\delta_1+f({\bf u}_1)+f({\bf u}_2)+f({\bf u}_3)+f({\bf u}_4).
\end{eqnarray}
Remarkably, we can also perform additional merge steps to obtain new bounds. Continuing with our $K=7$ example, if we merge ${\bf u}_3=(4,6,7)$ and ${\bf u}_4=(4,5,3,6,7)$ at $6$, then we obtain ${\bf u}_1'=(4,6)$, ${\bf u}_2'=(4,5,3,6)$, ${\bf u}_3'=(6,7)$, ${\bf u}_4'=(6,7)$. Bounds corresponding to ${\bf u}_1', {\bf u}_2', {\bf u}_3', {\bf u}_4'$ now replace the bounds \eqref{eq:ex3}, \eqref{eq:ex4}. Proceeding according to Lemma \ref{lemma3},
\begin{eqnarray}
{\bf u}_1=(1,2,3,4):&&\sum_{i\in\{2,3,4\}}d_i\leq \bar{H}(Y_1\mid W_1)+f((1,2,3,4))-\bar{H}(Y_4\mid W_{\{1,2,3,4\}})\label{eq:exm1}\\
{\bf u}_2=(1,2,5,4):&&\sum_{i\in\{2,5,4\}}d_i\leq \bar{H}(Y_1\mid W_1)+f((1,2,5,4))-\bar{H}(Y_4\mid W_{\{1,2,5,4\}})\label{eq:exm2}\\
{\bf u}_1'=(4,6):&&d_6\leq \bar{H}(Y_4\mid W_{\{1,2,3,4,5\}})+f((4,6))-\bar{H}(Y_6\mid W_{\{1,2,3,4,5,6\}})\\
{\bf u}_2'=(4,5,3,6):&&\sum_{i\in\{5,3,6\}}d_i\leq \bar{H}(Y_4\mid W_{1,2,4})+f((4,5,3,6))-\bar{H}(Y_6\mid W_{\{1,2,3,4,5,6\}})\\
{\bf u}_3'=(6,7):&&d_7\leq \bar{H}(Y_6\mid W_{\{1,2,3,4,5,6\}})+f((6,7))-\bar{H}(Y_7\mid W_{\{1,2,3,4,5,6,7\}})\label{eq:exm5}\\
{\bf u}_4'=(6,7):&&d_7\leq \bar{H}(Y_6\mid W_{\{1,2,3,4,5,6\}})+f((6,7))-\bar{H}(Y_7\mid W_{\{1,2,3,4,5,6,7\}})\label{eq:exm6}
\end{eqnarray}
Adding all $6$ inequalities, dropping the negative entropy terms in \eqref{eq:exm5}, \eqref{eq:exm6} and canceling the positive entropy terms in \eqref{eq:exm1}, \eqref{eq:exm2} with the corresponding negative entropy terms from Lemma \ref{lemma1}, i.e., $d_1\leq \delta_1-\bar{H}(Y_1|W_1)$, we obtain the bound 
\begin{eqnarray}
2\sum_{n=1}^7d_n&\leq& 2\delta_1+f({\bf u}_1)+f({\bf u}_2)+f({\bf u}_1')+f({\bf u}_2')+f({\bf u}_3')+f({\bf u}_4').
\end{eqnarray}
Proceeding in this manner, we can obtain potentially infinitely many bounds. We conjecture that only a finite number of these bounds will be non-redundant, but identifying the precise set of redundant bounds, or even proving that there are only finitely many of them, remains an open problem. We also conjecture that these bounds will be sufficient to identify a regime where SLS is optimal for the $K>3$ user setting, however, given the difficulty of this settling this question for $K=3$ with our current approach, the generalization to $K>3$ also remains open. What remains is to formalize the complete set of bounds that can be obtained through the application of Lemma \ref{lemma1}, Lemma \ref{lemma2} and Lemma \ref{lemma3} in the final theorem of this section. To this end, we need the following definition.

\begin{definition}\label{defmulti} {\normalfont ({\bf Bounding Pattern})} 
Let $A=\{{\bf p}_1, {\bf p}_2, \cdots, {\bf p}_m\}$, $B=\{{\bf q}_1, {\bf q}_2, \cdots, {\bf q}_n\}$ be \emph{multisets}\footnote{Unlike a set, multiple instances of elements are allowed in a multiset, e.g.,  $\{\bar{\bf p}_a,\bar{\bf p}_a, \bar{\bf p}_b\}$ and $\{\bar{\bf p}_a, \bar{\bf p}_b\}$ are different multisets although they are the same set.} of permutations of subsets of $\{0\}\cup [K]$. For compact notation, let us represent the tuple $(A, B)$ as
\begin{eqnarray}
(A,B)&=&\{{\bf p}_1, {\bf p}_2, \cdots, {\bf p}_m, \bar{\bf q}_1, \bar{\bf q}_2, \cdots, \bar{\bf q}_n\},
\end{eqnarray}
where we use the overhead bar to identify elements of $B$ separately from the elements of $A$. We say that $(A, B)$ is a  bounding pattern if it can be generated from the following three properties.
\begin{enumerate}
\item If ${\bf p}$ is a permutation of $S\subset [K]$, and $|{\bf p}|>1$, then 
$(A,B)=((0,p(1)),\bar{\bf p})$ is a  bounding pattern. For example, for $K\geq 4$, it follows that $(A,B)=\{(0,3), \overline{(3,2,4)}\}$ is a  bounding pattern.
\item If $(A_1,B_1)$ and $(A_2, B_2)$ are  bounding patterns, then $(A_1\uplus A_2, B_1\uplus B_2)$ is\footnote{For instance, $\{\bar{\bf p}_a,\bar{\bf p}_a,\bar{\bf p}_c\}\uplus\{\bar{\bf p}_a,\bar{\bf p}_b\}=\{\bar{\bf p}_a,\bar{\bf p}_a,\bar{\bf p}_a,\bar{\bf p}_b,\bar{\bf p}_c\}$.}  a  bounding pattern. For example, for $K\geq 4$, from the first property we know that $\{(0,3), \overline{(3,2,4)}\}$ and $\{(0,1), \overline{(1,2,3)}\}$ are valid bounding patterns. Then, from the second property it follows that $\{(0,3),(0,1), \overline{(3,2,4)}, \overline{(1,2,3)}\}$ is also a  bounding pattern.
\item If $(A,B)$ is a  bounding pattern with $A=\{{\bf p}_{1}, {\bf p}_{2}, \cdots, {\bf p}_{m}\}$ and $B=\{{\bf q}_{1}, {\bf q}_{2}, \cdots, {\bf q}_{n}\}$, and permutations ${\bf q}_{1}, {\bf q}_{2}$ can be merged to obtain ${\bf u}_1, {\bf u}_2, {\bf u}_3, {\bf u}_4$ as described in Definition \ref{defpp}, then $(A', B')$ is a  bounding pattern where 
\begin{eqnarray}
A'&=&\{{\bf p}_{1}, {\bf p}_{2},\cdots, {\bf p}_{m},{\bf u}_1, {\bf u}_2\}\\
B'&=&\{{\bf q}_{3}, {\bf q}_{4}, \cdots, {\bf q}_{n},{\bf u}_3, {\bf u}_4\}
\end{eqnarray}
For example, for $K\geq 4$, from the first two properties we know that $\{(0,3),(0,1), \overline{(3,2,4)},$ $\overline{(1,2,3)}\}$ is a  bounding pattern. We can merge $(3,2,4)$ and $(1,2,3)$ at $2$ to obtain ${\bf u}_1=(3,2)$, ${\bf u}_2=(1,2)$, ${\bf u}_3=(2)$, ${\bf u}_4=(2,3,4)$. Therefore, the third property implies that
$\{(0,3),(0,1), (3,2), (1,2), \overline{(2)},$ $\overline{(2,3,4)}\}$ is also a  bounding pattern.
\end{enumerate}
\end{definition}

\begin{theorem}\label{theorem:GDoF2} In a $K$ user MISO BC with $M$ antennas at the transmitter, if $(A, B)$ is a bounding pattern for $A=\{{\bf p}_1, {\bf p}_2, \cdots, {\bf p}_m\}$, $B=\{{\bf q}_1, {\bf q}_2, \cdots, {\bf q}_n\}$, then the GDoF region is bounded by, 
\begin{eqnarray}
\sum_{{\bf p}\in A\uplus B}\sum_{i=2}^{|{\bf p}|}d_{p(i)}\le \sum_{{\bf p}\in A\uplus B} f({\bf p}), \label{Bb6}
\end{eqnarray}
where for any permutation ${\bf p}$, $f({\bf p})$ is defined as,
\begin{eqnarray}
f({\bf p})&=&\left\{
\begin{array}{ll}
0,&\mbox{~if~} |{\bf p}|=1\\
\sum_{k=2}^{|{\bf p}|}{\delta}_{p(k),p(k-1)}, &\mbox{~if~} |{\bf p}|>1,p(1)\neq0\\ 
{\delta}_{p(2)},&\mbox{~if~} |{\bf p}|=2, p(1)=0\\
\end{array}\right.\label{f(p)}
\end{eqnarray}
and ${\delta}_{i,j}$ and ${\delta}_i$ are defined in Definition \ref{defdelta_}.
\end{theorem}

The proof of  Theorem \ref{theorem:GDoF2} is relegated to Appendix \ref{kconverse}.

\section{Proof of Theorem \ref{theorem:GDoF3}: Achievability}\label{achiev}
Since SLS is a simple achievable scheme, it is not difficult to characterize its achievable GDoF region.\footnote{ Note that, when  conditions \eqref{con1} and \eqref{con} are true the GDoF region given in \eqref{BB} does not depend on channel strengths of the links associated with the $m^{th}$ antenna for all $m>3$ and will remain the same if we remove all the transmit antennas except the first $3$. Therefore, it is sufficient to derive the achievability for the $3$ user MISO BC where only the first three antennas are present.} Recall that SLS allows arbitrary power control, as well as arbitrary partitioning of sub-messages across arbitrary decoding subsets of users. These choices are represented by auxiliary variables. In terms of these auxiliary variables a description of the SLS achievable GDoF region is straightforward. However, note that our GDoF outer bound does not involve any auxiliary variables, i.e., it represents a direct characterization of the GDoF region \emph{optimized} over all auxiliary variables. Eliminating the auxiliary variables from the achievable regions, and then proving that the union of those achievable regions matches the outer bound is the key technical challenge for proving the achievability result of Theorem \ref{theorem:GDoF3}. What is required is essentially a Fourier-Motzkin (FM) elimination, but the number of variables is large enough to make a direct application of the FM algorithm prohibitively complex. Recall that in \cite{Geng_TIN} the elimination of auxiliary power control variables was accomplished by the use of the Potential Theorem, in order to find a direct characterization of the achievable region of TIN. For SLS the potential theorem seems less useful due to the added complexity of layered rate-partitioning on top of power control. We will need a bit more tedious reasoning to navigate through this challenge. As it turns out, we need $12$ different specializations of SLS schemes. We will present two of them, leading to achievable GDoF regions labeled $\mathcal{\hat{D}}_{123}$ and $\mathcal{\hat{F}}_{123}$. The remaining $10$ cases are obtained from these two by switching indices.  We start with $\mathcal{\hat{D}}_{123}$.

\subsection{$\mathcal{\hat{D}}_{123}$}\label{ach1}
For this achievable scheme, we consider the parameter regime where
\begin{eqnarray}
\max_{k,m\in[3],k\neq m}\alpha_{km}\le\min(\alpha_{11},\alpha_{22}).
\end{eqnarray}
\subsubsection{SLS Coding}
\begin{figure}[h] 
\centering
\includegraphics[width=0.75\textwidth]{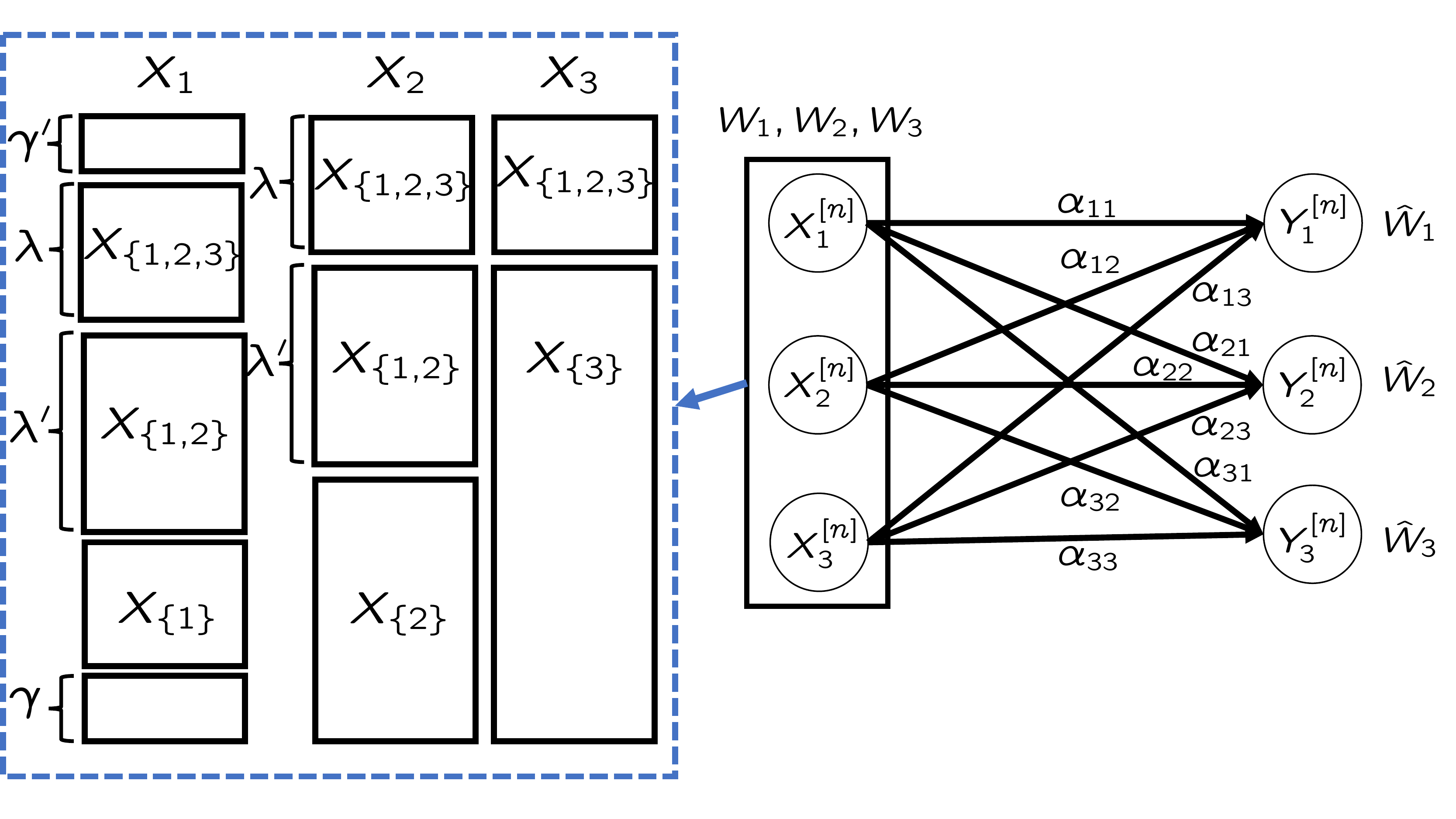}
\caption[]{SLS coding for $\mathcal{\hat{D}}_{123}$.}
\label{fig_sls1}
\end{figure}

Consider four non-negative values  $\lambda,\lambda',\gamma,\gamma'$, and five independent messages $\bar{W}_{\{1\}}$, $\bar{W}_{\{2\}}$, $\bar{W}_{\{3\}}$, $\bar{W}_{\{1,2\}}$, $\bar{W}_{\{1,2,3\}}$, carrying non-negative values of $d_{\{1\}},d_{\{2\}},d_{\{3\}},d_{\{1,2\}},d_{\{1,2,3\}}$ GDoF, respectively.  The messages $\bar{W}_{\{1\}},\bar{W}_{\{2\}},\bar{W}_{\{3\}},\bar{W}_{\{1,2\}},\bar{W}_{\{1,2,3\}}$ are encoded into independent Gaussian codebooks  ${X}_{\{1\}},{X}_{\{2\}},{X}_{\{3\}},{X}_{\{1,2\}},{X}_{\{1,2,3\}}$ with powers,
 \begin{eqnarray}
 E{|{X}_{\{1,2,3\}}|^2}&=&1-2P^{-\lambda}\label{m1}\\
 E{|{X}_{\{1,2\}}|^2}&=&P^{-\lambda}\\
 E{|{X}_{\{1\}}|^2}&=&P^{-\lambda-\lambda'}\\ 
 E{|{X}_{\{2\}}|^2}&=&P^{-\lambda-\lambda'}\\ 
  E{|{X}_{\{3\}}|^2}&=&P^{-\lambda}\label{m12}
 \end{eqnarray}
The transmitted and received signals are,
 \begin{eqnarray}
X_1&=&\bar{P}^{-\gamma'}({X}_{\{1,2,3\}}+{X}_{\{1,2\}}+{X}_{\{1\}})\label{m21}\\
X_2&=&{X}_{\{1,2,3\}}+{X}_{\{1,2\}}+{X}_{\{2\}}\label{m22}\\
X_3&=&{X}_{\{1,2,3\}}+{X}_{\{3\}}\label{m23}\\
Y_k&=&\sum_{m=1}^3\sqrt{P^{\alpha_{km}}}G_{km}X_m+Z_k, \forall k\in[3]\label{m24}
\end{eqnarray}
This SLS coding is illustrated in Figure \ref{fig_sls1}.
\subsubsection{Decoding}\label{const+}
The decoding proceeds as follows.
\begin{enumerate}
 \item{} At the first receiver, ${X}_{\{1,2,3\}},{X}_{\{1,2\}},{X}_{\{1\}}$ are decoded sequentially with successive interference cancellation while treating ${X}_{\{2\}}$ and ${X}_{\{3\}}$ as Gaussian noise.
 \item{} At the second receiver, ${X}_{\{1,2,3\}},{X}_{\{1,2\}},{X}_{\{2\}}$ are decoded sequentially with successive interference cancellation while treating ${X}_{\{1\}}$ and ${X}_{\{3\}}$ as noise.
  \item{}  At the third receiver, ${X}_{\{1,2,3\}},{X}_{\{3\}}$ are decoded sequentially with successive interference cancellation
 while treating ${X}_{\{1\}}$, ${X}_{\{2\}}$ and ${X}_{\{1,2\}}$ as noise.
\end{enumerate}
\subsubsection{Achievable Region $\mathcal{D}_{123}$}\label{a0}
As shown in Appendix \ref{sinr_app}, the following GDoF region is achievable.
 \begin{align}
 \mathcal{D}_{123}(\lambda,\lambda',\gamma,\gamma')=\Bigg\{(d_1,d_2,d_3): &\nonumber\\
d_1&=d_{\{1\}}+\mu_{1}d_{\{1,2\}}+\xi_1 d_{\{1,2,3\}}\label{eq:d}\\
d_2&=d_{\{2\}}+\mu_{2}d_{\{1,2\}}+\xi_2d_{\{1,2,3\}}\\
d_3&=d_{\{3\}}+\xi_3d_{\{1,2,3\}}\\
\mu_{1}+\mu_{2}&=1\\
\xi_{1}+\xi_{2}+\xi_{3}&=1\\
d_{\{1\}}&\le\alpha_{11}-\lambda-\lambda'-\gamma-\gamma'\label{eq:d1}\\
d_{\{2\}}&\le\alpha_{22}-\lambda-\lambda'\label{eq:d2}\\
d_{\{3\}}&\le\alpha_{33}-\lambda\label{eq:d3}\\
d_{\{1,2\}}&\le\lambda'\label{eq:d12}\\
d_{\{1,2,3\}}&\le\lambda\label{eq:d123}\\
 0&\le\mu_{1},\mu_{2},\xi_1,\xi_2,\xi_3, d_{\{1\}}, d_{\{2\}}, d_{\{3\}}, d_{\{1,2\}}, d_{\{1,2,3\}}\\
&\bigg\}\nonumber
 \end{align}
for all choices of $\lambda, \lambda', \gamma, \gamma'$ such that
 \begin{eqnarray}
\lambda+\lambda'+\gamma+\gamma'&\le&\alpha_{11}\label{llgg1}\\
\lambda+\lambda'&\le&\alpha_{22}\label{llgg2}\\
\lambda&\le&\alpha_{33}\label{llgg3}\\
\alpha_{12}&\le& \lambda+\lambda'+\gamma\label{llgg4}\\
\alpha_{13}&\le& \lambda+\gamma\label{llgg5}\\
\alpha_{21}&\le& \lambda+\lambda'+\gamma'\label{llgg6}\\
\alpha_{23}&\le& \lambda\label{llgg7}\\
\alpha_{31}&\le& \lambda+\gamma'\label{llgg8}\\
\alpha_{32}&\le& \lambda\label{llgg9}\\
0&\leq&\lambda,\lambda',\gamma,\gamma'\label{llgg10}
\end{eqnarray}
Note that this achievable region (which is one of $12$ different regions) involves $14$ auxiliary random variables that do not appear in the outer bound, namely, $ \mu_{1}$, $\mu_{2}$, $\xi_1,\xi_2,\xi_3$, $\lambda, \lambda'$, $\gamma, \gamma'$, $d_{\{1\}}$, $d_{\{2\}}$, $d_{\{3\}}$, $d_{\{1,2\}}$, $d_{\{1,2,3\}}$. The union over the regions corresponding to all feasible choices of these $14$ auxiliary variables is also achievable. Furthermore, there are $12$ such regions and their union gives us the overall achievable region. To show that the overall achievable region matches the outer bound we will need to eliminate the auxiliary variables. In the next step, we eliminate $ \mu_{1}$, $\mu_{2}$, $\xi_1,\xi_2,\xi_3$, $d_{\{1\}}$, $d_{\{2\}}$, $d_{\{3\}}$, $d_{\{1,2\}}$, $d_{\{1,2,3\}}$ from $\mathcal{{D}}_{123}$ to obtain the simplified region $\mathcal{\bar{D}}_{123}$.

\subsubsection{Achievable Region $\mathcal{\bar{D}}_{123}$ }\label{four++}
As shown in Appendix \ref{d_i}, elimination of $ \mu_{1}$, $\mu_{2}$, $\xi_1,\xi_2,\xi_3$,  $d_{\{1\}}$, $d_{\{2\}}$, $d_{\{3\}}$, $d_{\{1,2\}}$, $d_{\{1,2,3\}}$ gives us the following equivalent region ${\mathcal{\bar{D}}}_{123}$ which retains only $4$ auxiliary variables $\lambda,\lambda',\gamma,\gamma'$.
\begin{eqnarray}
\mathcal{\bar{D}}_{123}(\lambda,\lambda',\gamma,\gamma')=\Bigg\{(d_1,d_2,d_3)\in\mathbb{R}_+^3&:& \nonumber\\
d_1&\le& \alpha_{11}-\gamma-\gamma',\label{d1bar}\\
d_2&\le& \alpha_{22},\label{d2bar}\\
d_3&\le& \alpha_{33},\label{d3bar}\\
d_1+d_2&\le& \alpha_{11}+\alpha_{22}-\lambda-\lambda'-\gamma-\gamma',\label{d12bar}\\
d_1+d_3&\le& \alpha_{11}+\alpha_{33}-\lambda-\gamma-\gamma',\label{d13bar}\\
d_2+d_3&\le& \alpha_{22}+\alpha_{33}-\lambda,\label{d23bar}\\
d_1+d_2+d_3&\le&  \alpha_{11}+\alpha_{22}+\alpha_{33}-2\lambda-\lambda'-\gamma-\gamma'\Bigg\}\label{d123bar}
\end{eqnarray} 
such that $\lambda,\lambda',\gamma,\gamma'$ satisfy conditions \eqref{llgg1} to \eqref{llgg10}.
\subsubsection{Achievable Region $\hat{\mathcal{D}}_{123}$}\label{hatDfinal}
As shown in Appendix \ref{equi1}, the union of the regions ${\mathcal{\bar{D}}}_{123}$ over all possible choices of $\lambda,\lambda',\gamma,\gamma'$ gives us the following region ${\mathcal{\hat{D}}}_{123}$.
 \begin{align}
\hat{\mathcal{D}}_{123}=\Bigg\{(d_1,d_2,d_3)&\in\mathbb{R}_+^3:\nonumber\\
d_1&\le \alpha_{11},\label{d1hat}\\
d_2&\le \alpha_{22},\label{d2hat}\\
d_3&\le \alpha_{33},\label{d3hat}\\
d_1+d_2&\le \alpha_{11}+\alpha_{22}-\max_{l,m\in[3],l\neq m}\alpha_{lm},\label{d12hat}\\
d_1+d_3&\le \alpha_{11}+\alpha_{33}-\max(\alpha_{23},\alpha_{32},\alpha_{31},\alpha_{13}),\label{d13hat}\\
d_2+d_3&\le \alpha_{22}+\alpha_{33}-\max(\alpha_{23},\alpha_{32}),\label{d23hat}\\
d_1+d_2+d_3&\le\left.  \alpha_{11}+\alpha_{22}+\alpha_{33}-\max \left\{\begin{matrix}
\max_{l,m\in[3],l\neq m}\alpha_{lm}+\max(\alpha_{32},\alpha_{23}),\\ 
\alpha_{13}+\alpha_{21},\\ 
\alpha_{12}+\alpha_{31},\\ 
\alpha_{13}+\alpha_{31}
\end{matrix}\right\}
\right.\label{d123hat}\\
&&\Bigg\}\nonumber
\end{align} 

\subsection{$\mathcal{\hat{F}}_{123}$}
Assume that 
\begin{eqnarray}
\max_{k,m\in[3],k\neq m}\alpha_{km}\le\min(\alpha_{11},\alpha_{22}).
\end{eqnarray}
\subsubsection{SLS Coding}
\begin{figure}[h] 
\centering
\includegraphics[width=0.75\textwidth]{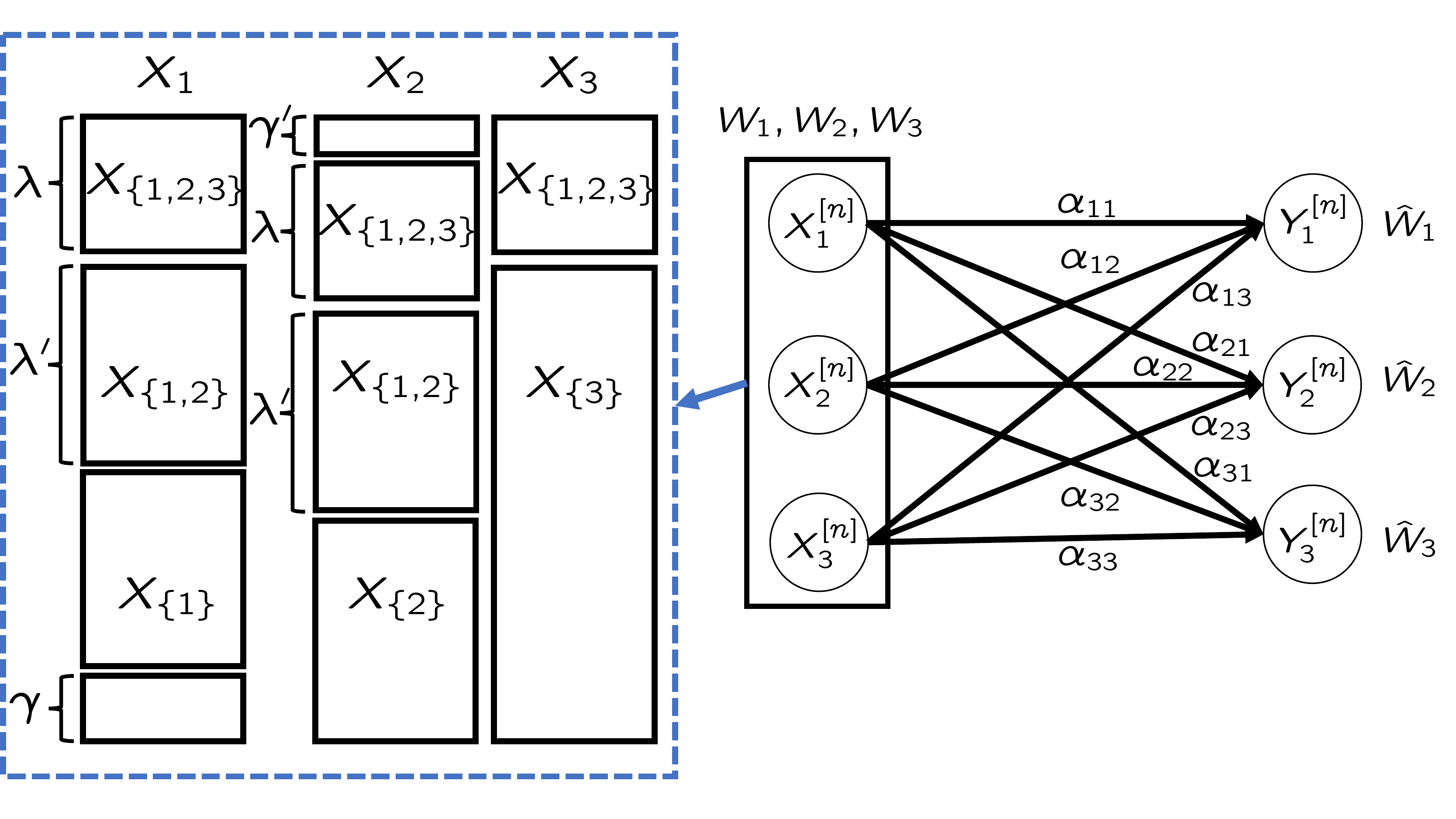}
\caption[]{SLS coding for $\mathcal{\hat{F}}_{123}$}
\label{fig_sls2}
\end{figure}

 Similar to~ \ref{ach1}, consider four non-negative values  $\lambda$, $\lambda'$, $\gamma$, $\gamma'$, and five independent messages $\bar{W}_{\{1\}}$, $\bar{W}_{\{2\}}$, $\bar{W}_{\{3\}}$, $\bar{W}_{\{1,2\}}$, $\bar{W}_{\{1,2,3\}}$ each carrying non-negative values of $d_{\{1\}}$, $d_{\{2\}}$, $d_{\{3\}}$, $d_{\{1,2\}}$, $d_{\{1,2,3\}}$ GDoF, respectively.   The messages $\bar{W}_{\{1\}}$, $\bar{W}_{\{2\}}$, $\bar{W}_{\{3\}}$, $\bar{W}_{\{1,2\}}$, $\bar{W}_{\{1,2,3\}}$ are encoded into independent Gaussian codebooks  ${X}_{\{1\}}$, ${X}_{\{2\}}$, ${X}_{\{3\}}$, ${X}_{\{1,2\}}$, ${X}_{\{1,2,3\}}$ with powers,
 \begin{eqnarray}
 E{|{X}_{\{1,2,3\}}|^2}&=&1-2P^{-\lambda}\label{n1}\\
 E{|{X}_{\{1,2\}}|^2}&=&P^{-\lambda}\\
 E{|{X}_{\{1\}}|^2}&=&P^{-\lambda-\lambda'}\\ 
 E{|{X}_{\{2\}}|^2}&=&P^{-\lambda-\lambda'}\\ 
  E{|{X}_{\{3\}}|^2}&=&P^{-\lambda}\label{n12}
 \end{eqnarray}
The transmitted and received signals are,
 \begin{eqnarray}
X_1&=&{X}_{\{1,2,3\}}+{X}_{\{1,2\}}+{X}_{\{1\}}\label{n21}\\
X_2&=&\bar{P}^{-\gamma'}({X}_{\{1,2,3\}}+{X}_{\{1,2\}}+{X}_{\{2\}})\label{n22}\\
X_3&=&{X}_{\{1,2,3\}}+{X}_{\{3\}}\label{n23}\\
Y_k&=&\sum_{j=1}^3\sqrt{P^{\alpha_{kj}}}G_{kj}X_j+Z_k, \forall k\in[3]\label{n24}
\end{eqnarray}
This SLS coding is illustrated in Figure \ref{fig_sls2}.

\subsubsection{Decoding}
The decoding proceeds similar to \ref{const+}.
\subsubsection{Achievable Region $\mathcal{F}_{123}$}
As shown in Appendix \ref{sinr_app2}, the following GDoF region is achievable.
 \begin{align}
 \mathcal{F}_{123}(\lambda,\lambda',\gamma,\gamma')=\Bigg\{(d_1,d_2,d_3):&\nonumber\\
d_1&=d_{\{1\}}+\mu_{1}d_{\{1,2\}}+\xi_1 d_{\{1,2,3\}},\\
d_2&=d_{\{2\}}+\mu_{2}d_{\{1,2\}}+\xi_2d_{\{1,2,3\}},\\
d_3&=d_{\{3\}}+\xi_3d_{\{1,2,3\}},\\
\mu_{1}+\mu_{2}&=1,\\
\xi_{1}+\xi_{2}+\xi_{3}&=1\\
d_{\{1\}}&\le\alpha_{11}-\lambda-\lambda'-\gamma,\label{+eq:d1}\\
d_{\{2\}}&\le\alpha_{22}-\lambda-\lambda'-\gamma',\label{+eq:d2}\\
d_{\{3\}}&\le\alpha_{33}-\lambda,\label{+eq:d3}\\
 d_{\{1,2\}}&\le\lambda',\label{+eq:d12}\\
d_{\{1,2,3\}}&\le\lambda,\label{+eq:d123}\\
0&\le  \mu_{1},\mu_{2},\xi_1,\xi_2,\xi_3,d_{\{1\}},d_{\{2\}},d_{\{3\}},d_{\{1,2\}},d_{\{1,2,3\}}\label{+eq:d}\\ 
&\bigg\}
 \end{align}
for all choices of $\lambda, \lambda', \gamma, \gamma'$ such that
 \begin{eqnarray}
\lambda+\lambda'+\gamma&\le&\alpha_{11}\label{,,1}\\
\lambda+\lambda'+\gamma'&\le&\alpha_{22}\label{,,2}\\
\lambda&\le&\alpha_{33}\label{,,3}\\
\alpha_{12}&\le& \lambda+\lambda'+\gamma+\gamma'\label{,,4}\\
\alpha_{13}&\le& \lambda+\gamma\label{,,5}\\
\alpha_{21}&\le& \lambda+\lambda'\label{,,6}\\
\alpha_{23}&\le& \lambda\label{,,7}\\
\alpha_{31}&\le& \lambda\label{,,8}\\
\alpha_{32}&\le& \lambda+\gamma'\label{,,9}\\
0&\leq&\lambda,\lambda',\gamma,\gamma'\label{,,10}
 \end{eqnarray}
Similar to \ref{a0}, this achievable region  involves $14$ auxiliary random variables that do not appear in the outer bound, namely, $ \mu_{1}$, $\mu_{2}$, $\xi_1,\xi_2,\xi_3$,  $\lambda, \lambda'$, $\gamma, \gamma'$, $d_{\{1\}}$, $d_{\{2\}}$, $d_{\{3\}}$, $d_{\{1,2\}}$, $d_{\{1,2,3\}}$. The union over the regions corresponding to all feasible choices of these $14$ auxiliary variables is also achievable.  In the next step, we eliminate $ \mu_{1}$, $\mu_{2}$, $\xi_1,\xi_2,\xi_3$,  $d_{\{1\}}$, $d_{\{2\}}$, $d_{\{3\}}$, $d_{\{1,2\}}$, $d_{\{1,2,3\}}$ from $\mathcal{{F}}_{123}$ to obtain the simplified region $\mathcal{\bar{F}}_{123}$.
\subsubsection{Achievable Region $\mathcal{\bar{F}}_{123}$ }\label{four++-}
Similar to \eqref{four++-}, elimination of $ \mu_{1}$, $\mu_{2}$, $\xi_1,\xi_2,\xi_3$,  $d_{\{1\}}$, $d_{\{2\}}$, $d_{\{3\}}$, $d_{\{1,2\}}$, $d_{\{1,2,3\}}$ gives us the following equivalent region ${\mathcal{\bar{F}}}_{123}$ which retains only $4$ auxiliary variables $\lambda,\lambda',\gamma,\gamma'$.\footnote{The proof follows similar to Appendix \ref{d_i}.}
\begin{eqnarray}
\mathcal{\bar{F}}_{123}(\lambda,\lambda',\gamma,\gamma')=\Bigg\{(d_1,d_2,d_3)\in\mathbb{R}_+^3&:& \nonumber\\
d_1&\le& \alpha_{11}-\gamma,\label{f1bar}\\
d_2&\le& \alpha_{22}-\gamma',\label{f2bar}\\
d_3&\le& \alpha_{33},\label{f3bar}\\
d_1+d_2&\le& \alpha_{11}+\alpha_{22}-\lambda-\lambda'-\gamma-\gamma',\label{f12bar}\\
d_1+d_3&\le& \alpha_{11}+\alpha_{33}-\lambda-\gamma,\\
d_2+d_3&\le& \alpha_{22}+\alpha_{33}-\lambda-\gamma',\\
d_1+d_2+d_3&\le&  \alpha_{11}+\alpha_{22}+\alpha_{33}-2\lambda-\lambda'-\gamma-\gamma'\Bigg\}\label{f123bar}
\end{eqnarray} 
such that $\lambda,\lambda',\gamma,\gamma'$ satisfy conditions \eqref{,,1} to \eqref{,,5}.
\subsubsection{Achievable Region $\hat{\mathcal{F}}_{123}$}
Similar to \ref{hatDfinal}, the union of the regions ${\mathcal{\bar{F}}}_{123}$ over all possible choices of $\lambda,\lambda',\gamma,\gamma'$ gives us the following region ${\mathcal{\hat{F}}}_{123}$.
 \begin{align}
 \hat{\mathcal{F}}_{123}=\Bigg\{(d_1,d_2,d_3)&\in\mathbb{R}_+^3:\nonumber\\
 d_1&\le \alpha_{11}, \label{vw1}\\
 d_2&\le \alpha_{22},\\
d_3&\le \alpha_{33},\\
d_1+d_2&\le \alpha_{11}+\alpha_{22}-\max_{l,m\in[3],l\neq m}\alpha_{lm},\\
d_1+d_3&\le \alpha_{11}+\alpha_{33}-\max(\alpha_{23},\alpha_{31},\alpha_{13}),\\
d_2+d_3&\le \alpha_{22}+\alpha_{33}-\max(\alpha_{23},\alpha_{31},\alpha_{32}),\\
d_1+d_2+d_3&\le \alpha_{11}+\alpha_{22}+\alpha_{33}-\max \left\{\begin{matrix}
\max_{l,m\in[3],l\neq m}\alpha_{lm}+\max(\alpha_{31},\alpha_{23}),\\ 
\alpha_{13}+\alpha_{21},\\ 
\alpha_{32}+\alpha_{21},\\ 
\alpha_{32}+\alpha_{13},\\ 
\frac{\alpha_{12}+\alpha_{13}+\alpha_{32}+\alpha_{21}}{2}
\end{matrix}\right\}\label{vw2}\\
&&\Bigg\}\nonumber
\end{align} 
The equivalence of $\hat{\mathcal{F}}_{123}=\cup_{\lambda,\lambda',\gamma,\gamma'}\bar{\mathcal{F}}_{123}(\lambda,\lambda',\gamma,\gamma')$ is proved similar to the equivalence of $\hat{\mathcal{D}}_{123}=\cup_{\lambda,\lambda',\gamma,\gamma'}\bar{\mathcal{D}}_{123}(\lambda,\lambda',\gamma,\gamma')$.

\subsection{All $12$  Achievable GDoF Regions}
By symmetry, switching  the indices, e.g., $(1,2,3)\rightarrow(2,3,1)$ in $(\eqref{d1hat}-\eqref{d123hat})$ and $(\eqref{vw1}-\eqref{vw2})$, ten other achievable regions are obtained. Therefore, the following region is achievable.
 \begin{eqnarray}
\mathcal{{D}}_a=\bigcup_{\{i,j,k\}{=}\{1,2,3\}}\left(\mathcal{\hat{D}}_{ijk}\cup\mathcal{\hat{F}}_{ijk}\right)\label{Dach}
\end{eqnarray} 
where for distinct values of $\{i,j,k\}=\{1,2,3\}$, $\mathcal{\hat{D}}_{ijk}$ and $\mathcal{\hat{F}}_{ijk}$ are defined as follows.
 \begin{eqnarray}
\mathcal{\hat{D}}_{ijk}=\bigg\{(d_i,d_j,d_k)&:& 0\le d_1\le \alpha_{11}, 0\le d_2\le \alpha_{22},0\le d_3\le \alpha_{33},\label{WW1}\\
d_i+d_j&\le& \alpha_{ii}+\alpha_{jj}-\max_{l,m\in[3],l\neq m}\alpha_{lm},\label{WW2}\\
d_i+d_k&\le& \alpha_{ii}+\alpha_{kk}-\max(\alpha_{jk},\alpha_{kj},\alpha_{ki},\alpha_{ik}),\label{WW3}\\
d_j+d_k&\le& \alpha_{jj}+\alpha_{kk}-\max(\alpha_{jk},\alpha_{kj}),\label{WW4}\\
d_1+d_2+d_3&\le&\left.  \alpha_{11}+\alpha_{22}+\alpha_{33}-\max \left\{\begin{matrix}
\max_{l,m\in[3],l\neq m}\alpha_{lm}+\max(\alpha_{jk},\alpha_{kj}),\\ 
\alpha_{ik}+\alpha_{ki},\\ 
\alpha_{ki}+\alpha_{ij},\\ 
\alpha_{ji}+\alpha_{ik}
\end{matrix}\right\}\right\}\label{WW5}
\end{eqnarray} 
if $\max_{l,m\in[3],l\neq m}\alpha_{lm}\le\min(\alpha_{ii},\alpha_{jj})$. Otherwise, we define $\mathcal{\hat{D}}_{ijk}=\varnothing$.
 \begin{eqnarray}
\mathcal{\hat{F}}_{ijk}=\bigg\{(d_i,d_j,d_k)&:& 0\le d_1\le \alpha_{11}, 0\le d_2\le \alpha_{22},0\le d_3\le \alpha_{33},\label{W1}\\
d_i+d_j&\le& \alpha_{ii}+\alpha_{jj}-\max_{l,m\in[3],l\neq m}\alpha_{lm},\label{W2}\\
d_i+d_k&\le& \alpha_{ii}+\alpha_{kk}-\max(\alpha_{jk},\alpha_{ki},\alpha_{ik}),\label{W3}\\
d_j+d_k&\le& \alpha_{jj}+\alpha_{kk}-\max(\alpha_{jk},\alpha_{ki},\alpha_{kj}),\label{W4}\\
d_1+d_2+d_3&\le& \left. \alpha_{11}+\alpha_{22}+\alpha_{33}-\max \left\{\begin{matrix}
\max_{l,m\in[3],l\neq m}\alpha_{lm}+\max(\alpha_{ki},\alpha_{jk}),\\ 
\alpha_{ik}+\alpha_{ji},\\ 
\alpha_{kj}+\alpha_{ji},\\ 
\alpha_{kj}+\alpha_{ik},\\ 
\frac{\alpha_{ij}+\alpha_{ik}+\alpha_{kj}+\alpha_{ji}}{2}
\end{matrix}\right\}\right\}\label{W5}
\end{eqnarray} 
if $\max_{l,m\in[3],l\neq m}\alpha_{lm}\le\min(\alpha_{ii},\alpha_{jj})$. Otherwise, we define $\mathcal{\hat{F}}_{ijk}=\varnothing$.

\section{Achievability Matches the Outer Bound}
Finally,  $\mathcal{D}_a$  is shown to produce region \eqref{BB}. Specifically, for each value of parameters $\alpha_{ij}$, we show that one of the $12$ regions $\mathcal{\hat{D}}_{ijk},\mathcal{\hat{F}}_{ijk},\forall \{i,j,k\}=\{1,2,3\}$ subsumes all others and matches \eqref{BB}. For example, in the $3$ user MISO BC illustrated at the top of Fig. \ref{fig_region}, it is not difficult to verify that the GDoF region \eqref{BB} turns out to be identical to the region $\mathcal{\hat{F}}_{123}$ described in $(\eqref{vw1}-\eqref{vw2})$.
In this section, we prove that the GDoF region $\mathcal{{D}}_a$ defined in \eqref{Dach} and  $\mathcal{{D}}$ defined in \eqref{BB} are equivalent.
\begin{enumerate}
\item{$\mathcal{{D}}_a\subset\mathcal{{D}}$.}\\
In order to show that $\mathcal{{D}}_a\subset\mathcal{{D}}$, we prove  $\mathcal{\hat{D}}_{ijk}\subset\mathcal{{D}}$ and $\mathcal{\hat{F}}_{ijk}\subset\mathcal{{D}}$ for any $\{i,j,k\}=\{1,2,3\}$. For instance, consider the region $\mathcal{\hat{D}}_{123}$. Any tuple $(d_1,d_2,d_3)\in\mathcal{\hat{D}}_{123}$ satisfies  the inequalities $(\eqref{d1hat}-\eqref{d123hat})$.  Comparing $(\eqref{d1hat}-\eqref{d123hat})$ and \eqref{BB},  it is verified that $(d_1,d_2,d_3)\in\mathcal{D}$. For instance from \eqref{d12hat} we have,
 \begin{eqnarray}
d_1+d_2&\le& \alpha_{11}+\alpha_{22}-\max_{l,m\in[3],l\neq m}\alpha_{lm},\\
&\le& \alpha_{11}+\alpha_{22}-\max(\alpha_{12},\alpha_{21})
\end{eqnarray}
Therefore, we conclude that $\mathcal{\hat{D}}_{123}\subset\mathcal{D}$. Similarly, $\mathcal{\hat{D}}_{ijk}\subset\mathcal{{D}}$ and $\mathcal{\hat{F}}_{ijk}\subset\mathcal{{D}}$ is concluded for any $\{i,j,k\}=\{1,2,3\}$. 
\item{$\mathcal{{D}}\subset\mathcal{{D}}_a$.}\\
Without loss of generality assume $\alpha_{12}$ is the largest of all cross links. $$\alpha_{12}=\max_{l,m\in[3],l\neq m}\alpha_{lm}.$$ Therefore, from \eqref{con1}, $\max_{l,m\in[3],l\neq m}\alpha_{lm}\le\min(\alpha_{11},\alpha_{22})$. Consider the following three cases.
\begin{enumerate}
\item{$\max(\alpha_{13},\alpha_{31})\le\alpha_{23}$.} Consider any tuple $(d_1,d_2,d_3)\in\mathcal{{D}}$. From \eqref{BB}, $\mathcal{{D}}$ is represented as
 \begin{eqnarray}
\mathcal{{D}}=\{(d_1,d_2,d_3)&:& 0\le d_1\le \alpha_{11}, 0\le d_2\le \alpha_{22},0\le d_3\le \alpha_{33},\\
d_1+d_2&\le& \alpha_{11}+\alpha_{22}-\alpha_{12},\\
d_1+d_3&\le& \alpha_{11}+\alpha_{33}-\max(\alpha_{13},\alpha_{31}),\\
d_2+d_3&\le& \alpha_{22}+\alpha_{33}-\max(\alpha_{23},\alpha_{32}),\\
d_1+d_2+d_3&\le&  \alpha_{11}+\alpha_{22}+\alpha_{33}-\max \left\{\alpha_{12}+\alpha_{23},
\alpha_{32}+\alpha_{21}\right\}
\end{eqnarray} 
On the other hand, from $(\eqref{WW1}-\eqref{WW5})$ as $\max_{l,m\in[3],l\neq m}\alpha_{lm}\le\min(\alpha_{11},\alpha_{22})$, $\mathcal{\hat{D}}_{213}$ is equal to,
 \begin{eqnarray}
\mathcal{\hat{D}}_{213}=\{(d_1,d_2,d_3)&:& 0\le d_1\le \alpha_{11}, 0\le d_2\le \alpha_{22},0\le d_3\le \alpha_{33},\\
d_1+d_2&\le& \alpha_{11}+\alpha_{22}-\alpha_{12},\\
d_1+d_3&\le& \alpha_{11}+\alpha_{33}-\max(\alpha_{13},\alpha_{31}),\\
d_2+d_3&\le& \alpha_{22}+\alpha_{33}-\max(\alpha_{13},\alpha_{31},\alpha_{23},\alpha_{32}),\\
d_1+d_2+d_3&\le&  \alpha_{11}+\alpha_{22}+\alpha_{33}-\max \left\{\begin{matrix}
\alpha_{12}+\max(\alpha_{13},\alpha_{31}),\\ 
\alpha_{23}+\alpha_{32},\\ 
\alpha_{32}+\alpha_{21},\\ 
\alpha_{12}+\alpha_{23}
\end{matrix}\right.
\end{eqnarray} 
In this case, $\mathcal{{D}}=\mathcal{\hat{D}}_{213}$ as $\max(\alpha_{13},\alpha_{31})\le\alpha_{23}$ and $\alpha_{12}$ is the biggest one among all cross links.

\item{$\max(\alpha_{23},\alpha_{32})\le\alpha_{31}$.} From \eqref{BB}, $(\eqref{d1hat}-\eqref{d123hat})$ and $\max_{l,m\in[3],l\neq m}\alpha_{lm}\le\min(\alpha_{11},\alpha_{22})$, we conclude that $\mathcal{{D}}=\mathcal{\hat{D}}_{123}$.
 \begin{eqnarray}
\mathcal{{D}}=\hat{\mathcal{D}}_{123}&=&\Bigg\{(d_1,d_2,d_3): 0\le d_1\le \alpha_{11}, 0\le d_2\le \alpha_{22},0\le d_3\le \alpha_{33},\\
d_1+d_2&\le& \alpha_{11}+\alpha_{22}-\alpha_{12},\\
d_1+d_3&\le& \alpha_{11}+\alpha_{33}-\max(\alpha_{31},\alpha_{13}),\\
d_2+d_3&\le& \alpha_{22}+\alpha_{33}-\max(\alpha_{23},\alpha_{32}),\\
d_1+d_2+d_3&\le&  \alpha_{11}+\alpha_{22}+\alpha_{33}-\max (\alpha_{13}+\alpha_{21}, 
\alpha_{12}+\alpha_{31})\bigg\}\nonumber\\
\end{eqnarray}

\item{$\alpha_{31}\le\max(\alpha_{23},\alpha_{32}),\alpha_{23}\le\max(\alpha_{13},\alpha_{31})$.}  In this case from \eqref{BB},  $\mathcal{{D}}$ is represented as
 \begin{eqnarray}
\mathcal{{D}}=\{(d_1,d_2,d_3)&:& 0\le d_1\le \alpha_{11}, 0\le d_2\le \alpha_{22},0\le d_3\le \alpha_{33},\\
d_1+d_2&\le& \alpha_{11}+\alpha_{22}-\alpha_{12},\\
d_1+d_3&\le& \alpha_{11}+\alpha_{33}-\max(\alpha_{13},\alpha_{31}),\\
d_2+d_3&\le& \alpha_{22}+\alpha_{33}-\max(\alpha_{23},\alpha_{32}),\\
d_1+d_2+d_3&\le&  \alpha_{11}+\alpha_{22}+\alpha_{33}-\max \left\{\begin{matrix}
\alpha_{12}+\max(\alpha_{23},\alpha_{31}),\\ 
\alpha_{13}+\alpha_{21},\\ 
\alpha_{32}+\alpha_{21},\\ 
\alpha_{13}+\alpha_{32},\\
\frac{\alpha_{12}+\alpha_{13}+\alpha_{32}+\alpha_{21}}{2}
\end{matrix}\right.
\end{eqnarray} 
Therefore, $\mathcal{{D}}=\mathcal{\hat{F}}_{123}$  from $(\eqref{vw1}-\eqref{vw2})$ and $\max_{l,m\in[3],l\neq m}\alpha_{lm}\le\min(\alpha_{11},\alpha_{22})$.

\end{enumerate}
\end{enumerate}
 
\section{Conclusion}
A broad regime of channel strength parameters is identified where simple layered superposition coding achieves the GDoF region of a $3$ user  MISO BC with $M$ antennas at the transmitter, under finite precision CSIT. The parameter regime is larger than the corresponding regime for the $3$ user IC where treating interference as noise (TIN) is shown to be GDoF-optimal, and reveals an interesting duality property in that the region remains unchanged if the roles of all transmit antennas and receive antennas are switched. {{}Extensions to $K\geq 4$ users for the MISO BC, 
is studied in Theorem \ref{theorem:GDoF2}.} The combination of simplicity, robustness and information theoretic optimality imparts this research avenue the potential for both theoretical and practical impact.

\appendix
\section{Proof of Sum GDoF Bound of $15/4$ in the Three User Cyclic $(1,2,2)$ MISO BC}\label{154}
Consider the three user cyclic $(1,2,2)$ MISO BC in Fig. \ref{figcyclic}. From the deterministic model in Section \ref{DM_1},  the following input-output relationship holds,
\begin{eqnarray}
\bar{Y}_{k}(t)&=&\left\lfloor G_{kk}(t)\left\lfloor\bar{P}^{-1} \bar{X}_k(t)\right\rfloor\right\rfloor+\sum_{m\in[3],m\neq k}\lfloor G_{km}(t)\bar{X}_m(t)\rfloor\\
&=&\left\lfloor G_{kk}(t)\left(\bar{X}_k(t)\right)_1^2\right\rfloor+\sum_{m\in[3],m\neq k}\lfloor G_{km}(t)\bar{X}_m(t)\rfloor
\end{eqnarray}
where $\bar{X}_{k}(t)\in\mathcal{X}_{2}$ for all $k\in[3],t\in[T]$. Define the random variables $\bar{Y'}_{k}(t)$ as,
\begin{eqnarray}
\bar{Y'}_{1}(t)&=&\sum_{m\in[3]}\lfloor G'_{1m}(t)\bar{X}_m(t)\rfloor\\
\bar{Y'}_{2}(t)&=&\sum_{m\in[3]}\lfloor G'_{2m}(t)\bar{X}_m(t)\rfloor\\
\bar{Y'}_{3}(t)&=&\left\lfloor G'_{33}(t)\lfloor\bar{P}^{-1} \bar{X}_3(t)\rfloor\right\rfloor+\sum_{m\in[3],m\neq3}\lfloor G'_{3m}(t)\bar{X}_m(t)\rfloor\\
&=&\left\lfloor G'_{33}(t)(\bar{X}_3(t))_1^2\right\rfloor+\sum_{m\in[3],m\neq3}\lfloor G'_{3m}(t)\bar{X}_m(t)\rfloor
\end{eqnarray}
where  for all $k,m\in[3]$, $G'_{km}(t)$ are distinct random variables chosen from $\mathcal{G}$ and are different from the random variables $G_{km}(t), \forall k,m\in[3]$.  Writing Fano's inequality for all  three users, we obtain the following bounds,\footnote{Suppressing $o(T)$ terms for simplicity, we have
\begin{eqnarray}
TR_1+TR_2&\le&I(\bar{Y}_{1}^{[T]};W_1\mid \mathcal{G})+I(\bar{Y}_{2}^{[T]};W_2\mid \mathcal{G})\\
&\le& I(\bar{Y}_{1}^{[T]},\bar{Y}_{2}^{[T]};W_1,W_2\mid \mathcal{G})\label{jg1}\\
&\le& I(\bar{Y}_{1}^{[T]},\bar{Y}_{2}^{[T]};W_1,W_2\mid W_3,\mathcal{G})\label{jg2}
\end{eqnarray}
\eqref{jg1} and \eqref{jg2} follow from the facts that $I(A;B)+I(D;C)\le I(A,D;B,C)$ and $I(A;B)\le I(A;B\mid C)$ if $B$ and $C$ are independent of each other.}
\begin{eqnarray}
TR_1+TR_2&\le& I(\bar{Y}_{1}^{[T]},\bar{Y}_{2}^{[T]};W_1,W_2\mid W_3,\mathcal{G})\label{fa1}\\
TR_3&\le& I(\bar{Y}_{3}^{[T]};W_3\mid \mathcal{G})\label{fa2}
\end{eqnarray}
From \eqref{fa2}, we have,
\begin{eqnarray}
2TR_3&\le&2 I(\bar{Y}_{3}^{[T]};W_3\mid \mathcal{G})\nonumber\\
&\le&4T\log{\bar{P}}-2H(\bar{Y}_{3}^{[T]}\mid W_3, \mathcal{G})+T~o~(\log{\bar{P}})\label{fa3}\\
&=&4T\log{\bar{P}}-H(\bar{Y}_{3}^{[T]}\mid W_3, \mathcal{G})-H(\bar{Y'}_{3}^{[T]}\mid W_3, \mathcal{G})+T~o~(\log{\bar{P}})\label{fa4}
\end{eqnarray}
where \eqref{fa3} is true as similar to \eqref{fn5} we have  $H(\bar{Y}_{3}^{[T]}\mid \mathcal{G})\le 2T\log{\bar{P}}+T~o~(\log{\bar{P}})$. In order to check whether \eqref{fa4} is true or not observe that, $\bar{Y'}_{3}(t)$ is a bounded density copy of $\bar{Y}_{3}(t)$. So, we expect that 
\begin{eqnarray}
\mid H(\bar{Y}_{3}^{[T]}\mid W_3, \mathcal{G})-H(\bar{Y'}_{3}^{[T]}\mid W_3, \mathcal{G})\mid&\le& T~o~(\log{\bar{P}})
\end{eqnarray}
 which is true\footnote{Note that from  \eqref{deltaij} we have,
\begin{eqnarray}
H(\bar{Y}_{3}^{[T]}\mid W_3, \mathcal{G})-H(\bar{Y'}_{3}^{[T]}\mid W_3, \mathcal{G})&\le&T~o~(\log{\bar{P}})\\
H(\bar{Y'}_{3}^{[T]}\mid W_3, \mathcal{G})-H(\bar{Y}_{3}^{[T]}\mid W_3, \mathcal{G})&\le&T~o~(\log{\bar{P}})
\end{eqnarray}} from \eqref{deltaij}. Summing over \eqref{fa1}  and \eqref{fa4}, we have,
\begin{eqnarray}
&&TR_1+TR_2+2TR_3 \nonumber\\
&\le&4T\log{\bar{P}}+\bigg(H(\bar{Y}_{1}^{[T]},\bar{Y}_{2}^{[T]}\mid W_3,\mathcal{G})-H(\bar{Y}_{3}^{[T]},\bar{Y'}_{3}^{[T]}\mid W_3,\mathcal{G})\bigg)+T~o~(\log{\bar{P}})\label{5/4c}
\end{eqnarray}
With the aid of Lemma \ref{lemmaxc}, let us prove that
\begin{eqnarray}
H(\bar{Y}_{1}^{[T]},\bar{Y}_{2}^{[T]}\mid W_3,\mathcal{G})-H(\bar{Y}_{3}^{[T]},\bar{Y'}_{3}^{[T]}\mid W_3,\mathcal{G})
&\le&T\log{\bar{P}}+T~o~(\log{\bar{P}})\label{5/4}
\end{eqnarray}
\eqref{5/4} is proved in the following three steps.
\begin{enumerate}
\item{}Consider the random variables $\bar{Y}_{1}(t)$ and $\bar{Y'}_{1}(t)$.  $\bar{Y}_{1}(t)$ is a bounded density linear combination of  $\lfloor\bar{P}^{-1}\bar{X}_1(t)\rfloor = \left(\bar{X}_1(t)\right)_1^2,\bar{X}_2(t),\bar{X}_3(t)$ while $\bar{Y'}_{1}(t)$ is a bounded density linear combination of  $\bar{X}_1(t),\bar{X}_2(t),\bar{X}_3(t)$. Now, compare the terms $H(\bar{Y}_{1}^{[T]}\mid \bar{Y}_{2}^{[T]},W_3,\mathcal{G})$ and $H(\bar{Y'}_{1}^{[T]}\mid \bar{Y}_{2}^{[T]},W_3,\mathcal{G})$. Due to the bounded density assumption, we expect that
\begin{eqnarray}
&&H(\bar{Y}_{1}^{[T]},\bar{Y}_{2}^{[T]}\mid W_3,\mathcal{G})-H(\bar{Y'}_{1}^{[T]},\bar{Y}_{2}^{[T]}\mid W_3,\mathcal{G})\nonumber\\
&=&H(\bar{Y}_{1}^{[T]}\mid \bar{Y}_{2}^{[T]},W_3,\mathcal{G})-H(\bar{Y'}_{1}^{[T]}\mid \bar{Y}_{2}^{[T]},W_3,\mathcal{G})\nonumber\\
&\le&T~o~(\log{\bar{P}})\label{fa5}
\end{eqnarray}
which is true from \eqref{deltaij}.
\item{} Similarly, we have
\begin{eqnarray}
&&H(\bar{Y'}_{1}^{[T]},\bar{Y}_{2}^{[T]}\mid W_3,\mathcal{G})-H(\bar{Y'}_{1}^{[T]},\bar{Y'}_{2}^{[T]}\mid W_3,\mathcal{G})\nonumber\\
&=&H(\bar{Y}_{2}^{[T]}\mid \bar{Y'}_{1}^{[T]},W_3,\mathcal{G})-H(\bar{Y'}_{2}^{[T]}\mid \bar{Y'}_{1}^{[T]},W_3,\mathcal{G})\nonumber\\
&\le&T~o~(\log{\bar{P}})\label{fa6}
\end{eqnarray}
where \eqref{fa6} follows from \eqref{deltaij} similar to \eqref{fa5}.
\item{} Now, let us prove the following inequality.
\begin{eqnarray}
H(\bar{Y'}_{1}^{[T]},\bar{Y'}_{2}^{[T]}\mid W_3,\mathcal{G})-H(\bar{Y}_{3}^{[T]},\bar{Y'}_{3}^{[T]}\mid W_3,\mathcal{G})
&\le&T\log{\bar{P}}+T~o~(\log{\bar{P}})\label{fa7}
\end{eqnarray}
 To apply Lemma \ref{lemmaxc}, set $l=3$, $N=2$ and $\eta=2$. The random variables  $\bar{ U}_{11}^{[T]}$, $\bar{ U}_{12}^{[T]}$, $\bar{ U}_{21}^{[T]}$, $\bar{ U}_{22}^{[T]}$, $\bar{ V}_1^{[T]}$, $\bar{ V}_2^{[T]}$ and $\bar{ V}_3^{[T]}$ are interpreted as $\bar{ Y'}_1^{[T]}$, $\bar{ Y'}_2^{[T]}$, $\bar{ Y}_3^{[T]}$,$\bar{ Y'}_3^{[T]}$, $\bar{ X}_1^{[T]}$, $\bar{ X}_2^{[T]}$ and $\bar{ X}_3^{[T]}$, respectively.  Thus, from Lemma \ref{lemmaxc} we conclude \eqref{fa7} as  $(\lambda_{11}-\lambda_{21})^+=1$,  $(\lambda_{12}-\lambda_{22})^+=(\lambda_{13}-\lambda_{23})^+=0$.
\end{enumerate}
 \eqref{5/4} is concluded by summing \eqref{fa5}, \eqref{fa6} and \eqref{fa7}. By symmetry from \eqref{5/4c} and \eqref{5/4} we have,
\begin{eqnarray}
TR_1+TR_2+2TR_3&\le&5T\log{\bar{P}}+T~o~(\log{\bar{P}})\label{fa8}\\
TR_1+2TR_2+TR_3&\le&5T\log{\bar{P}}+T~o~(\log{\bar{P}})\label{fa9}\\
2TR_1+TR_2+TR_3&\le&5T\log{\bar{P}}+T~o~(\log{\bar{P}})\label{fa10}
\end{eqnarray}
Summing \eqref{fa8}, \eqref{fa9} and \eqref{fa10}  and applying the GDoF limit, we conclude that $d_1+d_2+d_3\le 15/4$.

\section{Proof of Theorem \ref{theorem:GDoF2}} \label{kconverse}
Consider  a $K$ user MISO BC with $M$ antennas at the transmitter. Our goal is to prove that,  if $(A, B)$ is a bounding pattern for $A=\{{\bf p}_1, {\bf p}_2, \cdots, {\bf p}_m\}$, $B=\{{\bf q}_1, {\bf q}_2, \cdots, {\bf q}_n\}$, then the GDoF region is bounded by, 
\begin{eqnarray}
\sum_{{\bf p}\in A\uplus B}\sum_{i=2}^{|{\bf p}|}d_{p(i)}\le \sum_{{\bf p}\in A\uplus B} f({\bf p}), \label{Bb6+}
\end{eqnarray}
where $f({\bf p})$ is defined in \eqref{f(p)}. The first step of the proof is the transformation into a deterministic setting which is the same as the one in \ref{proof3}.
\subsection{Deterministic Model}\label{DM_1K}
Similar to \ref{DM_1}, the following relationship is assumed between the transmitted and received signals,
\begin{align}
\bar{Y}_{j}(t)=&L_{j}^b(t)\big((\bar{X}_1(t))^{\max_{k\in[K]}\alpha_{k1}}_{\max_{k\in[K]}\alpha_{k1}-\alpha_{j1}},(\bar{X}_2(t))^{\max_{k\in[K]}\alpha_{k2}}_{\max_{k\in[K]}\alpha_{k2}-\alpha_{j2}},\cdots,(\bar{X}_M(t))^{\max_{k\in[K]}\alpha_{kM}}_{\max_{k\in[K]}\alpha_{kM}-\alpha_{jM}}\big)\label{dm1K}
\end{align}
for all $j\in[K],t\in[T]$.   Moreover, we assume $\bar{X}_{m}(t)\in\mathcal{X}_{\max_{k\in[K]}\alpha_{km}}$, $\forall m\in[M],t\in[T]$.
\subsection{Some Observations}
 In order to prove \eqref{Bb6+}, consider some arbitrary bounding pattern 
$(A, B)$ where $A=\{{\bf p}_1, {\bf p}_2, \cdots, {\bf p}_m\}$, $B=\{{\bf q}_1, {\bf q}_2, \cdots, {\bf q}_n\}$.
Consider two permutations ${\bf p}=({ p}(1), { p}(2), \cdots, { p}(|{\bf p}|))$ and ${\bf q}=({ q}(1), { q}(2), \cdots, { q}(|{\bf q}|))$ with non-zero elements. From Lemma \ref{lemma2}, we have
\begin{eqnarray}
d_{{ p}(2)}+d_{{ p}(3)}+\cdots+d_{{ p}(|{\bf p}|)}&\leq& \bar{H}(Y_{{ p}(1)}\mid W_{{ p}(1)}, W_{S_{{\bf p}}'})+f({\bf p}) - \bar{H}(Y_{{ p}(|{\bf p}|)}\mid W_{S_{{\bf p}}}, W_{S_{{\bf p}}'})\label{fano_per1}\\
d_{{ q}(2)}+d_{{ q}(3)}+\cdots+d_{{ q}(|{\bf q}|)}&\leq& \bar{H}(Y_{{ q}(1)}\mid W_{{ q}(1)}, W_{S_{{\bf q}}^c})+f({\bf q})\label{fano_per2}
\end{eqnarray}
where $W_{S_{{\bf p}}}$, $W_{S_{{\bf p}}'}$ and $W_{S_{{\bf q}}^c}$ satisfy
\begin{eqnarray}
W_{S_{{\bf p}}}=\{W_i;i\in[K],i\in\{{ p}(1), { p}(2), \cdots, { p}(|{\bf p}|)\}\}\\
W_{S_{{\bf p}}'}\subset\{W_i;i\in[K],i\notin\{{ p}(1), {\bf p}(2), \cdots, { p}(|{\bf p}|)\}\}\\
W_{S_{{\bf q}}^c}=\{W_i;i\in[K],i\notin\{{ q}(1), { q}(2), \cdots, { q}(|{\bf q}|)\}\}
\end{eqnarray}
Consider any permutation ${\bf r}=(0,{ r}(2))\in A$. Similarly, we have
\begin{eqnarray}
d_{{ r}(2)}&\le& f({\bf 	r})-\bar{H}({Y}_{{ r}(2)}\mid W_{{ r}(2)})\label{fano_per3}
\end{eqnarray}
We choose the sets $W_{S_{{\bf p}}'}$ in a way that the following condition is satisfied for the bounding pattern $(A, B)$.
\begin{align}
&\sum_{{\bf p}\in A, { p}(k)\neq 0, \forall k\in[|{\bf p}|]} \{ \bar{H}(Y_{{ p}(1)}\mid W_{{ p}(1)}, W_{S_{{\bf p}}'})- \bar{H}(Y_{{ p}(|{\bf p}|)}\mid W_{S_{{\bf p}}}, W_{S_{{\bf p}}'})\}\nonumber\\
&-\sum_{{\bf r}\in A, { r}(1)= 0} \{ \bar{H}({Y}_{{ r}(2)}\mid W_{{ r}(2)})\}+\sum_{{\bf q}\in B}  \{\bar{H}(Y_{{ q}(1)}\mid W_{{ q}(1)}, W_{S_{{\bf q}}^c})\}\le 0\label{condition_1}
\end{align}
 Summing  \eqref{fano_per1}, \eqref{fano_per2}, \eqref{fano_per3} for all permutations in $A\uplus B$  and \eqref{condition_1} we conclude \eqref{Bb6+} as follows.
 \begin{eqnarray}
\sum_{{\bf p}\in A\uplus B}\sum_{i=2}^{|{\bf p}|}d_{p(i)}\le \sum_{{\bf p}\in A\uplus B} f({\bf p}) \label{Bb6++}
\end{eqnarray}

\subsection{ Proof of \eqref{condition_1}}\label{con_pqr}
 Our goal is to choose $W_{S_{{\bf p}}'}$ for any ${\bf p}\in A$ in a way that \eqref{condition_1}  is satisfied. From Definition \ref{defmulti}, any bounding pattern $A\uplus B$ satisfy the three conditions  specified in Definition \ref{defmulti}. We prove the bound \eqref{condition_1} for any bounding pattern $A\uplus B$ by induction over  $|A\uplus B|$.\footnote{Note that, for any $A\uplus B$, $|A\uplus B|$ is an even number from Definition \ref{defmulti}.}
\subsubsection{$|s|=2$}
From Definition \ref{defmulti}, any $A\uplus B$ where $|A\uplus B|=2$ is of the form of $\left\{\bar{\bf p},{\bf p}'\right\}$ where  $\bar{\bf p}=\overline{({p}(1),{p}(2),\cdots,{p}(l(\bar{\bf p})))}$ is a permutation of some subset of $[K]$  and ${\bf p}'=(0,{p}(1))$.  In this case, \eqref{condition_1} is simplified as follows by choosing $W_{S_{{\bf p}}'}=\{W_i;i\in[K],i\notin\{{p}(2),\cdots,{p}(l(\bar{\bf p}))\}\}$.
\begin{eqnarray}
 -\bar{H}(Y_{{ p}(1)}\mid W_{{ p}(1)})+\bar{H}(Y_{{ p}(1)}\mid W_{{ p}(1)},W_{S_{\bar{\bf p}}^c})\le0
\end{eqnarray}
which is true as conditioning decreases the entropy.
\subsubsection{$|A\uplus B|=2c$ for all $2\le c$}\label{c52}
Let us assume that the bound \eqref{condition_1} is true for any $A\uplus B$ where $|A\uplus B|\le 2c$  and prove \eqref{condition_1}  for  $|A\uplus B|= 2c$. Consider some arbitrary multiset $A\uplus B$ where $|s|=2c$. Two cases are possible for this multiset. It is either created by multiset sum in Definition \ref{defmulti},  or by {\it merging}  two permutations of a multiset. 
\begin{enumerate}
\item{ $(A=A_1\uplus A_2, B=B_1\uplus B_2)$.}\\
As $(A_1,B_1)$ and $(A_2,B_2)$ are bounding patterns, we have
\begin{align}
&\sum_{{\bf p}\in A_1, { p}(k)\neq 0, \forall k\in[|{\bf p}|]} \{ \bar{H}(Y_{{ p}(1)}\mid W_{{ p}(1)}, W_{S_{{\bf p}}'})- \bar{H}(Y_{{ p}(|{\bf p}|)}\mid W_{S_{{\bf p}}}, W_{S_{{\bf p}}'})\}\nonumber\\
&-\sum_{{\bf r}\in A_1, { r}(1)= 0} \{ \bar{H}({Y}_{{ r}(2)}\mid W_{{ r}(2)})\}+\sum_{{\bf q}\in B_1}  \{\bar{H}(Y_{{ q}(1)}\mid W_{{ q}(1)}, W_{S_{{\bf q}}^c})\}\le 0\label{jjo1}\\
&\sum_{{\bf p}\in A_2, { p}(k)\neq 0, \forall k\in[|{\bf p}|]} \{ \bar{H}(Y_{{ p}(1)}\mid W_{{ p}(1)}, W_{S_{{\bf p}}'})- \bar{H}(Y_{{ p}(|{\bf p}|)}\mid W_{S_{{\bf p}}}, W_{S_{{\bf p}}'})\}\nonumber\\
&-\sum_{{\bf r}\in A_2, { r}(1)= 0} \{ \bar{H}({Y}_{{ r}(2)}\mid W_{{ r}(2)})\}+\sum_{{\bf q}\in B_2}  \{\bar{H}(Y_{{ q}(1)}\mid W_{{ q}(1)}, W_{S_{{\bf q}}^c})\}\le 0\label{jjo2}
\end{align}
Summing \eqref{jjo1} and \eqref{jjo2}, \eqref{condition_1} is concluded for $(A=A_1\uplus A_2, B=B_1\uplus B_2)$  as  $\sum_{x\in s_1\uplus s_2}f(x)=\sum_{x\in s_1}f(x)+\sum_{x\in s_2}f(x)$ for any function $f(x)$.

\item{$A\uplus B$ is obtained from {\it merging} two permutations.}  \\
Consider a bounding pattern $A\uplus B$ obtained from  {\it merging} two permutations of bounding pattern $(A',B')$, i.e.,
\begin{eqnarray}
A&=&\{{\bf p}_{1}, {\bf p}_{2},\cdots, {\bf p}_{m},{\bf u}_1, {\bf u}_2\}\\
B&=&\{{\bf q}_{3}, {\bf q}_{4}, \cdots, {\bf q}_{n},{\bf u}_3, {\bf u}_4\}\\
A'&=&\{{\bf p}_{1}, {\bf p}_{2},\cdots, {\bf p}_{m}\}\\
B'&=&\{{\bf q}_{1}, {\bf q}_{2}, \cdots, {\bf q}_{n}\}
\end{eqnarray}
where two permutations ${\bf q}_{1}, {\bf q}_{2}$ are merged to obtain ${\bf u}_1, {\bf u}_2, {\bf u}_3, {\bf u}_4$ as described in Definition \ref{defpp}. From the induction assumption \eqref{condition_1} is true for the multiset $A'\uplus B'$ as $|A'\uplus B'|=2c-2$, i.e.,  for any ${\bf p}\in A'$ there exists $W_{S_{{\bf p}}'}$ that the following condition is satisfied for the bounding pattern $(A', B')$.
\begin{align}
&\sum_{{\bf p}\in A', { p}(k)\neq 0, \forall k\in[|{\bf p}|]} \{ \bar{H}(Y_{{ p}(1)}\mid W_{{ p}(1)}, W_{S_{{\bf p}}'})- \bar{H}(Y_{{ p}(|{\bf p}|)}\mid W_{S_{{\bf p}}}, W_{S_{{\bf p}}'})\}\nonumber\\
&-\sum_{{\bf r}\in A', { r}(1)= 0} \{ \bar{H}({Y}_{{ r}(2)}\mid W_{{ r}(2)})\}+\sum_{{\bf q}\in B'}  \{\bar{H}(Y_{{ q}(1)}\mid W_{{ q}(1)}, W_{S_{{\bf q}}^c})\}\le 0\label{jjo3}
\end{align}
On the other hand, writing \eqref{condition_1} for the multiset $A\uplus B$ we need to prove that
\begin{align}
&\sum_{{\bf p}\in A, { p}(k)\neq 0, \forall k\in[|{\bf p}|]} \{ \bar{H}(Y_{{ p}(1)}\mid W_{{ p}(1)}, W_{S_{{\bf p}}'})- \bar{H}(Y_{{ p}(|{\bf p}|)}\mid W_{S_{{\bf p}}}, W_{S_{{\bf p}}'})\}\nonumber\\
&-\sum_{{\bf r}\in A, { r}(1)= 0} \{ \bar{H}({Y}_{{ r}(2)}\mid W_{{ r}(2)})\}+\sum_{{\bf q}\in B}  \{\bar{H}(Y_{{ q}(1)}\mid W_{{ q}(1)}, W_{S_{{\bf q}}^c})\}\le 0\label{jjo4}
\end{align}
Decreasing \eqref{jjo3} from \eqref{jjo4}, it is sufficient to prove the following bound.
\begin{align}
&-\bar{H}(Y_{{ q}_1(1)}\mid W_{{ q}_1(1)}, W_{S_{{\bf q}_1}^c})-\bar{H}(Y_{{ q}_2(1)}\mid W_{{ q}_2(1)}, W_{S_{{\bf q}_2}^c})+\bar{H}(Y_{{ u}_3(1)}\mid W_{{ u}_3(1)}, W_{S_{{\bf u}_3}^c})\nonumber\\
&+\bar{H}(Y_{{ u}_4(1)}\mid W_{{ u}_4(1)}, W_{S_{{\bf u}_4}^c})+\bar{H}(Y_{{ u}_1(1)}\mid W_{{ u}_1(1)}, W_{S_{{\bf u}_1}'})- \bar{H}(Y_{{ u}_1(|{\bf u}_1|)}\mid W_{S_{{\bf u}_1}}, W_{S_{{\bf u}_1}'})\nonumber\\
&+\bar{H}(Y_{{ u}_2(1)}\mid W_{{ u}_2(1)}, W_{S_{{\bf u}_2}'})- \bar{H}(Y_{{ u}_2(|{\bf u}_2|)}\mid W_{S_{{\bf u}_2}}, W_{S_{{\bf u}_2}'})\le 0
\end{align}
Note that from Definition \ref{defpp}, ${ q}_1(1)={ u}_1(1)$, ${ q}_2(1)={ u}_2(1)$ and ${ u}_1(|{\bf u}_1|)={ u}_2(|{\bf u}_2|)={ u}_3(1)={ u}_4(1)$. Let us choose $W_{S_{{\bf u}_1}'}=W_{S_{{\bf q}_1}^c}$ and $W_{S_{{\bf u}_2}'}=W_{S_{{\bf q}_2}^c}$. Therefore, we have
\begin{eqnarray}
\bar{H}(Y_{{ q}_1(1)}\mid W_{{ q}_1(1)}, W_{S_{{\bf q}_1}^c})&=&\bar{H}(Y_{{ u}_1(1)}\mid W_{{ u}_1(1)}, W_{S_{{\bf u}_1}'})\\
\bar{H}(Y_{{ q}_2(1)}\mid W_{{ q}_2(1)}, W_{S_{{\bf q}_2}^c})&=&\bar{H}(Y_{{ u}_2(1)}\mid W_{{ u}_2(1)}, W_{S_{{\bf u}_2}'})
\end{eqnarray}
\begin{eqnarray}
&&\bar{H}(Y_{{ u}_3(1)}\mid W_{{ u}_3(1)}, W_{S_{{\bf u}_3}^c})+\bar{H}(Y_{{ u}_4(1)}\mid W_{{ u}_4(1)}, W_{S_{{\bf u}_4}^c})\nonumber\\
&&- \bar{H}(Y_{{ u}_1(|{\bf u}_1|)}\mid W_{S_{{\bf u}_1}}, W_{S_{{\bf u}_1}'})- \bar{H}(Y_{{ u}_2(|{\bf u}_2|)}\mid W_{S_{{\bf u}_2}}, W_{S_{{\bf u}_2}'})\nonumber\\
&=&\bar{H}(Y_{{ u}_3(1)}\mid W_{{ u}_3(1)}, W_{S_{{\bf u}_3}^c})+\bar{H}(Y_{{ u}_3(1)}\mid W_{{ u}_3(1)}, W_{S_{{\bf u}_4}^c})\nonumber\\
&&- \bar{H}(Y_{{ u}_3(1)}\mid W_{S_{{\bf u}_1}}, W_{S_{{\bf q}_1}^c})- \bar{H}(Y_{{ u}_3(1)}\mid W_{S_{{\bf u}_2}}, W_{S_{{\bf q}_2}^c})\nonumber\\
&\le& 0\label{jjo7}
\end{eqnarray}
\eqref{jjo7} follows similar to proof of Lemma \ref{lemma3}  using the submodularity property of entropy, $H(X|A)+H(X|B)\geq H(X|A\cup B)+H(X|A\cap B)$ as follows.  Consider the {\it merge} of two permutations $\bar{\bf q}_1=({ p}(1),\cdots,{ p}(k))$ and  $\bar{\bf q}_2=({ q}(1),\cdots,{ q}(l))$ for the two numbers $k'$ and $l'$, i.e.,
\begin {eqnarray}
{p}(k')&=&{ q}(l')\\
{\bf u}_1&=&({ p}(1),\cdots,{ p}(k'))\\
{\bf u}_2&=&({ q}(1),\cdots,{ q}(l'))\\
\bar{\bf u}_3&=&({ p}(k'),i_1,i_2,\cdots,i_{|{\bf p}_{+}\cap {\bf q}_{+}|}),\\
&&\{i_1,\cdots,i_{|{\bf p}_{+}\cap {\bf q}_{+}|}\}={\bf p}_{+}\cap {\bf q}_{+}\nonumber\\
\bar{\bf u}_4&=&({ q}(l'),j_1,j_2,\cdots,j_{|{\bf p}_{+}\cup {\bf q}_{+})|},\\
&&\{j_1,\cdots,j_{|{\bf p}_{+}\cup {\bf q}_{+}|}\}={\bf p}_{+}\cup {\bf q}_{+}\nonumber
\end{eqnarray}
where ${\bf p}_{+}$ and ${\bf q}_{+}$ are defined as $\{{ p}(k'+1),\cdots,{ p}(k)\}$ and $\{{ q}(l'+1),\cdots,{ q}(l)\}$, respectively. Remember that 
 \begin{eqnarray}
W_{S_{{\bf u}_1}}, W_{S_{{\bf q}_1}^c}&=&\{W_i;i\in[K],i\notin {\bf p}_{+}\}\label{jjo8}\\
W_{S_{{\bf u}_2}}, W_{S_{{\bf q}_2}^c}&=&\{W_i;i\in[K],i\notin  {\bf q}_{+}\}\label{jjo9}\\
W_{{ u}_3(1)}, W_{S_{{\bf u}_3}^c}&=&\{W_i;i\in[K],i\notin{\bf p}_{+}\cap {\bf q}_{+}\}\label{jjo10}\\
W_{{ u}_3(1)}, W_{S_{{\bf u}_4}^c}&=&\{W_i;i\in[K],i\notin{\bf p}_{+}\cup {\bf q}_{+}\}\label{jjo11}
\end{eqnarray}
From \eqref{jjo8}, \eqref{jjo9}, \eqref{jjo10}, \eqref{jjo11} and the submodularity property of entropy, $H(X|A)+H(X|B)\geq H(X|A\cup B)+H(X|A\cap B)$, \eqref{jjo7} is concluded. 

\end{enumerate}

\section{Proof of Lemma \ref{lemmaxc} (\cite{Arash_Jafar_KMIMOIC}, Lemma $1$)}\label{lemmaxcp}
 Before proceeding to prove \eqref{lemmamimox5}, note that for any $e\times1$ vector discrete random variable ${\bf V}$ and $e\times e$ matrix $A$,
\begin{eqnarray}
 H({\bf V})=H(A{\bf V})\text{~~if~~} |A|\neq0.
\end{eqnarray}
Since multiplying  a vector discrete random variable with an invertible matrix does not change its entropy, it is sufficient to prove \eqref{lemmamimox5} for the random variables $\breve{\bf U}_{1}$ and $\breve{\bf U}_{2}$ which are defined as,
\begin{eqnarray}
\breve{\bf U}_1&=&\left(\breve{U}_{11}^{[T]},\breve{U}_{12}^{[T]},\cdots,\breve{U}_{1N}^{[T]}\right)\label{lemmamimox7}\\
\breve{\bf U}_2&=&\left(\breve{U}_{21}^{[T]},\breve{U}_{22}^{[T]},\cdots,\breve{U}_{2N}^{[T]}\right)\label{lemmamimox8}
\end{eqnarray}
where for any $i\in[2],t\in[T]$, $\breve{U}_{in}(t)$ are defined as,
\begin{eqnarray}
\breve{U}_{in}(t)&=& L_{in}^b(t)\left((\bar{{V}}_l(t))^{\eta}_{\eta-\lambda_{il}}, n\le l\le M\right)\label{lemmamimox9}
\end{eqnarray}
Thus, we have,
\begin{eqnarray}
&&H({\breve{\bf U}}_2\mid W,\mathcal{G})-H({\breve{\bf U}}_1\mid W,\mathcal{G})\nonumber\\
&=&H(\{\breve{U}_{2i}^{[T]},i\in[N]\}\mid W,\mathcal{G})-H(\{\breve{U}_{1i}^{[T]},i\in[N]\}\mid W,\mathcal{G})\label{sw1}\\
&=&\sum_{n=1}^{N}\big(H(\{\breve{U}_{1i'}^{[T]},\breve{U}_{2i}^{[T]},~\forall i,i'\in[N],i'<n\le i\}\mid W,\mathcal{G})\nonumber\\
&&-H(\{\breve{U}_{1i'}^{[T]},\breve{U}_{2i}^{[T]},~\forall i,i'\in[N],i'\le n<i\}\mid W,\mathcal{G})\big)\label{sw2}\\
&=&\sum_{n=1}^{N}\big(H(\breve{U}_{1n}^{[T]}\mid W,W_n,\mathcal{G})-H(\breve{U}_{2n}^{[T]}\mid W,W_n,\mathcal{G})\big)\label{sw3}\\
&\le&T\big(\sum_{n=1}^N(\lambda_{1n}-\lambda_{2n})^+\big)\log{\bar{P}}+T~o~(\log{\bar{P}})\label{lemmamimox5+}
\end{eqnarray}
where $W_n$ is defined as the set of random variables $\{\breve{U}_{1i'}^{[T]},\breve{U}_{2i}^{[T]},i,i'\in[N],i'<n<i\}$. \eqref{sw2} follows from definition of $\breve{U}_{in}(t)$ and \eqref{sw3} is a result of the chain rule. \eqref{lemmamimox5+} is true as for any $n\in[N]$ we have,
\begin{eqnarray}
H(\breve{U}_{1n}^{[T]}\mid W,W_n,\mathcal{G})-H(\breve{U}_{2n}^{[T]}\mid W,W_n,\mathcal{G})
&\le&T(\lambda_{1n}-\lambda_{2n})^+\log{\bar{P}} +NT~o~(\log{\bar{P}}) \label{lemma3final}
\end{eqnarray}
\subsection{Proof of \eqref{lemma3final}}
Without loss of generality, let us prove \eqref{lemma3final} for $n=2$  as \eqref{lemma3final} follows for the other $n\in[N]$ similarly.
\begin{eqnarray}
&&H(\breve{U}_{12}^{[T]}\mid W,W_2,\mathcal{G})-H(\breve{U}_{22}^{[T]}\mid W,W_2,\mathcal{G})\nonumber\\
&\le&H(\breve{U}_{32}^{[T]}\mid W,W_2,\mathcal{G})-H(\breve{U}_{22}^{[T]}\mid W,W_2,\mathcal{G})\label{sw5}\\
&=&H((\breve{U}_{32}^{[T]})^{\max_{i\in[l],i\neq1}\lambda_{2i}+{(\lambda_{12}-\lambda_{22})^+}}_{{(\lambda_{12}-\lambda_{22})^+}},(\breve{U}_{32}^{[T]})_{(\lambda_{12}-\lambda_{22})^+}\mid W,W_2,\mathcal{G})-H(\breve{U}_{22}^{[T]}\mid W,W_2,\mathcal{G})\label{sw6}\\
&=&H(\breve{U}_{22}^{[T]},(\breve{U}_{32}^{[T]})_{(\lambda_{12}-\lambda_{22})^+}\mid W,W_2,\mathcal{G})-H(\breve{U}_{22}^{[T]}\mid W,W_2,\mathcal{G})\label{sw7}\\
&=&H((\breve{U}_{32}^{[T]})_{(\lambda_{12}-\lambda_{22})^+}\mid \breve{U}_{22}^{[T]},W,W_2,\mathcal{G})\label{sw8}\\
&\le&T(\lambda_{12}-\lambda_{22})^+\log{\bar{P}}+n~o~(\log{\bar{P}})\label{sw9}
\end{eqnarray}
where  for any $t\in[T]$, $\breve{U}_{32}(t)$ is defined as,
\begin{eqnarray}
\breve{U}_{32}(t)&=& L_{32}^b(t)\big((\bar{{V}}_2(t))^{\max(\lambda_{12},\lambda_{22})},(\bar{{V}}_3(t))^{\max(\lambda_{13},\lambda_{23})}\bar{P}^{(\lambda_{12}-\lambda_{22})^+-(\lambda_{13}-\lambda_{23})^+},\nonumber\\
&&\cdots,(\bar{{V}}_M(t))^{\max(\lambda_{1M},\lambda_{2M})}\bar{P}^{(\lambda_{12}-\lambda_{22})^+-(\lambda_{1M}-\lambda_{2M})^+}\big)
\end{eqnarray}
\eqref{sw6} follows from Definition \ref{powerlevel} and \eqref{sw7} is true from definition of the random variable $\breve{U}_{32}(t)$. Note that $(\breve{U}_{32}^{[T]})^{\max_{i\in[l],i\neq1}\lambda_{2i}+{(\lambda_{12}-\lambda_{22})^+}}_{{(\lambda_{12}-\lambda_{22})^+}}$ captures the top $\max_{i\in[l],i\neq1}\lambda_{2i}$ level of $\breve{U}_{32}^{[T]}$ which is equal to $\breve{U}_{22}^{[T]}$. \eqref{sw8} yields from chain rule and \eqref{sw9}  is true as the entropy of a discrete random variable is bounded by logarithm of the cardinality of it. \eqref{sw5} follows from the following observation.
\subsubsection{An observation} 
Setting  $M=1$, $l_1=1,K=1$ and $I_{11}=\{1\}$ in Theorem $4$ in \cite{Arash_Jafar_sumset}, we have
\begin{theorem}[Theorem $4$ in \cite{Arash_Jafar_sumset}] \label{Theorem AIS04}
Consider   non-negative number $\lambda$  and random variables $X_j (t) \in \mathcal{X}_{\lambda}$, $j\in[N]$, $t\in[T]$,  independent of $\mathcal{G}$,  and define 
\begin {eqnarray}
Z(t)&=&L^b(t)(X_1(t),X_2(t), \cdots, X_N(t))\label{mn1}\\
Z'(t)&=& {L'}(t)({(X_1(t))}_{\kappa_1}^{\gamma_1},{(X_2(t))}_{\kappa_2}^{\gamma_2}, \cdots, {(X_N(t))}_{\kappa_N}^{\gamma_N})\label{mn2}
\end{eqnarray}
where   $\kappa_i,\gamma_i$ are arbitrary non-negative real valued constants. The channel uses are indexed by $t\in[T]$. Then,
\begin {eqnarray}
H(Z^{[T]}\mid W,\mathcal{G})&\geq& H({Z'}^{[T]}\mid W)+T~o(\log{\bar{P}})\label{dssd4}
\end{eqnarray}
\end{theorem}
 Note that, from  \eqref{dssd4}, $H(\breve{U}_{12}^{[T]}\mid {W},W_2,\mathcal{G})$ can be bounded by $H(\breve{U}_{32}^{[T]}\mid {W},W_2,\mathcal{G})$ from above with the penalty equal to $T~o~\log{\bar{P}}$, i.e., 
\begin{eqnarray}
H(\breve{U}_{12}^{[T]}\mid W,W_2,\mathcal{G})-H(\breve{U}_{32}^{[T]}\mid W,W_2,\mathcal{G})&\le&T~o~\log{\bar{P}} \label{lemmamimox7}
\end{eqnarray}

\section{Achievability of $\mathcal{{D}}_{123}$}\label{sinr_app}
\begin{enumerate}
 \item{} ${X}_{\{1\}},{X}_{\{1,2\}},{X}_{\{1,2,3\}}$ are decoded with successive interference cancellation
 at the first receiver treating ${X}_{\{2\}}$ and ${X}_{\{3\}}$ as noise.
 \begin{enumerate}
 \item{} The SINR for decoding ${X}_{\{1,2,3\}}$ at the first receiver treating the other signals as white Gaussian noise is equal to
\begin{eqnarray}
&&\frac{P^{\alpha_{11}}{P}^{-\gamma'}(1-2P^{-\lambda}){|G_{11}|}^2}{1+P^{\alpha_{11}}{P}^{-\gamma'}P^{-\lambda}{|G_{11}|}^2+P^{\alpha_{12}}P^{-\lambda}{|G_{12}|}^2+P^{\alpha_{13}}P^{-\lambda}{|G_{13}|}^2}\nonumber\\
&\approx& P^{\min(\lambda,\alpha_{11}-\gamma',\lambda+\alpha_{11}-\gamma'-\alpha_{12},\lambda+\alpha_{11}-\gamma'-\alpha_{13})}
\end{eqnarray}
The codeword  ${X}_{\{1,2,3\}}$ which carries $d_{\{1,2,3\}}$ GDoF is  decoded successfully if  
\begin{eqnarray}
d_{\{1,2,3\}}&\leq&\min(\lambda,\alpha_{11}-\gamma',\lambda+\alpha_{11}-\gamma'-\alpha_{12},\lambda+\alpha_{11}-\gamma'-\alpha_{13})\label{eq:sinr123}
\end{eqnarray}
 From \eqref{llgg1} and \eqref{llgg10} we have $\lambda\leq \alpha_{11}-\gamma'$. Adding \eqref{llgg1} and  \eqref{llgg4} we have $\alpha_{11}-\gamma'-\alpha_{12}\geq 0$, and similarly, adding \eqref{llgg1} and \eqref{llgg5} we have $\alpha_{11}-\gamma'-\alpha_{13}\geq \lambda'\geq 0$. Therefore, the RHS of \eqref{eq:sinr123} is equal to $\lambda$. From \eqref{eq:d123} we have $d_{\{1,2,3\}}\leq \lambda$, therefore \eqref{eq:sinr123} holds and ${X}_{\{1,2,3\}}$ is successfully decoded at Receiver $1$.
 \item{}  After decoding the  messages $\bar{W}_{\{1,2,3\}}$, the first receiver reconstructs the codeword  ${X}_{\{1,2,3\}}$ and subtracts its contribution from the
received signal. The SINR for decoding ${X}_{\{1,2\}}$ at the first receiver while treating the other signals as white Gaussian noise is equal to
\begin{eqnarray}
&&\frac{P^{\alpha_{11}}{P}^{-\gamma'}P^{-\lambda}{|G_{11}|}^2}{1+P^{\alpha_{11}}{P}^{-\gamma'}P^{-\lambda-\lambda'}{|G_{11}|}^2+P^{\alpha_{12}}P^{-\lambda-\lambda'}{|G_{12}|}^2+P^{\alpha_{13}}P^{-\lambda}{|G_{13}|}^2}\nonumber\\
&\approx& P^{\min(\lambda',\alpha_{11}-\lambda-\gamma',\lambda'+\alpha_{11}-\gamma'-\alpha_{12},\alpha_{11}-\gamma'-\alpha_{13})}\label{sinr1}
\end{eqnarray}
 The codeword  ${X}_{\{1,2\}}$  which carries $d_{\{1,2\}}$ GDoF  is  decoded successfully if 
 \begin{eqnarray}
d_{\{1,2\}}&\leq&\min(\lambda',\alpha_{11}-\lambda-\gamma',\lambda'+\alpha_{11}-\gamma'-\alpha_{12},\alpha_{11}-\gamma'-\alpha_{13})\label{eq:sinr12}
\end{eqnarray}
 From \eqref{llgg1}, \eqref{llgg4}, \eqref{llgg5} it is easy to verify that the RHS of \eqref{eq:sinr12} is equal to $\lambda'$. However, from \eqref{eq:d12} we have $d_{\{1,2\}}\leq \lambda'$, therefore \eqref{eq:sinr12} holds and ${X}_{\{1,2\}}$ is successfully decoded at Receiver $1$.
\item{} After decoding the  messages $\bar{W}_{\{1,2\}}$, the first receiver reconstructs the codeword  ${X}_{\{1,2\}}$ and subtracts its contribution from the
received signal. The SINR for decoding ${X}_{\{1\}}$ at the first receiver  while treating the other signals as white Gaussian noise is equal to
\begin{eqnarray}
&&\frac{P^{\alpha_{11}}{P}^{-\gamma'}P^{-\lambda-\lambda'}{|G_{11}|}^2}{1+P^{\alpha_{12}}P^{-\lambda-\lambda'}{|G_{12}|}^2+P^{\alpha_{13}}P^{-\lambda}{|G_{13}|}^2}\nonumber\\
&\approx& P^{\min(\alpha_{11}-\lambda-\lambda'-\gamma',\alpha_{11}-\alpha_{12}-\gamma',\alpha_{11}-\alpha_{13}-\gamma'-\lambda')}\nonumber
\end{eqnarray}
The message  ${X}_{\{1\}}$ which carries $d_{\{1\}}$ GDoF is  decoded successfully  if
\begin{eqnarray}
d_{\{1\}}&\leq&\min(\alpha_{11}-\lambda-\lambda'-\gamma-\gamma',\alpha_{11}-\alpha_{12}-\gamma',\alpha_{11}-\alpha_{13}-\gamma'-\lambda')\label{eq:sinr1}
\end{eqnarray}
 From \eqref{llgg4} and \eqref{llgg5},  we conclude that the RHS of \eqref{eq:sinr1} is equal to $\alpha_{11}-\lambda-\lambda'-\gamma-\gamma'$. However, from \eqref{eq:d1} we have $d_{\{1\}}\leq\alpha_{11}-\lambda-\lambda'-\gamma-\gamma'$, therefore \eqref{eq:sinr1} holds and ${X}_{\{1\}}$ is successfully decoded at Receiver $1$.
  \end{enumerate}
 \item{} ${X}_{\{2\}},{X}_{\{1,2\}},{X}_{\{1,2,3\}}$ are decoded with successive interference cancellation
at the second receiver treating ${X}_{\{1\}}$ and ${X}_{\{3\}}$ as noise.
\begin{enumerate}
\item{} The SINR for decoding ${X}_{\{1,2,3\}}$ at the second receiver treating the other signals as noise is equal to
\begin{eqnarray}
\frac{P^{\alpha_{22}}(1-2P^{-\lambda}){|G_{22}|}^2}{1+P^{\alpha_{21}}{P}^{-\gamma'}P^{-\lambda}{|G_{21}|}^2+P^{\alpha_{22}}P^{-\lambda}{|G_{22}|}^2+P^{\alpha_{23}}P^{-\lambda}{|G_{23}|}^2}\approx P^{\lambda}\label{sinr+-}
\end{eqnarray}
 \eqref{sinr+-} follows as $\max(\alpha_{ji},\alpha_{ij})\le\alpha_{ii}$ is true for all $i,j,k\in[3]$ from \eqref{con1}. Therefore, the message  ${X}_{\{1,2,3\}}$ which carries $d_{\{1,2,3\}}$ GDoF is  decoded successfully at the second  receiver. 
\item{} After decoding the  messages $\bar{W}_{\{1,2,3\}}$, the second receiver reconstructs the codeword  ${X}_{\{1,2,3\}}$ and subtracts its contribution from the
received signal. The SINR for decoding ${X}_{\{1,2\}}$ at the second receiver treating the other signals as noise is equal to
\begin{eqnarray}
&&\frac{P^{\alpha_{22}}P^{-\lambda}{|G_{22}|}^2}{1+P^{\alpha_{21}}{P}^{-\gamma'}P^{-\lambda-\lambda'}{|G_{21}|}^2+P^{\alpha_{22}}P^{-\lambda-\lambda'}{|G_{22}|}^2+P^{\alpha_{23}}P^{-\lambda}{|G_{23}|}^2}\nonumber\\
&\approx& P^{\min(\lambda',\alpha_{22}-\lambda,\alpha_{22}+\lambda'+\gamma'-\alpha_{21},\alpha_{22}-\alpha_{23})}\label{sinr1}
\end{eqnarray}
The message  ${X}_{\{1,2\}}$ which carries $d_{\{1,2\}}$ GDoF is  decoded successfully if
\begin{eqnarray}
d_{\{1,2\}}&\leq&\min(\lambda',\alpha_{22}-\lambda,\alpha_{22}+\lambda'+\gamma'-\alpha_{21},\alpha_{22}-\alpha_{23})\label{eq:sinr12+}
\end{eqnarray}
 Adding \eqref{llgg2} and  \eqref{llgg9} we have $\lambda'\le\alpha_{22}-\alpha_{23}$. From \eqref{llgg2} and \eqref{llgg6},  the RHS of \eqref{eq:sinr12+} is equal to $\lambda'$. Moreover, from \eqref{eq:d12} we have $d_{\{1,2\}}\leq\lambda'$, therefore \eqref{eq:sinr12+} holds and ${X}_{\{1,2\}}$ is successfully decoded at Receiver $2$.
\item{}After decoding the  messages $\bar{W}_{\{1,2\}}$, the second receiver reconstructs the codeword  ${X}_{\{1,2\}}$ and subtracts its contribution from the
received signal.  SINR for decoding ${X}_{\{2\}}$ at the second receiver  is equal to
\begin{eqnarray}
&&\frac{P^{\alpha_{22}}P^{-\lambda-\lambda'}{|G_{22}|}^2}{1+P^{\alpha_{21}}{P}^{-\gamma'}P^{-\lambda-\lambda'}{|G_{21}|}^2+P^{\alpha_{23}}P^{-\lambda}{|G_{23}|}^2}\nonumber\\
&\approx& P^{\min(\alpha_{22}-\lambda-\lambda',\alpha_{22}-\alpha_{21}+\gamma',\alpha_{22}-\alpha_{23}-\lambda')}
\end{eqnarray}
Thus, the message  ${X}_{\{2\}}$ which carries $d_{\{2\}}$ GDoF is  decoded successfully if 
\begin{eqnarray}
d_{\{2\}}&\leq&\min(\alpha_{22}-\lambda-\lambda',\alpha_{22}-\alpha_{21}+\gamma',\alpha_{22}-\alpha_{23}-\lambda')\label{eq:sinr2+}
\end{eqnarray}
From \eqref{llgg6} and \eqref{llgg7}, we conclude that the RHS of \eqref{eq:sinr2+} is equal to $\alpha_{22}-\lambda-\lambda'$. However, from \eqref{eq:d2} we have $d_{\{2\}}\leq\alpha_{22}-\lambda-\lambda'$, therefore \eqref{eq:sinr2+} holds and ${X}_{\{2\}}$ is successfully decoded at Receiver $2$.
\end{enumerate}
  \item{} ${X}_{\{3\}},{X}_{\{1,2,3\}}$ are decoded with successive interference cancellation
 at the third receiver treating ${X}_{\{1\}}$, ${X}_{\{2\}}$ and ${X}_{\{1,2\}}$ as noise.
\begin{enumerate}
\item{} The SINR for decoding ${X}_{\{1,2,3\}}$ at the  third receiver treating the other signals as noise is equal to
\begin{eqnarray}
\frac{P^{\alpha_{33}}(1-2P^{-\lambda}){|G_{33}|}^2}{1+P^{\alpha_{31}}{P}^{-\gamma'}P^{-\lambda}{|G_{31}|}^2+P^{\alpha_{32}}P^{-\lambda}{|G_{32}|}^2+P^{\alpha_{33}}P^{-\lambda}{|G_{33}|}^2}\approx P^{\lambda}\label{+sinr}
\end{eqnarray}
where \eqref{+sinr} follows as from \eqref{con1} we have $\max(\alpha_{ji},\alpha_{ij})\le\alpha_{ii}$ for all $i,j,k\in[3]$. Therefore, the message  ${X}_{\{1,2,3\}}$ which carries $d_{\{1,2,3\}}$ GDoF is  decoded successfully at the   third receiver. 
\item{} Finally, the third receiver decodes ${X}_{\{3\}}$  treating ${X}_{\{1\}},{X}_{\{2\}}$ as noise with  SINR  equal to,
\begin{eqnarray}
\frac{P^{\alpha_{33}}P^{-\lambda}{|G_{33}|}^2}{1+P^{\alpha_{31}}{P}^{-\gamma'}P^{-\lambda}{|G_{31}|}^2+P^{\alpha_{32}}P^{-\lambda}{|G_{32}|}^2}&\approx& P^{\min(\alpha_{33}-\lambda,\alpha_{33}-\alpha_{31}+\gamma',\alpha_{33}-\alpha_{32})}
\end{eqnarray}
Therefore, the message  ${X}_{\{3\}}$ which carries $d_{\{3\}}$ GDoF is  decoded successfully if
\begin{eqnarray}
d_{\{3\}}&\leq&\min(\alpha_{33}-\lambda,\alpha_{33}-\alpha_{31}+\gamma',\alpha_{33}-\alpha_{32})\label{eq:sinr3+}
\end{eqnarray}
From \eqref{llgg8} and \eqref{llgg9}, we conclude that the RHS of \eqref{eq:sinr3+} is equal to $\alpha_{33}-\lambda$. However, from \eqref{eq:d3} we have $d_{\{3\}}\leq\alpha_{33}-\lambda$, therefore \eqref{eq:sinr3+} holds and ${X}_{\{3\}}$ is successfully decoded at Receiver $3$.
\end{enumerate}
\end{enumerate}

\section{$\mathcal{D}_{123}(\lambda,\lambda',\gamma,\gamma')=\mathcal{\bar{D}}_{123}(\lambda,\lambda',\gamma,\gamma')$}\label{d_i}
Consider some arbitrary quadruple  $(\lambda,\lambda',\gamma,\gamma')$ satisfying $(\eqref{llgg1}-\eqref{llgg10})$ and the regions   $\mathcal{{D}}_{123}(\lambda,\lambda',\gamma,\gamma')$ and $\mathcal{\bar{D}}_{123}(\lambda,\lambda',\gamma,\gamma')$  given in $(\eqref{eq:d}-\eqref{eq:d123})$ and $(\eqref{d1bar}-\eqref{d123bar})$. In order to show $\mathcal{D}_{123}(\lambda,\lambda',\gamma,\gamma')=\mathcal{\bar{D}}_{123}(\lambda,\lambda',\gamma,\gamma')$, let us prove $\mathcal{D}_{123}(\lambda,\lambda',\gamma,\gamma')\subset\mathcal{\bar{D}}_{123}(\lambda,\lambda',\gamma,\gamma')$ and  $\mathcal{\bar{D}}_{123}(\lambda,\lambda',\gamma,\gamma')\subset\mathcal{D}_{123}(\lambda,\lambda',\gamma,\gamma')$ separately as follows.

\subsection{$\mathcal{D}_{123}(\lambda,\lambda',\gamma,\gamma')\subset\mathcal{\bar{D}}_{123}(\lambda,\lambda',\gamma,\gamma')$}
Consider some arbitrary triple $(d_1,d_2,d_3)\in \mathcal{D}_{123}(\lambda,\lambda',\gamma,\gamma')$. From $(\eqref{eq:d}-\eqref{eq:d123})$ there exists some tuple $(d_{\{1\}},d_{\{2\}},d_{\{3\}},d_{\{1,2\}},d_{\{1,2,3\}}, \mu_{1},\mu_{2},\xi_1,\xi_2,\xi_3)$ where
\begin{eqnarray}
d_{\{1\}}&\le&\alpha_{11}-\lambda-\lambda'-\gamma-\gamma'\label{x1}\\
d_{\{2\}}&\le&\alpha_{22}-\lambda-\lambda'\\
d_{\{3\}}&\le&\alpha_{33}-\lambda\\
 d_{\{1,2\}}&\le&\lambda'\\
d_{\{1,2,3\}}&\le&\lambda\\
d_1&=&d_{\{1\}}+\mu_{1}d_{\{1,2\}}+\xi_1 d_{\{1,2,3\}}\\
d_2&=&d_{\{2\}}+\mu_{2}d_{\{1,2\}}+\xi_2d_{\{1,2,3\}}\\
d_3&=&d_{\{3\}}+\xi_3d_{\{1,2,3\}}\\
0&\le& \mu_{1},\mu_{2},\xi_1,\xi_2,\xi_3\\
\mu_{1}+\mu_{2}&=&1\\
\xi_{1}+\xi_{2}+\xi_{3}&=&1\label{x2}
 \end{eqnarray}
 Now, we claim that $(d_1,d_2,d_3)\in \mathcal{\bar{D}}_{123}(\lambda,\lambda',\gamma,\gamma')$. In order to prove the claim, it is sufficient to check the following bounds.  
 \begin{eqnarray}
0\le d_1&\le& \alpha_{11}-\gamma-\gamma', 0\le d_2\le \alpha_{22},0\le d_3\le \alpha_{33},\\
d_1+d_2&\le& \alpha_{11}+\alpha_{22}-\lambda-\lambda'-\gamma-\gamma',\\
d_1+d_3&\le& \alpha_{11}+\alpha_{33}-\lambda-\gamma-\gamma',\\
d_2+d_3&\le& \alpha_{22}+\alpha_{33}-\lambda,\\
d_1+d_2+d_3&\le&  \alpha_{11}+\alpha_{22}+\alpha_{33}-2\lambda-\lambda'-\gamma-\gamma'
\end{eqnarray} 
which are true from $(\eqref{x1}-\eqref{x2})$ as follows. 
\begin{eqnarray}
d_1&\le&d_{\{1\}}+d_{\{1,2\}}+d_{\{1,2,3\}}\le \alpha_{11}-\gamma-\gamma'\\
d_2&\le&d_{\{2\}}+d_{\{1,2\}}+d_{\{1,2,3\}}\le \alpha_{22}\\
d_3&\le&d_{\{3\}}+d_{\{1,2,3\}}\le \alpha_{33}\\
d_1+d_2&\le&d_{\{1\}}+d_{\{2\}}+d_{\{1,2\}}+d_{\{1,2,3\}}\\
&\le&(\alpha_{11}-\lambda-\lambda'-\gamma-\gamma')+(\alpha_{22}-\lambda-\lambda')+\lambda'+\lambda\\
&=&\alpha_{11}+\alpha_{22}-\lambda-\lambda'-\gamma-\gamma'\\
d_1+d_3&\le&d_{\{1\}}+d_{\{3\}}+d_{\{1,2\}}+d_{\{1,2,3\}}\\
&\le&(\alpha_{11}-\lambda-\lambda'-\gamma-\gamma')+(\alpha_{33}-\lambda)+\lambda'+\lambda\\
&=&\alpha_{11}+\alpha_{33}-\lambda-\gamma-\gamma'\\
d_2+d_3&\le&d_{\{2\}}+d_{\{3\}}+d_{\{1,2\}}+d_{\{1,2,3\}}\\
&\le&(\alpha_{22}-\lambda-\lambda')+(\alpha_{33}-\lambda)+\lambda'+\lambda\\
&=&\alpha_{22}+\alpha_{33}-\lambda\\
d_1+d_2+d_3&\le&d_{\{1\}}+d_{\{2\}}+d_{\{3\}}+d_{\{1,2\}}+d_{\{1,2,3\}}\\
&\le&(\alpha_{11}-\lambda-\lambda'-\gamma-\gamma')+(\alpha_{22}-\lambda-\lambda')+(\alpha_{33}-\lambda)+\lambda'+\lambda\\
&=&\alpha_{11}+\alpha_{22}+\alpha_{33}-2\lambda-\lambda'-\gamma-\gamma'
\end{eqnarray}
Therefore, for any triple $(d_1,d_2,d_3)\in \mathcal{D}_{123}(\lambda,\lambda',\gamma,\gamma')$ we conclude that $(d_1,d_2,d_3)\in \mathcal{\bar{D}}_{123}(\lambda,\lambda',\gamma,\gamma')$. Thus, $\mathcal{D}_{123}(\lambda,\lambda',\gamma,\gamma')\subset\mathcal{\bar{D}}_{123}(\lambda,\lambda',\gamma,\gamma')$.

\subsection{$\mathcal{\bar{D}}_{123}(\lambda,\lambda',\gamma,\gamma')$ and $\mathcal{D}_{123}(\lambda,\lambda',\gamma,\gamma')$ are convex sets}\label{convexity}
The set $\mathcal{\bar{D}}_{123}(\lambda,\lambda',\gamma,\gamma')$ is a convex polyhedron  by definition.\footnote{A convex polyhedron in $\mathbb{R}^n$ is defined as $\{x\mid Ax\leq b\}$, $A\in\mathbb{R}^{m\times n}$, $b\in\mathbb{R}^{m\times 1}$.} In fact, since it is bounded, it is a convex polytope \cite{Schrijver}.

Let us consider the set $\mathcal{D}_{123}(\lambda,\lambda',\gamma,\gamma')$ and two arbitrary members of it, e.g., $(d_1,d_2,d_3)$ and $(d'_1,d'_2,d'_3)$. From $(\eqref{eq:d}-\eqref{eq:d123})$,   there exists $(d_{\{1\}},d_{\{2\}},d_{\{3\}},d_{\{1,2\}},d_{\{1,2,3\}}, \mu_{1},\mu_{2},\xi_1,\xi_2,\xi_3)$ where $(\eqref{eq:d}-\eqref{eq:d123})$ are satisfied for $(d_1,d_2,d_3)$ and there exists $(d'_{\{1\}},d'_{\{2\}},d'_{\{3\}},d'_{\{1,2\}},d'_{\{1,2,3\}}, \mu'_{1},\mu'_{2},\xi'_1,\xi'_2,\xi'_3)$ where $(\eqref{eq:d}-\eqref{eq:d123})$ are satisfied for $(d'_1,d'_2,d'_3)$. Consider $0< \zeta<1$. Let us prove that $(\zeta d_1+(1-\zeta)d'_1$, $\zeta d_2+(1-\zeta)d'_2,\zeta d_3+(1-\zeta)d'_3)$ $\in\mathcal{D}_{123}(\lambda,\lambda',\gamma,\gamma')$. In order to do so, we  derive the variables $(d''_{\{1\}}$, $d''_{\{2\}}$, $d''_{\{3\}}$, $d''_{\{1,2\}}$, $d''_{\{1,2,3\}}$, $\mu''_{1}$, $\mu''_{2}$, $\xi''_1$, $\xi''_2$, $\xi''_3)$ satisfying $(\eqref{eq:d}-\eqref{eq:d123})$ for the point $(\zeta d_1+(1-\zeta)d'_1,\zeta d_2+(1-\zeta)d'_2,\zeta d_3+(1-\zeta)d'_3)$ as follows.
\begin{eqnarray}
d''_{\{k\}}&=&\zeta d_{\{k\}}+(1-\zeta)d'_{\{k\}},\forall k\in[3]\label{kjh1}\\
d''_{\{1,2\}}&=&\zeta d_{\{1,2\}}+(1-\zeta)d'_{\{1,2\}}\\
d''_{\{1,2,3\}}&=&\zeta d_{\{1,2,3\}}+(1-\zeta)d'_{\{1,2,3\}}\\
\mu''_{1}&=&\frac{\zeta \mu_{1}d_{\{1,2\}}+(1-\zeta)\mu'_{1}d'_{\{1,2\}}}{d''_{\{1,2\}}}\\
\xi''_{1}&=&\frac{\zeta \xi_{1}d_{\{1,2,3\}}+(1-\zeta)\xi'_{1}d'_{\{1,2,3\}}}{d''_{\{1,2,3\}}}\\
\xi''_{2}&=&\frac{\zeta \xi_{2}d_{\{1,2,3\}}+(1-\zeta)\xi'_{2}d'_{\{1,2,3\}}}{d''_{\{1,2,3\}}}\\
\mu''_{2}&=&1-\mu''_{1}\\
\xi''_{3}&=&1-\xi''_{1}-\xi''_{2}\label{kjh2}
\end{eqnarray}
Note that, the variables $(d''_{\{1\}},d''_{\{2\}},d''_{\{3\}},d''_{\{1,2\}},d''_{\{1,2,3\}}, \mu''_{1},\mu''_{2},\xi''_1,\xi''_2,\xi''_3)$ satisfy $(\eqref{eq:d}-\eqref{eq:d123})$. For instance,
 \begin{eqnarray}
d''_1&=&\zeta d_1+(1-\zeta)d'_1\nonumber\\
&=&(\zeta d_{\{1\}}+(1-\zeta)d'_{\{1\}})+(\zeta \mu_{1}d_{\{1,2\}}+(1-\zeta) \mu'_{1}d'_{\{1,2\}})+(\zeta \xi_{1}d_{\{1,2,3\}}+(1-\zeta) \xi'_{1}d'_{\{1,2,3\}})\nonumber\\
&=&d''_{\{1\}}+\mu''_{1}d''_{\{1,2\}}+\xi''_{1}d''_{\{1,2,3\}} \end{eqnarray}
Therefore, $\mathcal{D}_{123}(\lambda,\lambda',\gamma,\gamma')$ is a convex set.

\subsection{$\mathcal{\bar{D}}_{123}(\lambda,\lambda',\gamma,\gamma')\subset\mathcal{D}_{123}(\lambda,\lambda',\gamma,\gamma')$}
Consider a feasible quadruple  $(\lambda,\lambda',\gamma,\gamma')$. It is sufficient to show that all the corner points of the convex polytope $\mathcal{\bar{D}}_{123}(\lambda,\lambda',\gamma,\gamma')$ reside in the convex set $\mathcal{D}_{123}(\lambda,\lambda',\gamma,\gamma')$.\footnote{For any  convex polytope $A$ and a convex set $B$, $A\subset B$ if and only if the vertices of $A$ are members of $B$. While the statement is  obvious, a short proof is included in Appendix \ref{convexpolyhedron} for the sake of completeness.}
\begin{enumerate}
\item{$d_1= \alpha_{11}-\gamma-\gamma'$.} Consider the  hyperplane $d_1=\alpha_{11}-\gamma-\gamma'$ and the set of
all the points contained in this hyperplane which satisfy all the
other inequalities $(\eqref{d1bar}-\eqref{d123bar})$, i.e., 
\begin{eqnarray}
\mathbb{S}_1&=&\{(d_1,d_2,d_3);d_1= \alpha_{11}-\gamma-\gamma',0\le d_2\le \alpha_{22}-\lambda-\lambda',0\le d_3\le \alpha_{33}-\lambda\}
\end{eqnarray}
In order to check that $\mathbb{S}_1\subset\mathcal{D}_{123}(\lambda,\lambda',\gamma,\gamma')$, ~it is sufficient to prove~ that~ for~ any~ $(d_1,d_2,d_3)\in\mathbb{S}_1$~ there~ exists~ some ~tuple~ $(d_{\{1\}},d_{\{2\}},d_{\{3\}},d_{\{1,2\}},d_{\{1,2,3\}}, \mu_{1},\mu_{2},\xi_1,\xi_2,\xi_3)$ where $(\eqref{eq:d}-\eqref{eq:d123})$ are satisfied. This is true by choosing  $(d_{\{1\}},d_{\{2\}},d_{\{3\}},d_{\{1,2\}},d_{\{1,2,3\}}, \mu_{1},\mu_{2},\xi_1,\xi_2,\xi_3)$ as follows.
\begin{eqnarray}
&&(d_{\{1\}},d_{\{2\}},d_{\{3\}},d_{\{1,2\}},d_{\{1,2,3\}}, \mu_{1},\mu_{2},\xi_1,\xi_2,\xi_3)\nonumber\\
&=&(\alpha_{11}-\lambda-\lambda'-\gamma-\gamma',d_2,d_3,\lambda',\lambda, 1,0,1,0,0)
\end{eqnarray}
\item{$d_2= \alpha_{22}$.} Consider the  hyperplane $d_2=\alpha_{22}$ and the set of
all the points contained in this hyperplane which satisfy all the
other inequalities $(\eqref{d1bar}-\eqref{d123bar})$, i.e., 
\begin{eqnarray}
\mathbb{S}_2&=&\{(d_1,d_2,d_3);0\le d_1\le \alpha_{11}-\lambda-\lambda'-\gamma-\gamma',d_2=\alpha_{22},0\le d_3\le \alpha_{33}-\lambda\}
\end{eqnarray}
Similarly, we show that $\mathbb{S}_2\subset\mathcal{D}_{123}(\lambda,\lambda',\gamma,\gamma')$  as for~ any~ $(d_1,d_2,d_3)\in\mathbb{S}_2$~ there~ exists~ some ~tuple~ $(d_{\{1\}},d_{\{2\}},d_{\{3\}},d_{\{1,2\}},d_{\{1,2,3\}}, \mu_{1},\mu_{2},\xi_1,\xi_2,\xi_3)$ where $(\eqref{eq:d}-\eqref{eq:d123})$ are satisfied. This is true by choosing  $(d_{\{1\}},d_{\{2\}},d_{\{3\}},d_{\{1,2\}},d_{\{1,2,3\}}, \mu_{1},\mu_{2},\xi_1,\xi_2,\xi_3)$ as follows.
\begin{eqnarray}
&&(d_{\{1\}},d_{\{2\}},d_{\{3\}},d_{\{1,2\}},d_{\{1,2,3\}}, \mu_{1},\mu_{2},\xi_1,\xi_2,\xi_3)\nonumber\\
&=&(d_1,\alpha_{22}-\lambda-\lambda',d_3,\lambda',\lambda, 0,1,0,1,0)
\end{eqnarray}
\item{$d_3= \alpha_{33}$.} Consider the  hyperplane $d_3=\alpha_{33}$ and the hyperplane $\mathbb{S}_3$ as follows. 
\begin{eqnarray}
\mathbb{S}_3&=&\bigg\{(d_1,d_2,d_3);d_3=\alpha_{33},~0\le d_1\le~ \alpha_{11}-\lambda-\gamma-\gamma',\nonumber\\
&&0\le d_2\le~ \alpha_{22}-\lambda,d_1+d_2\le\alpha_{11}+\alpha_{22}-2\lambda-\lambda'-\gamma-\gamma'\bigg\}
\end{eqnarray}
In order to prove that $\mathbb{S}_3\subset\mathcal{D}_{123}(\lambda,\lambda',\gamma,\gamma')$, it is sufficient to show that the line $l_3\subset\mathcal{D}_{123}(\lambda,\lambda',\gamma,\gamma')$ where $l_3$ is defined as follows.\footnote{To see why $l_3\subset\mathcal{D}_{123}(\lambda,\lambda',\gamma,\gamma')$ results in $\mathbb{S}_3\subset\mathcal{D}_{123}(\lambda,\lambda',\gamma,\gamma')$, consider the tuples  $(d'_1,d'_2,d'_3)$ and  $(d_1,d_2,d_3)$ where $0\le d'_i\le d_i$ for any $i\in[3]$. From $(\eqref{eq:d}-\eqref{eq:d123})$, we have
\begin{eqnarray}
(d_1,d_2,d_3)\in\mathcal{D}_{123}(\lambda,\lambda',\gamma,\gamma')\Rightarrow(d'_1,d'_2,d'_3)\in\mathcal{D}_{123}(\lambda,\lambda',\gamma,\gamma')
\end{eqnarray}
Therefore, if $l_3\subset\mathcal{D}_{123}(\lambda,\lambda',\gamma,\gamma')$ then we conclude that $\mathbb{S}_3\subset\mathcal{D}_{123}(\lambda,\lambda',\gamma,\gamma')$.}
\begin{eqnarray}
l_3&=&\bigg\{(d_1,d_2,d_3);d_3=\alpha_{33},~0\le d_1\le~ \alpha_{11}-\lambda-\gamma-\gamma',\nonumber\\
&&0\le d_2\le~ \alpha_{22}-\lambda,d_1+d_2=\alpha_{11}+\alpha_{22}-2\lambda-\lambda'-\gamma-\gamma'\bigg\}
\end{eqnarray}
Now, let us prove that ~$l_3\subset\mathcal{D}_{123}(\lambda,\lambda',\gamma,\gamma')$, i.e., for any $(d_1,d_2,d_3)\in l_3$ there exists some tuple $(d_{\{1\}}$, $d_{\{2\}}$, $d_{\{3\}}$, $d_{\{1,2\}}$, $d_{\{1,2,3\}}$, $\mu_{1}$, $\mu_{2}$, $\xi_1$, $\xi_2$, $\xi_3)$ where $(\eqref{eq:d}-\eqref{eq:d123})$ are satisfied. This is verified to be true by choosing  $(d_{\{1\}},d_{\{2\}},d_{\{3\}},d_{\{1,2\}},d_{\{1,2,3\}}, \mu_{1},\mu_{2},\xi_1,\xi_2,\xi_3)$ as follows.
\begin{align}
&(d_{\{1\}},d_{\{2\}},d_{\{3\}},d_{\{1,2\}},d_{\{1,2,3\}}, \mu_{1},\mu_{2},\xi_1,\xi_2,\xi_3)\nonumber\\
=&(\alpha_{11}-\lambda-\lambda'-\gamma-\gamma',\alpha_{22}-\lambda-\lambda',\alpha_{33}-\lambda,\lambda',\lambda, \frac{d_1-d_{\{1\}}}{\lambda'},\frac{d_2-d_{\{2\}}}{\lambda'},0,0,1)
\end{align}
\item{} It is trivial to verify that the corner point $(0,0,0)\in\mathcal{D}_{123}(\lambda,\lambda',\gamma,\gamma')$ by choosing $d_{\{1\}}=d_{\{2\}}=d_{\{3\}}=d_{\{1,2\}}=d_{\{1,2,3\}}=0$.

\item{}Surprisingly, all  the corner points are already considered in the previous cases. For instance consider the point $s$ obtained from the intersection of the following three facets.\footnote{We are considering the points not considered in the previous cases. Thus, we assume that
\begin{eqnarray}
d_1&<& \alpha_{11}-\gamma-\gamma',\\
d_2&<& \alpha_{22},\\
d_3&<& \alpha_{33}.
\end{eqnarray}}
\begin{eqnarray}
d_1+d_3&=& \alpha_{11}+\alpha_{33}-\lambda-\gamma-\gamma'\label{xx1}\\
d_2+d_3&=& \alpha_{22}+\alpha_{33}-\lambda\label{xx2}\\
d_1+d_2+d_3&=&  \alpha_{11}+\alpha_{22}+\alpha_{33}-2\lambda-\lambda'-\gamma-\gamma'\label{xx3}
\end{eqnarray}
From $(\eqref{xx1}-\eqref{xx3})$, we have $d_3=\alpha_{33}+\lambda'$ which contradicts the condition $d_3<\alpha_{33}$. 
\end{enumerate}

\section{$\hat{\mathcal{D}}_{123}=\cup_{\lambda,\lambda',\gamma,\gamma'}\bar{\mathcal{D}}_{123}(\lambda,\lambda',\gamma,\gamma')$}\label{equi1}
Let us prove that $\cup_{\lambda,\lambda',\gamma,\gamma'}\bar{\mathcal{D}}_{123}(\lambda,\lambda',\gamma,\gamma')\subset\hat{\mathcal{D}}_{123}$ and  $\hat{\mathcal{D}}_{123}\subset\cup_{\lambda,\lambda',\gamma,\gamma'}\bar{\mathcal{D}}_{123}(\lambda,\lambda',\gamma,\gamma')$ separately.

\subsection{$\cup_{\lambda,\lambda',\gamma,\gamma'}\bar{\mathcal{D}}_{123}(\lambda,\lambda',\gamma,\gamma')\subset\hat{\mathcal{D}}_{123}$}
In order to prove that $\cup_{\lambda,\lambda',\gamma,\gamma'}\bar{\mathcal{D}}_{123}(\lambda,\lambda',\gamma,\gamma')\subset\hat{\mathcal{D}}_{123}$, it is sufficient to prove that $\mathcal{\bar{D}}_{123}(\lambda,\lambda',\gamma,\gamma')\subset\hat{\mathcal{D}}_{123}$ for any $\lambda,\lambda',\gamma,\gamma'$ satisfying $(\eqref{llgg1}-\eqref{llgg10})$. In other words, we need to prove that \eqref{llgg1}-\eqref{llgg10} and \eqref{d1bar}-\eqref{d123bar} together imply \eqref{d1hat}-\eqref{d123hat}. But this is easily verified as follows.
\begin{eqnarray}
\eqref{llgg10},\eqref{d1bar}&\Rightarrow&\eqref{d1hat}\\
\eqref{d2bar}&\Rightarrow&\eqref{d2hat}\\
\eqref{d3bar}&\Rightarrow&\eqref{d3hat}\\
\eqref{llgg4},\eqref{llgg5},\eqref{llgg6},\eqref{llgg7},\eqref{llgg8},\eqref{llgg9}, \eqref{d12bar}&\Rightarrow&\eqref{d12hat}\\
\eqref{llgg5}, \eqref{llgg7},\eqref{llgg8},\eqref{llgg9}, \eqref{llgg10}, \eqref{d13bar}&\Rightarrow&\eqref{d13hat}\\
\eqref{llgg7},\eqref{llgg9}, \eqref{d23bar}&\Rightarrow&\eqref{d23hat}\\
\eqref{llgg4},\eqref{llgg5},\eqref{llgg6},\eqref{llgg7}, \eqref{llgg8},\eqref{llgg9},\eqref{llgg10}, \eqref{d123bar}&\Rightarrow&\eqref{d123hat}
\end{eqnarray}

\subsection{$\hat{\mathcal{D}}_{123}$ and $\cup_{\lambda,\lambda',\gamma,\gamma'}\bar{\mathcal{D}}_{123}(\lambda,\lambda',\gamma,\gamma')$ are convex sets}
Similar to \ref{convexity}, $\hat{\mathcal{D}}_{123}$ is a convex polytope by definition. Next we have $\cup_{\lambda,\lambda',\gamma,\gamma'}\bar{\mathcal{D}}_{123}(\lambda,\lambda',\gamma,\gamma')$. Consider two members of it, e.g., $(\hat{d}_1,\hat{d}_2,\hat{d}_3)$ and $(\bar{d}_1,\bar{d}_2,\bar{d}_3)$. 
\begin{enumerate}
\item{} As $(\hat{d}_1,\hat{d}_2,\hat{d}_3)\in\cup_{\lambda,\lambda',\gamma,\gamma'}\bar{\mathcal{D}}_{123}(\lambda,\lambda',\gamma,\gamma')$, there exists $(\hat{\lambda},\hat{\lambda}',\hat{\gamma},\hat{\gamma}')$ where $(\hat{d}_1,\hat{d}_2,\hat{d}_3)\in\bar{\mathcal{D}}_{123}(\hat{\lambda},\hat{\lambda}',\hat{\gamma},\hat{\gamma}')$, i.e.,
\begin{eqnarray}
\hat{d}_1&\le& \alpha_{11}-\hat{\gamma}-\hat{\gamma}'\\
\hat{d}_2&\le& \alpha_{22}\\
\hat{d}_3&\le& \alpha_{33},\\
\hat{d}_1+\hat{d}_2&\le& \alpha_{11}+\alpha_{22}-\hat{\lambda}-\hat{\lambda}'-\hat{\gamma}-\hat{\gamma}',\\
\hat{d}_1+\hat{d}_3&\le& \alpha_{11}+\alpha_{33}-\hat{\lambda}-\hat{\gamma}-\hat{\gamma}',\\
\hat{d}_2+\hat{d}_3&\le& \alpha_{22}+\alpha_{33}-\hat{\lambda},\\
\hat{d}_1+\hat{d}_2+\hat{d}_3&\le&  \alpha_{11}+\alpha_{22}+\alpha_{33}-2\hat{\lambda}-\hat{\lambda}'-\hat{\gamma}-\hat{\gamma}'
\end{eqnarray} 
such that $\hat{\lambda},\hat{\lambda}',\hat{\gamma},\hat{\gamma}'$ satisfy conditions \eqref{llgg1} to \eqref{llgg10}.
\item{} Similarly, as $(\bar{d}_1,\bar{d}_2,\bar{d}_3)\in\cup_{\lambda,\lambda',\gamma,\gamma'}\bar{\mathcal{D}}_{123}(\lambda,\lambda',\gamma,\gamma')$, there exists $(\bar{\lambda},\bar{\lambda}',\bar{\gamma},\bar{\gamma}')$ where $(\bar{d}_1,\bar{d}_2,\bar{d}_3)\in\bar{\mathcal{D}}_{123}(\bar{\lambda},\bar{\lambda}',\bar{\gamma},\bar{\gamma}')$, i.e.,
\begin{eqnarray}
\bar{d}_1&\le& \alpha_{11}-\bar{\gamma}-\bar{\gamma}'\\
\bar{d}_2&\le& \alpha_{22}\\
\bar{d}_3&\le& \alpha_{33},\\
\bar{d}_1+\bar{d}_2&\le& \alpha_{11}+\alpha_{22}-\bar{\lambda}-\bar{\lambda}'-\bar{\gamma}-\bar{\gamma}',\\
\bar{d}_1+\bar{d}_3&\le& \alpha_{11}+\alpha_{33}-\bar{\lambda}-\bar{\gamma}-\bar{\gamma}',\\
\bar{d}_2+\bar{d}_3&\le& \alpha_{22}+\alpha_{33}-\bar{\lambda},\\
\bar{d}_1+\bar{d}_2+\bar{d}_3&\le&  \alpha_{11}+\alpha_{22}+\alpha_{33}-2\bar{\lambda}-\bar{\lambda}'-\bar{\gamma}-\bar{\gamma}'
\end{eqnarray} 
such that $\bar{\lambda},\bar{\lambda}',\bar{\gamma},\bar{\gamma}'$ satisfy conditions \eqref{llgg1} to \eqref{llgg10}.
\end{enumerate}
Now, consider the point $d=(\zeta \hat{d}_1+(1-\zeta)\bar{d}_1,\zeta \hat{d}_2+(1-\zeta)\bar{d}_2,\zeta \hat{d}_3+(1-\zeta)\bar{d}_3)$. We claim that  $d\in\bar{\mathcal{D}}_{123}(\breve{\lambda},\breve{\lambda}',\breve{\gamma},\breve{\gamma}')$ where  $\breve{\lambda},\breve{\lambda}',\breve{\gamma},\breve{\gamma}'$ are defined as,\footnote{Note that, 
$\breve{\lambda},\breve{\lambda}',\breve{\gamma},\breve{\gamma}'$ satisfy conditions \eqref{llgg1} to \eqref{llgg10}. For instance, \eqref{llgg1} is verified as,
\begin{eqnarray}
&&\breve{\lambda}+\breve{\lambda}'+\breve{\gamma}+\breve{\gamma}'\nonumber\\
&=&\zeta(\hat{\lambda}+\hat{\lambda}'+\hat{\gamma}+\hat{\gamma}')+(1-\zeta)(\bar{\lambda}+\bar{\lambda}'+\bar{\gamma}+\bar{\gamma}')\nonumber\\
&\le&\zeta\alpha_{11}+(1-\zeta)\alpha_{11}=\alpha_{11}
\end{eqnarray}
All the other conditions \eqref{llgg1} to \eqref{llgg10} are also true as they are linear combinations of $\breve{\lambda},\breve{\lambda}',\breve{\gamma},\breve{\gamma}'$.}
\begin{eqnarray}
\breve{\lambda}&=&\zeta\hat{\lambda}+(1-\zeta)\bar{\lambda}\\
\breve{\lambda}'&=&\zeta\hat{\lambda}'+(1-\zeta)\bar{\lambda}'\\
\breve{\gamma}&=&\zeta\hat{\gamma}+(1-\zeta)\bar{\gamma}\\
\breve{\gamma}'&=&\zeta\hat{\gamma}'+(1-\zeta)\bar{\gamma}'
\end{eqnarray} 
This is verified by checking $(\eqref{d1bar}-\eqref{d123bar})$. For instance, we check the inequalities $\zeta \hat{d}_1+(1-\zeta)\bar{d}_1+\zeta \hat{d}_2+(1-\zeta)\bar{d}_2\le \alpha_{11}+\alpha_{22}-\breve{\lambda}-\breve{\lambda}'-\breve{\gamma}-\breve{\gamma}'$ and $\zeta \hat{d}_1+(1-\zeta)\bar{d}_1+\zeta \hat{d}_2+(1-\zeta)\bar{d}_2+\zeta \hat{d}_3+(1-\zeta)\bar{d}_3\le \alpha_{11}+\alpha_{22}+\alpha_{33}-2\breve{\lambda}-\breve{\lambda}'-\breve{\gamma}-\breve{\gamma}'$ as follows.
\begin{eqnarray}
\lefteqn{\zeta \hat{d}_1+(1-\zeta)\bar{d}_1+\zeta \hat{d}_2+(1-\zeta)\bar{d}_2}\nonumber\\
&=&\zeta( \hat{d}_1+ \hat{d}_2)+(1-\zeta)(\bar{d}_1+\bar{d}_2)\\
&\le&\zeta( \alpha_{11}+\alpha_{22}-\hat{\lambda}-\hat{\lambda}'-\hat{\gamma}-\hat{\gamma}')+(1-\zeta)( \alpha_{11}+\alpha_{22}-\bar{\lambda}-\bar{\lambda}'-\bar{\gamma}-\bar{\gamma}')\\
&=&\alpha_{11}+\alpha_{22}-\breve{\lambda}-\breve{\lambda}'-\breve{\gamma}-\breve{\gamma}'\\
\lefteqn{\zeta \hat{d}_1+(1-\zeta)\bar{d}_1+\zeta \hat{d}_2+(1-\zeta)\bar{d}_2+\zeta \hat{d}_3+(1-\zeta)\bar{d}_3}\nonumber\\
&=&\zeta( \hat{d}_1+ \hat{d}_2+ \hat{d}_3)+(1-\zeta)(\bar{d}_1+\bar{d}_2+\bar{d}_3)\\
&\le&\zeta( \alpha_{11}+\alpha_{22}+\alpha_{33}-2\hat{\lambda}-\hat{\lambda}'-\hat{\gamma}-\hat{\gamma}')+(1-\zeta)( \alpha_{11}+\alpha_{22}+\alpha_{33}-2\bar{\lambda}-\bar{\lambda}'-\bar{\gamma}-\bar{\gamma}')\\
&=&\alpha_{11}+\alpha_{22}+\alpha_{33}-2\breve{\lambda}-\breve{\lambda}'-\breve{\gamma}-\breve{\gamma}'
\end{eqnarray}
Therefore, as $d\in\bar{\mathcal{D}}_{123}(\breve{\lambda},\breve{\lambda}',\breve{\gamma},\breve{\gamma}')$, we conclude that $d\in\cup_{\lambda,\lambda',\gamma,\gamma'}\bar{\mathcal{D}}_{123}(\lambda,\lambda',\gamma,\gamma')$. This proves that the set $\cup_{\lambda,\lambda',\gamma,\gamma'}\bar{\mathcal{D}}_{123}(\lambda,\lambda',\gamma,\gamma')$ is convex.

\subsection{$\hat{\mathcal{D}}_{123}\subset\cup_{\lambda,\lambda',\gamma,\gamma'}\bar{\mathcal{D}}_{123}(\lambda,\lambda',\gamma,\gamma')$}
In this section, let us use the compact notation
\begin{eqnarray}
\hat{\mathcal{D}}'_{123}&\define&\cup_{\lambda,\lambda',\gamma,\gamma'}\bar{\mathcal{D}}_{123}(\lambda,\lambda',\gamma,\gamma')
\end{eqnarray}
It is sufficient to show that all the corner points of the convex polytope $\hat{\mathcal{D}}_{123}$ reside in the convex set $\hat{\mathcal{D}}'_{123}$.
\begin{enumerate}
\item consider the tuples  $(d'_1,d'_2,d'_3)$ and  $(d_1,d_2,d_3)$ where $0\le d'_i\le d_i$ for any $i\in[3]$. We claim that, if $(d_1,d_2,d_3)\in\hat{\mathcal{D}}'_{123}$ then $(d'_1,d'_2,d'_3)\in\hat{\mathcal{D}}'_{123}$. This is true as from $(\eqref{d1bar}-\eqref{d123bar})$, we have
\begin{eqnarray}
(d_1,d_2,d_3)\in\bar{\mathcal{D}}_{123}(\lambda,\lambda',\gamma,\gamma')\Rightarrow(d'_1,d'_2,d'_3)\in\bar{\mathcal{D}}_{123}(\lambda,\lambda',\gamma,\gamma')
\end{eqnarray}
Therefore, as $\hat{\mathcal{D}}'_{123}\define\cup_{\lambda,\lambda',\gamma,\gamma'}\bar{\mathcal{D}}_{123}(\lambda,\lambda',\gamma,\gamma')$ we have
\begin{eqnarray}
&&(d_1,d_2,d_3)\in\hat{\mathcal{D}}'_{123}\Rightarrow \exists (\lambda,\lambda',\gamma,\gamma') \mbox{~s.t.~} (d_1,d_2,d_3)\in\bar{\mathcal{D}}_{123}(\lambda,\lambda',\gamma,\gamma')\label{in+1}\\
&\Rightarrow& (d'_1,d'_2,d'_3)\in\bar{\mathcal{D}}_{123}(\lambda,\lambda',\gamma,\gamma') \Rightarrow(d'_1,d'_2,d'_3)\in \hat{\mathcal{D}}'_{123}\label{in+2}
\end{eqnarray}
Therefore, if $(d_1,d_2,d_3)\in\hat{\mathcal{D}}'_{123}$ then $(d'_1,d'_2,d'_3)\in\hat{\mathcal{D}}'_{123}$.

\item{$d_1=\alpha_{11}$.} Consider the  hyperplane $d_1=\alpha_{11}$ and the set of
all the points contained in this hyperplane which satisfy all the
other inequalities $(\eqref{d1hat}-\eqref{d123hat})$, i.e., 
\begin{eqnarray}
\mathcal{S}_1&=&\{(d_1,d_2,d_3);d_1=\alpha_{11},~0\le d_2\le~ \alpha_{22}-\max_{l,m\in[3],l\neq m}\alpha_{lm},\nonumber\\
&&0\le d_3\le~ \alpha_{33}-\max(\alpha_{23},\alpha_{32},\alpha_{31},\alpha_{13})\}
\end{eqnarray}
Consider the following corner point.
\begin{eqnarray}
A&=&(d_1=\alpha_{11},d_2= \alpha_{22}-\max_{l,m\in[3],l\neq m}\alpha_{lm},d_3=\alpha_{33}-\max(\alpha_{23},\alpha_{32},\alpha_{31},\alpha_{13}))
\end{eqnarray}
$\mathcal{S}_1\subset\hat{\mathcal{D}}'_{123}$ since $A\in\mathcal{\bar{D}}_{123}(\lambda,\lambda',\gamma,\gamma')$  where,\footnote{This is true from \eqref{in+1} and \eqref{in+2}.} 
 \begin{align}
 (\lambda,\lambda',\gamma,\gamma')&=(\max(\alpha_{23},\alpha_{32},\alpha_{31},\alpha_{13}),\max_{l,m\in[3],l\neq m}\alpha_{lm}-\max(\alpha_{23},\alpha_{32},\alpha_{31},\alpha_{13}),0,0)
  \end{align}  
\item{$d_2=\alpha_{22}$.} The hyperplane $d_2=\alpha_{22}$ is represented as,
\begin{eqnarray}
\mathcal{S}_2&=&\bigg\{(d_1,d_2,d_3);d_2=\alpha_{22},~0\le d_1\le~ \alpha_{11}-\max_{l,m\in[3],l\neq m}\alpha_{lm},\nonumber\\
&&0\le d_3\le~ \alpha_{33}-\max(\alpha_{23},\alpha_{32}),d_1+d_3\le\alpha_{11}+\alpha_{33}-\alpha\bigg\}
\end{eqnarray}
where $\alpha$ is equal to
\begin{eqnarray}
\alpha&=&\max \left\{\begin{matrix}
\max_{l,m\in[3],l\neq m}\alpha_{lm}+\max(\alpha_{32},\alpha_{23}),\\ 
\alpha_{13}+\alpha_{21},\\ 
\alpha_{12}+\alpha_{31},\\ 
\alpha_{13}+\alpha_{31}
\end{matrix}\right\}\label{def_alpha}
\end{eqnarray}
In order to show that $\mathcal{S}_2\subset\hat{\mathcal{D}}'_{123}$, it is sufficient to show that the two corner points $B$ and $C$ belong to the set $\hat{\mathcal{D}}'_{123}$ where $B$ and $C$ are equal to,\footnote{This is true from \eqref{in+1} and \eqref{in+2}.}
\begin{eqnarray}
B&=&(d_1=\alpha_{11}-\alpha+\max(\alpha_{23},\alpha_{32}),d_2=\alpha_{22},d_3=\alpha_{33}-\max(\alpha_{23},\alpha_{32}))\\
C&=&(d_1=\alpha_{11}-\max_{l,m\in[3],l\neq m}\alpha_{lm},d_2=\alpha_{22},d_3=\alpha_{33}+\max_{l,m\in[3],l\neq m}\alpha_{lm}-\alpha)
\end{eqnarray}
\begin{enumerate}
 \item{} In order to prove that $B\in\hat{\mathcal{D}}'_{123}$, it is sufficient to show that $B\in\mathcal{\bar{D}}_{123}(\lambda,\lambda',\gamma,\gamma')$ for a quadruple $(\lambda,\lambda',\gamma,\gamma')$ satisfying $(\eqref{llgg1}-\eqref{llgg10})$. Let us show how the variables $(\lambda,\lambda',\gamma,\gamma')$ are derived. First of all, note that as $B\in\mathcal{\bar{D}}_{123}(\lambda,\lambda',\gamma,\gamma')$ and as $(\lambda,\lambda',\gamma,\gamma')$ satisfies $(\eqref{llgg1}-\eqref{llgg10})$, we have
 \begin{eqnarray}
d_2=\alpha_{22},d_3=\alpha_{33}-\max(\alpha_{23},\alpha_{32}),\eqref{d23bar} &\Rightarrow& \lambda\le\max(\alpha_{32},\alpha_{23})\label{kl01}\\
\eqref{llgg7},\eqref{llgg9}&\Rightarrow& \lambda\ge\max(\alpha_{32},\alpha_{23})\label{kl1}\\
d_1+d_2+d_3=\alpha_{11}+\alpha_{22}+\alpha_{33}-\alpha,\eqref{d123bar}&\Rightarrow& 2 \lambda+\lambda'+\gamma+\gamma'\le\alpha\label{kl011}\\
\eqref{llgg4},\eqref{llgg5},\eqref{llgg6},\eqref{llgg7},\eqref{llgg8},\eqref{llgg9},\eqref{def_alpha}&\Rightarrow&2 \lambda+\lambda'+\gamma+\gamma'\ge\alpha\label{kl11}
 \end{eqnarray}
 Therefore, we conclude that
 \begin{eqnarray}
\lambda&=&\max(\alpha_{32},\alpha_{23})\label{kkk1+}\\
2 \lambda+\lambda'+\gamma+\gamma'&=&\alpha\label{kkk2+}
 \end{eqnarray} 
Next, $(\gamma,\gamma',\lambda')$ satisfying $(\eqref{llgg1}-\eqref{llgg10})$, \eqref{kkk1+} and \eqref{kkk2+}  are obtained as follows.
 \begin{enumerate}
 \item{}If $\alpha_{13}+\alpha_{21}=\alpha$, then
 \begin{eqnarray}
 (\gamma,\gamma',\lambda')&=&(\alpha_{13}-\max(\alpha_{32},\alpha_{23}),\alpha_{31}-\max(\alpha_{32},\alpha_{23}),\alpha_{21}-\alpha_{31})\label{kl12}
 \end{eqnarray}
 \item{}If $\alpha_{13}+\alpha_{31}=\alpha$, then
 \begin{eqnarray}
 (\gamma,\gamma',\lambda') &=&(\alpha_{13}-\max(\alpha_{32},\alpha_{23}),\alpha_{31}-\max(\alpha_{32},\alpha_{23}),0)
  \end{eqnarray}
 \item{}If $\alpha_{12}+\alpha_{31}=\alpha$, then
 \begin{eqnarray}
 (\gamma,\gamma',\lambda') &=&(\alpha_{13}-\max(\alpha_{32},\alpha_{23}),\alpha_{31}-\max(\alpha_{32},\alpha_{23}),\alpha_{12}-\alpha_{13})
 \end{eqnarray} 
   \item{}If $\max_{l,m\in[3],l\neq m}\alpha_{lm}+\lambda=\alpha$ and $\max_{l,m\in[3],l\neq m}\alpha_{lm}=\alpha_{12}$, then \begin{eqnarray}
 (\gamma,\gamma',\lambda') &=&(\alpha_{12}-\max(\alpha_{21},\alpha_{32},\alpha_{23}),0,\max(\alpha_{21},\alpha_{32},\alpha_{23})-\lambda)
 \end{eqnarray}
    \item{}If $\max_{l,m\in[3],l\neq m}\alpha_{lm}+\lambda=\alpha$ and $\max_{l,m\in[3],l\neq m}\alpha_{lm}=\alpha_{13}$, then \begin{eqnarray}
  (\gamma,\gamma',\lambda') &=&(\alpha_{13}-\max(\alpha_{32},\alpha_{23}),0,0)
 \end{eqnarray}
     \item{}If $\max_{l,m\in[3],l\neq m}\alpha_{lm}+\lambda=\alpha$ and $\max_{l,m\in[3],l\neq m}\alpha_{lm}=\alpha_{21}$, then \begin{eqnarray}
   (\gamma,\gamma',\lambda') &=&(0,\alpha_{21}-\max(\alpha_{12},\alpha_{23},\alpha_{32}),\max(\alpha_{12},\alpha_{23},\alpha_{32})-\lambda)
 \end{eqnarray}
     \item{}If $\max_{l,m\in[3],l\neq m}\alpha_{lm}+\lambda=\alpha$ and $\max_{l,m\in[3],l\neq m}\alpha_{lm}=\alpha_{31}$, then \begin{eqnarray}
    (\gamma,\gamma',\lambda') &=&(0,\alpha_{31}-\max(\alpha_{32},\alpha_{23}),0)
 \end{eqnarray}
      \item{}If $\max_{l,m\in[3],l\neq m}\alpha_{lm}+\lambda=\alpha$ and $\max_{l,m\in[3],l\neq m}\alpha_{lm}=\max(\alpha_{23},\alpha_{32})$, then \begin{eqnarray}
    (\gamma,\gamma',\lambda') &=&(0,0,0)\label{kl2}
 \end{eqnarray}
  \end{enumerate}
   \item{}  $C\in\hat{\mathcal{D}}'_{123}$ since $C\in\mathcal{\bar{D}}_{123}(\lambda,\lambda',\gamma,\gamma')$ for a quadruple $(\lambda,\lambda',\gamma,\gamma')$ where
\begin{eqnarray}
\lambda&=&\alpha-\max_{l,m\in[3],l\neq m}\alpha_{lm}
 \end{eqnarray}
and $(\gamma,\gamma',\lambda')$ is represented as follows. 
 \begin{enumerate}
 \item{}If $\max_{l,m\in[3],l\neq m}\alpha_{lm}=\alpha_{12}$, then 
 \begin{eqnarray}
 (\gamma,\gamma',\lambda')&=&(\alpha_{12}-\max(\alpha_{21},\alpha_{32},\alpha_{23}),0,\max(\alpha_{21},\alpha_{32},\alpha_{23})-\lambda)
 \end{eqnarray} 
 \item{}If $\max_{l,m\in[3],l\neq m}\alpha_{lm}=\alpha_{21}$, then 
 \begin{eqnarray}
 (\gamma,\gamma',\lambda')&=&(0,\alpha_{21}-\max(\alpha_{12},\alpha_{32},\alpha_{23}),\max(\alpha_{12},\alpha_{32},\alpha_{23})-\lambda)
 \end{eqnarray}  
 \item{}If $\max_{l,m\in[3],l\neq m}\alpha_{lm}=\alpha_{31}$, then 
 \begin{eqnarray}
 (\gamma,\gamma',\lambda')&=&(0,\alpha_{31}-\lambda,0)
 \end{eqnarray}   
 \item{}If $\max_{l,m\in[3],l\neq m}\alpha_{lm}=\alpha_{13}$, then 
 \begin{eqnarray}
 (\gamma,\gamma',\lambda')&=&(\alpha_{13}-\lambda,0,0)
 \end{eqnarray}   
 \item{}If $\max_{l,m\in[3],l\neq m}\alpha_{lm}=\max(\alpha_{23},\alpha_{32})$, then 
 \begin{eqnarray}
 (\gamma,\gamma',\lambda')&=&(0,0,0)
 \end{eqnarray}   
 
  \end{enumerate}
 \end{enumerate}
 \item{$d_3=\alpha_{33}$.} The hyperplane $d_3=\alpha_{33}$ is represented as,
 \begin{eqnarray}
\mathcal{S}_3&=&\Bigg\{(d_1,d_2,d_3):  d_3=\alpha_{33},d_1\le \alpha_{11}-\max(\alpha_{23},\alpha_{32},\alpha_{31},\alpha_{13}),\\
d_2&\le& \alpha_{22}-\max(\alpha_{23},\alpha_{32}),d_1+d_2\le  \alpha_{11}+\alpha_{22}-\alpha\Bigg\}
\end{eqnarray} 
Let us consider the two corner points $D$ and $E$,
\begin{eqnarray}
D&=&(d_1=\alpha_{11}-\max(\alpha_{23},\alpha_{32},\alpha_{31},\alpha_{13}),d_2=\alpha_{22}+\max(\alpha_{23},\alpha_{32},\alpha_{31},\alpha_{13})-\alpha,d_3=\alpha_{33})\nonumber\\
E&=&(d_1=\alpha_{11}-\alpha+\max(\alpha_{23},\alpha_{32}),d_2=\alpha_{22}-\max(\alpha_{23},\alpha_{32}),d_3=\alpha_{33})
\end{eqnarray}
\begin{enumerate}
 \item{}   $D\in\hat{\mathcal{D}}'_{123}$ since $D\in\mathcal{\bar{D}}_{123}(\lambda,\lambda',\gamma,\gamma')$ for the following quadruple $(\lambda,\lambda',\gamma,\gamma')$.
 \begin{enumerate}
 \item{}If $\max(\alpha_{23},\alpha_{32},\alpha_{31},\alpha_{13})=\alpha_{13}$, then 
 \begin{eqnarray}
 (\lambda,\gamma,\gamma',\lambda')&=&(\max(\alpha_{23},\alpha_{32},\alpha_{31}),\alpha_{13}-\lambda,0,\alpha-\alpha_{13}-\lambda)
 \end{eqnarray} 
  \item{}If $\max(\alpha_{23},\alpha_{32},\alpha_{31},\alpha_{13})=\alpha_{31}$, then 
 \begin{eqnarray}
 (\lambda,\gamma,\gamma',\lambda')&=&(\max(\alpha_{23},\alpha_{32},\alpha_{13}),0,\alpha_{31}-\lambda,\alpha-\alpha_{31}-\lambda)
 \end{eqnarray} 
   \item{}If $\max(\alpha_{23},\alpha_{32},\alpha_{31},\alpha_{13})=\max(\alpha_{23},\alpha_{32})$, then 
 \begin{eqnarray}
 (\lambda,\gamma,\gamma',\lambda')&=&(\max(\alpha_{23},\alpha_{32}),0,0,\alpha-2\lambda)
 \end{eqnarray} 
 Note that, $\max_{l,m\in[3],l\neq m}\alpha_{lm}\le\min(\alpha_{11},\alpha_{22})$ is assumed in \ref{ach1}.
 \end{enumerate}
 \item{} $E\in\hat{\mathcal{D}}'_{123}$  since $E\in\mathcal{\bar{D}}_{123}(\lambda,\lambda',\gamma,\gamma')$ for the quadruple $(\lambda,\lambda',\gamma,\gamma')$  given in $(\eqref{kl12}-\eqref{kl2})$.
 \end{enumerate}
\item{} Consider the point $F$ obtained from the intersection of the following three facets.
\begin{eqnarray}
d_1+d_2&=& \alpha_{11}+\alpha_{22}-\max_{l,m\in[3],l\neq m}\alpha_{lm},\\
d_1+d_3&=& \alpha_{11}+\alpha_{33}-\max(\alpha_{23},\alpha_{32},\alpha_{31},\alpha_{13}),\\
d_1+d_2+d_3&=& \alpha_{11}+\alpha_{22}+\alpha_{33}- \alpha
\end{eqnarray}
$F\in\hat{\mathcal{D}}'_{123}$ as  $F\in\mathcal{\bar{D}}_{123}(\lambda,\lambda',\gamma,\gamma')$ for the following quadruple $(\lambda,\lambda',\gamma,\gamma')$.
\begin{eqnarray}
 \lambda&=&\alpha-\max_{l,m\in[3],l\neq m}\alpha_{lm}\\
  \lambda'&=&\max_{l,m\in[3],l\neq m}\alpha_{lm}-\max(\alpha_{23},\alpha_{32},\alpha_{31},\alpha_{13})\\
\gamma&=&\max(\alpha_{13}-\lambda,\alpha_{12}-\lambda-\lambda',0)\\
\gamma'&=&\max_{l,m\in[3],l\neq m}\alpha_{lm}+\max(\alpha_{23},\alpha_{32},\alpha_{31},\alpha_{13})-\alpha-\gamma
 \end{eqnarray}
 \item{}Trivially, the corner point $(0,0,0)\in\mathcal{\bar{D}}_{123}(\lambda,\lambda',\gamma,\gamma')$ by choosing $\lambda=\lambda'=\gamma=\gamma'=\max_{i,j\in[3]}\alpha_{ij}$. 
 \item{}Note that all the corner points are already considered in the previous cases.\footnote{We are considering the points not considered in the previous cases. Thus, we assume that
\begin{eqnarray}
d_1&<& \alpha_{11}\\
d_2&<& \alpha_{22}\\
d_3&<& \alpha_{33}
\end{eqnarray}} For instance, consider the point $s$ obtained from the intersection of the following three facets.
\begin{eqnarray}
d_1+d_2&=& \alpha_{11}+\alpha_{22}-\max_{l,m\in[3],l\neq m}\alpha_{lm}\label{xy1}\\
d_2+d_3&=& \alpha_{22}+\alpha_{33}-\max(\alpha_{23},\alpha_{32})\label{xy2}\\
d_1+d_2+d_3&=&  \alpha_{11}+\alpha_{22}+\alpha_{33}-\max \left\{\begin{matrix}
\max_{l,m\in[3],l\neq m}\alpha_{lm}+\max(\alpha_{32},\alpha_{23}),\\ 
\alpha_{13}+\alpha_{21},\\ 
\alpha_{12}+\alpha_{31},\\ 
\alpha_{13}+\alpha_{31}
\end{matrix}\right\}\label{xy3}
\end{eqnarray}
From $(\eqref{xy1}-\eqref{xy3})$, we have $\alpha_{22}\le d_2$ which contradicts  the condition $d_2<\alpha_{22}$. Thus, all the corner points  are already considered in the previous cases.
 \end{enumerate}

\section{Achievability of $\mathcal{F}_{123}$}\label{sinr_app2}
As the derivation of $(\eqref{,,1}-\eqref{,,10})$ is similar to the derivation of $(\eqref{llgg1}-\eqref{llgg10})$ in Appedix \ref{sinr_app}, we briefly go over it.
\begin{enumerate}
 \item{} ${X}_{\{1\}},{X}_{\{1,2\}},{X}_{\{1,2,3\}}$ are decoded with successive interference cancellation
 at the first receiver treating ${X}_{\{2\}}$ and ${X}_{\{3\}}$ as noise.
 \begin{enumerate}
 \item{} The SINR for decoding ${X}_{\{1,2,3\}}$ at the first receiver treating the other signals as noise is equal to
\begin{eqnarray}
&&\frac{P^{\alpha_{11}}(1-2P^{-\lambda}){|G_{11}|}^2}{1+P^{\alpha_{11}}P^{-\lambda}{|G_{11}|}^2+P^{\alpha_{12}}{P}^{-\gamma'}P^{-\lambda}{|G_{12}|}^2+P^{\alpha_{13}}P^{-\lambda}{|G_{13}|}^2}\nonumber\\
&\approx& P^{\min(\alpha_{11},\lambda,\lambda+\alpha_{11}+\gamma'-\alpha_{12},\lambda+\alpha_{11}-\alpha_{13})}
\end{eqnarray}
The codeword  ${X}_{\{1,2,3\}}$ which carries $d_{\{1,2,3\}}$ GDoF is  decoded successfully if  
\begin{eqnarray}
d_{\{1,2,3\}}&\leq&\min(\alpha_{11},\lambda,\lambda+\alpha_{11}+\gamma'-\alpha_{12},\lambda+\alpha_{11}-\alpha_{13})\label{+eq:sinr123}
\end{eqnarray}
which is true as we have $\lambda\le\alpha_{11}$ from \eqref{,,1}. From \eqref{+eq:d123} we have $d_{\{1,2,3\}}\leq \lambda$, therefore \eqref{+eq:sinr123} holds and ${X}_{\{1,2,3\}}$ is successfully decoded at Receiver $1$.
 \item{}  After decoding the  messages $\bar{W}_{\{1,2,3\}}$, the first receiver reconstructs the codeword  ${X}_{\{1,2,3\}}$ and subtracts its contribution from the
received signal. The SINR for decoding ${X}_{\{1,2\}}$ at the first receiver treating the other signals as noise is equal to
\begin{eqnarray}
&&\frac{P^{\alpha_{11}}P^{-\lambda}{|G_{11}|}^2}{1+P^{\alpha_{11}}P^{-\lambda-\lambda'}{|G_{11}|}^2+P^{\alpha_{12}}{P}^{-\gamma'}P^{-\lambda-\lambda'}{|G_{12}|}^2+P^{\alpha_{13}}P^{-\lambda}{|G_{13}|}^2}\nonumber\\
&\approx& P^{\min(\lambda',\alpha_{11}-\lambda,\lambda'+\alpha_{11}+\gamma'-\alpha_{12},\alpha_{11}-\alpha_{13})}\label{sinr1}
\end{eqnarray}
 Therefore,  ${X}_{\{1,2\}}$  which carries $d_{\{1,2\}}$ GDoF  is  decoded successfully  if
\begin{eqnarray}
d_{\{1,2\}}&\leq&\min(\lambda',\alpha_{11}-\lambda,\lambda'+\alpha_{11}+\gamma'-\alpha_{12},\alpha_{11}-\alpha_{13})\label{+eq:sinr12}
\end{eqnarray}
which is true from \eqref{+eq:d12}, \eqref{,,1}, \eqref{,,4} and \eqref{,,5}. Therefore, ${X}_{\{1,2\}}$ is successfully decoded at Receiver $1$.
\item{} After decoding the  messages $\bar{W}_{\{1,2\}}$, the first receiver reconstructs the codeword  ${X}_{\{1,2\}}$ and subtracts its contribution from the
received signal. SINR for decoding ${X}_{\{1\}}$ is equal to,
\begin{eqnarray}
&&\frac{P^{\alpha_{11}}P^{-\lambda-\lambda'}{|G_{11}|}^2}{1+P^{\alpha_{12}}{P}^{-\gamma'}P^{-\lambda-\lambda'}{|G_{12}|}^2+P^{\alpha_{13}}P^{-\lambda}{|G_{13}|}^2}\nonumber\\
&\approx& P^{\min(\alpha_{11}-\lambda-\lambda',\alpha_{11}-\alpha_{12}+\gamma',\alpha_{11}-\alpha_{13}-\lambda')}\nonumber
\end{eqnarray}
Therefore,  ${X}_{\{1\}}$  which carries $d_{\{1\}}$ GDoF  is  decoded successfully  if
\begin{eqnarray}
d_{\{1\}}&\leq&\min(\alpha_{11}-\lambda-\lambda',\alpha_{11}-\alpha_{12}+\gamma',\alpha_{11}-\alpha_{13}-\lambda')\label{+eq:sinr1}
\end{eqnarray}
which is true from  \eqref{+eq:d1}, \eqref{,,1}, \eqref{,,4} and \eqref{,,5}. Therefore, ${X}_{\{1\}}$ is successfully decoded at Receiver $1$.
 \end{enumerate}
 \item{} ${X}_{\{2\}},{X}_{\{1,2\}},{X}_{\{1,2,3\}}$ are decoded with successive interference cancellation
at the second receiver treating ${X}_{\{1\}}$ and ${X}_{\{3\}}$ as noise.
\begin{enumerate}\item{} The SINR for decoding ${X}_{\{1,2,3\}}$ at the second receiver treating the other signals as noise is equal to
\begin{eqnarray}
&&\frac{P^{\alpha_{22}}{P}^{-\gamma'}(1-2P^{-\lambda}){|G_{22}|}^2}{1+P^{\alpha_{21}}P^{-\lambda}{|G_{21}|}^2+P^{\alpha_{22}}{P}^{-\gamma'}P^{-\lambda}{|G_{22}|}^2+P^{\alpha_{23}}P^{-\lambda}{|G_{23}|}^2}\nonumber\\
&\approx& P^{\min(\alpha_{22}-\gamma',\lambda,\lambda+\alpha_{22}-\gamma'-\alpha_{21},\lambda+\alpha_{22}-\gamma'-\alpha_{23})}
\end{eqnarray}
Therefore,  ${X}_{\{1,2,3\}}$  which carries ${d}_{\{1,2,3\}}$ GDoF  is  decoded successfully  if
\begin{eqnarray}
d_{\{1,2,3\}}&\leq&\min(\alpha_{22}-\gamma',\lambda,\lambda+\alpha_{22}-\gamma'-\alpha_{21},\lambda+\alpha_{22}-\gamma'-\alpha_{23})
\end{eqnarray}
which is true from  \eqref{+eq:d123}, \eqref{,,2}, \eqref{,,6} and \eqref{,,7}. Therefore, ${X}_{\{1,2,3\}}$ is successfully decoded at Receiver $2$.

 \item{}  After decoding the  messages $\bar{W}_{\{1,2,3\}}$, the second receiver reconstructs the codeword  ${X}_{\{1,2,3\}}$ and subtracts its contribution from the
received signal. The SINR for decoding ${X}_{\{1,2\}}$ at the second receiver treating the other signals as noise is equal to
\begin{eqnarray}
&&\frac{P^{\alpha_{22}}{P}^{-\gamma'}P^{-\lambda}{|G_{22}|}^2}{1+P^{\alpha_{21}}P^{-\lambda-\lambda'}{|G_{21}|}^2+P^{\alpha_{22}}{P}^{-\gamma'}P^{-\lambda-\lambda'}{|G_{22}|}^2+P^{\alpha_{23}}P^{-\lambda}{|G_{23}|}^2}\nonumber\\
&\approx& P^{\min(\lambda',\alpha_{22}-\gamma'-\lambda,\lambda'+\alpha_{22}-\gamma'-\alpha_{21},\alpha_{22}-\gamma'-\alpha_{23})}\label{sinr1}
\end{eqnarray}
 Therefore,  ${X}_{\{1,2\}}$  which carries ${d}_{\{1,2\}}$ GDoF  is  decoded successfully  if
\begin{eqnarray}
d_{\{1,2\}}&\leq&\min(\lambda',\alpha_{22}-\gamma'-\lambda,\lambda'+\alpha_{22}-\gamma'-\alpha_{21},\alpha_{22}-\gamma'-\alpha_{23})
\end{eqnarray}
which is true from  \eqref{+eq:d12}, \eqref{,,2}, \eqref{,,6} and \eqref{,,7}. Therefore, ${X}_{\{1,2\}}$ is successfully decoded at Receiver $2$.

\item{} After decoding the  messages $\bar{W}_{\{1,2\}}$, the second receiver reconstructs the codeword  ${X}_{\{1,2\}}$ and subtracts its contribution from the
received signal. SINR for decoding ${X}_{\{1\}}$ is equal to,
\begin{eqnarray}
&&\frac{P^{\alpha_{22}}{P}^{-\gamma'}P^{-\lambda-\lambda'}{|G_{22}|}^2}{1+P^{\alpha_{21}}P^{-\lambda-\lambda'}{|G_{21}|}^2+P^{\alpha_{23}}P^{-\lambda}{|G_{23}|}^2}\nonumber\\
&\approx& P^{\min(\alpha_{22}-\lambda-\lambda'-\gamma',\alpha_{22}-\alpha_{21}-\gamma',\alpha_{22}-\alpha_{23}-\lambda'-\gamma')}
\end{eqnarray}
 Therefore,  ${X}_{\{2\}}$  which carries ${d}_{\{2\}}$ GDoF  is  decoded successfully  if
\begin{eqnarray}
d_{\{2\}}&\leq&\min(\alpha_{22}-\lambda-\lambda'-\gamma',\alpha_{22}-\alpha_{21}-\gamma',\alpha_{22}-\alpha_{23}-\lambda'-\gamma')
\end{eqnarray}
which is true from  \eqref{+eq:d2}, \eqref{,,2}, \eqref{,,6} and \eqref{,,7}. Therefore, ${X}_{\{2\}}$ is successfully decoded at Receiver $2$.
  \end{enumerate}

  \item{} ${X}_{\{3\}},{X}_{\{1,2,3\}}$ are decoded with successive interference cancellation
 at the third receiver treating ${X}_{\{1\}}$, ${X}_{\{2\}}$ and ${X}_{\{1,2\}}$ as noise.
\begin{enumerate}
\item{} The SINR for decoding ${X}_{\{1,2,3\}}$ at the  third receiver treating the other signals as noise is equal to
\begin{eqnarray}
\frac{P^{\alpha_{33}}(1-2P^{-\lambda}){|G_{33}|}^2}{1+P^{\alpha_{31}}P^{-\lambda}{|G_{31}|}^2+P^{\alpha_{32}}{P}^{-\gamma'}P^{-\lambda}{|G_{32}|}^2+P^{\alpha_{33}}P^{-\lambda}{|G_{33}|}^2}\approx P^{\lambda}\label{+sinr3dd}
\end{eqnarray}
\eqref{+sinr3dd} follows from \eqref{con1}, i.e.,  $\max(\alpha_{im},\alpha_{ki})\le\alpha_{ii}$ is true for all $i,k,m\in[3]$. Therefore,  ${X}_{\{1,2,3\}}$  which carries ${d}_{\{1,2,3\}}$ GDoF  is  decoded successfully as from  \eqref{+eq:d123} we have ${d}_{\{1,2,3\}}\le \lambda$.

\item{} Finally, the third receiver decodes ${X}_{\{3\}}$  treating ${X}_{\{1\}},{X}_{\{2\}}$ as noise with  SINR  equal to,
\begin{eqnarray}
\frac{P^{\alpha_{33}}P^{-\lambda}{|G_{33}|}^2}{1+P^{\alpha_{31}}P^{-\lambda}{|G_{31}|}^2+P^{\alpha_{32}}{P}^{-\gamma'}P^{-\lambda}{|G_{32}|}^2}&\approx& P^{\min(\alpha_{33}-\lambda,\alpha_{33}-\alpha_{31},\alpha_{33}-\alpha_{32}+\gamma')}
\end{eqnarray}
 Therefore,  ${X}_{\{3\}}$  which carries ${d}_{\{3\}}$ GDoF  is  decoded successfully  if
\begin{eqnarray}
d_{\{3\}}&\leq&\min(\alpha_{33}-\lambda,\alpha_{33}-\alpha_{31},\alpha_{33}-\alpha_{32}+\gamma')
\end{eqnarray}
which is true from  \eqref{+eq:d3}, \eqref{,,8} and \eqref{,,9}. Therefore, ${X}_{\{3\}}$ is successfully decoded at Receiver $3$.
\end{enumerate}
\end{enumerate}

\section{Convex Polyhedron}\label{convexpolyhedron}
\begin{lemma}\label{lemmaconvex}
Consider a compact convex polyhedron $A$ and a convex set $B$. Define $\mathcal{U}$ as the set of all vertices of $A$. Then, the following statement is true.\\

$A\subset B$ if and only if $\mathcal{U}\subset B$.
\end{lemma}
\subsection{Proof of Lemma \ref{lemmaconvex}}
If $A\subset B$, then $\mathcal{U}\subset B$ as $\mathcal{U}\subset A$. So, let us prove the converse part i.e., $A\subset B$ if $\mathcal{U}\subset B$. Note that, $A$ is the convex hull of $\mathcal{U}$ as it is a compact convex polyhedron. On the other hand, the convex hull of a given set $\mathcal{U}$  is defined as the set of all convex combinations of points in $\mathcal{U}$ (the union of all simplices with points in $\mathcal{U}$). Consider $m\in A$. Let us prove that $m\in B$. As $m\in A$, there exist the coefficients $0\le c_v$ where
\begin{eqnarray}
m&=&\sum_{v\in \mathcal{U}}c_v v \mbox{~~~~s.t.~~~~} \sum_{v\in \mathcal{U}}c_v=1
\end{eqnarray}
As $\mathcal{U}\subset B$, we infer that $\sum_{v\in \mathcal{U}}c_v v\in B$ as $B$ is a convex set. Thus, $m\in B$ is concluded.

\bibliographystyle{IEEEtran}
\bibliography{Thesis}
\end{document}